\documentclass[12pt,a4paper,english,amssymb,nofootinbib,superscriptaddress]{revtex4}
\usepackage{lmodern}
\usepackage[T1]{fontenc}
\usepackage[latin9]{inputenc}
\usepackage{color}
\usepackage{babel}
\usepackage{amsmath}
\usepackage{amssymb}
\usepackage{esint}
\usepackage[unicode=true,pdfusetitle,
 bookmarks=true,bookmarksnumbered=true,bookmarksopen=false,
 breaklinks=false,pdfborder={0 0 1},backref=false,colorlinks=true]
 {hyperref}

\makeatletter

\pdfpageheight\paperheight
\pdfpagewidth\paperwidth

\@ifundefined{textcolor}{}
{%
 \definecolor{BLACK}{gray}{0}
 \definecolor{WHITE}{gray}{1}
 \definecolor{RED}{rgb}{1,0,0}
 \definecolor{GREEN}{rgb}{0,1,0}
 \definecolor{BLUE}{rgb}{0,0,1}
 \definecolor{CYAN}{cmyk}{1,0,0,0}
 \definecolor{MAGENTA}{cmyk}{0,1,0,0}
 \definecolor{YELLOW}{cmyk}{0,0,1,0}
}

\hypersetup{linkcolor=blue,citecolor=blue,urlcolor=blue}

\@ifundefined{definecolor}{\usepackage{color}}{}
\usepackage{babel}

\usepackage{empheq}

\newcommand*\widefbox[1]{\fbox{\hspace{2em}#1\hspace{2em}}}

\usepackage{latexsym}\usepackage{bm}

\makeatother

\begin{document}

\title{Born-Infeld Gravity with a Massless Graviton in Four Dimensions}

\author{\.{I}brahim G\"{u}ll\"{u} }

\email{ibrahimgullu2002@gmail.com}

\affiliation{Department of Physics,\\
 Middle East Technical University, 06800, Ankara, Turkey}

\author{Tahsin \c{C}a\u{g}r\i\, \c{S}i\c{s}man}

\email{tahsin.c.sisman@gmail.com}

\affiliation{Centro de Estudios Cient\'ificos (CECS), Casilla 1469, Valdivia,
Chile}

\affiliation{Department of Astronautical Engineering,\\
University of Turkish Aeronautical Association, 06790, Ankara, Turkey}

\author{Bayram Tekin}

\email{btekin@metu.edu.tr}

\affiliation{Department of Physics,\\
 Middle East Technical University, 06800, Ankara, Turkey}

\date{\today}
\begin{abstract}
We construct Born-Infeld (BI) type gravity theories which describe
tree-level unitary (non-ghost and non-tachyonic) massless spin-2 modes
around their maximally symmetric vacua in four dimensions. Building
unitary BI actions around flat vacuum is straightforward; but, this
is a complicated task around (anti)-de Sitter backgrounds. In this
work, we solve the issue and give details of constructing perturbatively
viable determinantal BI theories. It is interesting that the Gauss-Bonnet
combination, which is a total derivative in four dimensions, plays
an important role in the construction of viable BI theories. \tableofcontents{}
\end{abstract}
\maketitle

\section{Introduction}

It is well known that Einstein's gravity, otherwise an excellent theory
in the ``intermediate'' scales, needs to be modified both at large
and small scales. At large distances, the galaxy rotation curves and
the accelerated expansion of the universe are somewhat urgent problems
which could be provisionally solved by keeping Einstein's theory intact
but introducing large amounts of dark matter and dark energy, or alternatively
by modifying gravity at long distances such as upgrading it to massive
gravity. {[}Of course, it could happen that one may need to both modify
gravity and also add dark matter and dark energy to solve these long
distance problems. This possibility should not be ruled out.{]} At
small scales or at high energies, the problem is more complicated,
even if phenomenologically less urgent, due to lack of data. It is
more complicated because what one really needs is a quantum theory
of gravity whose basic degrees of freedom, symmetries, and even principles
as applied to spacetime are unknown. Namely, the geometric nature
of spacetime, even its number of dimensions at small distances is
not clear. Even though there are candidates such as string theory,
loop quantum gravity, or asymptotically safe theories, it is fair
to say that we are still far away from a consistent theory of quantum
gravity. 

In the absence of guiding principles for a renormalizable theory of
gravity, one is forced to introduce effective theories which work
better than Einstein's gravity at small distances and hopefully also
at large distances. One such attempt is, emulating pre-quantum-electrodynamics
era electromagnetism, to write Born-Infeld (BI) type gravity theories
\cite{Deser_Gibbons} which were inspired by the work of Eddington
\cite{Eddington} who used the idea of ``generalized volume'' suggested
actions of the form 
\begin{equation}
I=\int d^{4}x\sqrt{\det\; R_{\mu\nu}\left(\Gamma\right)},
\end{equation}
and assumed the metric and the connection to be independent variables.
{[}A note about history: Eddington's work in gravity precedes the
works of Born and Infeld \cite{BI} in electrodynamics, but it is
actually difficult to find this action in Eddington's book in one
compact form even though the discussion is scattered in the book.
Schrödinger attributes this theory to Eddington on page 113 of his
book \cite{Schrodinger}.{]} After all, good ideas never disappear:
Eddington's idea was resuscitated recently in a number of works \cite{Banados_Eddington,Fiorini,Delsate_Steinhoff,Ferreira,Heisenberg}
which led to interesting results such as singularity free cosmology. 

In analogy with the \emph{minimal} electromagnetic BI theory, in this
current work, we shall take the more conventional path of assuming
the metric to be the only independent variable following Deser and
Gibbons \cite{Deser_Gibbons} who gave a jump-start to the BI gravity
theories. This line of reasoning recently \cite{Gullu-BINMG,Gullu-cfunc,Gullu-UniBI,Gullu-Horava}
bore much fruit in the lower dimensional setting  where we have found
a BI type action which reads 
\begin{equation}
I_{\text{BINMG}}=-\frac{4m^{2}}{\kappa^{2}}\int d^{3}x\,\left[\sqrt{-\det\left(g_{\mu\nu}-\frac{1}{m^{2}}G_{\mu\nu}\right)}-\left(\frac{\Lambda_{0}}{2m^{2}}+1\right)\sqrt{-\det g}\right],\label{eq:CBI_action}
\end{equation}
where $G_{\mu\nu}$ is the Einstein tensor without a cosmological
constant. This theory is called the Born-Infeld New Massive Gravity
(BINMG) theory with the following remarkable properties:
\begin{enumerate}
\item For $\Lambda_{0}\ne0$, unlike any generic finite order theory besides
the cosmological Einstein's theory, it has a unique maximally symmetric
vacuum with an effective cosmological constant $\Lambda=\Lambda_{0}\left(1+\frac{\Lambda_{0}}{4m^{2}}\right)$
for $\Lambda_{0}>-2m^{2}$ \cite{Nam,Gullu-cfunc}. Flat space is
the unique vacuum when $\Lambda_{0}=0$.
\item It has a unitary spin-2 massive degree of freedom with $M^{2}=m^{2}+\Lambda$
about the flat ($\Lambda=0$) and AdS backgrounds. This provides an
infinite order extension of the quadratic NMG%
\footnote{There are in fact two extensions of NMG, the second one being 
\[
I=-\frac{4m^{2}}{\kappa^{2}}\int d^{3}x\,\left\{ \sqrt{-\det\left[g_{\mu\nu}+\frac{1}{m^{2}}\left(R_{\mu\nu}-\frac{1}{6}g_{\mu\nu}R\right)\right]}-\left(\frac{\Lambda_{0}}{2m^{2}}+1\right)\sqrt{-\det g}\right\} ,
\]
which has the same perturbative properties. It was conjectured in
\cite{Jatkar} that this action could appear as a $\text{dS}_{4}$
counterterm. This theory is yet to be explored further.%
} \cite{BHT-PRL,BHT-PRD,Gullu-Canonical}.
\item It reproduces, up to desired order in the curvature expansion, the
extended NMG theories that are consistent with the AdS/CFT duality
and that have a $c$-function \cite{Sinha,Gullu-cfunc,Paulos}.
\item The BINMG action appears as a counterterm in $\text{AdS}{}_{4}$ \cite{Jatkar}.
\end{enumerate}
In addition to these properties, the existence of a supersymmetric
extension to the cubic order truncation of the theory suggests that
a supersymmetric extension presumably exists for the full theory \cite{Ozkan}.

All these virtues of the three-dimensional BI gravity led us to search
for similar theories beyond three dimensions and especially in the
more relevant $3+1$ dimensions. In this work, this is the task that
we take on. Some of our computations, especially in the context of
general formalism, will be in generic $n$ dimensions, but in most
of the current work we shall focus on $n=3+1$ dimensions since it
has rather distinctive features compared to the $n>4$ cases. The
generic $n$ dimensional theory is somewhat more complicated and deserves
a separate attention \cite{Gullu_n}. As we shall see, the set of
viable Lagrangians is larger in four dimensions compared to the three
dimensional case, where there are only two theories as mentioned above,
since vanishing of the Weyl tensor and the linear theory having no
propagating degrees of freedom by itself in three dimensions make
$n=2+1$ rather simple and unique. One should not expect such a simplicity
in four dimensions and beyond.\textcolor{red}{{} } 

In constructing viable BI-type gravity theories, the important point
is to find the physical constraints that one imposes on the theory.
Here, the constraints we shall assume are:
\begin{enumerate}
\item In small curvature expansion, the theory at the lowest order reduces
to (cosmological) Einstein's gravity,
\item The theory admits flat or (A)dS vacuum,
\item The theory describes only massless spin-2 excitations around its flat
or the (A)dS vacuum, and these excitations are non-ghost and non-tachyonic
(namely, the theory is tree-level unitary) as a full theory (thus,
infinitely many terms in the curvature expansion contribute to the
propagator of the theory).
\item On the top of the previous condition, the theory is tree-level unitary
at any finite \emph{truncated} order in the curvature expansion. 
\end{enumerate}
Let us briefly explain why these conditions are imperative for a healthy
theory. We require that in small curvature expansion the theory reduces
to the (cosmological) Einstein theory which is a natural condition
to reproduce the plethora of data explained by Einstein's theory.
The second requirement is sort of self-explanatory since one needs
a maximally symmetric vacuum with vanishing conserved quantities such
as energy and angular momentum. The third requirement is also somewhat
obvious both in the context of the stability of the vacuum and perturbative
viability of the quantum version of the theory. The fourth condition
is extremely difficult to satisfy in the curvature expansion for (A)dS
backgrounds. Observe that we require not only that the theory is tree-level
unitary as a whole (condition 3); but, it is also tree-level unitary
at any truncated order in the curvature expansion (condition 4).\textcolor{red}{{}
}To the best of our knowledge, the theories that we shall present
are the only ones that satisfy this requirement in four dimensions.
{[}There are Lovelock theories \cite{Lovelock} in higher dimensions
that also satisfy this requirement; but, they reduce to Einstein's
theory in four dimensions.{]} Lastly, note that viable BI-type gravity
theories were also studied in \cite{Comelli} where similar constraints
were required; however, the additional massive spin-0 degree of freedom
was allowed besides massless spin-2 mode and the unitarity of finite
truncations were not considered.

As we shall find out, these conditions still leave a large set of
viable theories. Of course, one can additionally impose that there
be no dimensionless or dimensionful parameters save, the Newton's
and, perhaps, the BI parameter, which highly constrains the viable
theories.\textcolor{red}{{} }As we shall see, the most ``minimal''
BI theory also has a unique vacuum. This is actually quite important,
since, once Einstein's theory is augmented with additional powers
of curvature, immediately one undesired feature arises that is the
non-uniqueness of the maximally symmetric vacuum. Since, asymptotic
structures of spacetimes with different cosmological constants are
different, their energy properties are not comparable. Therefore,
there is no way to choose one vacuum over the other if there are more
than one viable vacua. Hence, it would be highly desirable to have
a theory with a unique vacuum.\textcolor{red}{{} }

To see that four dimensional BI theories are somewhat special, let
us start with the following $n$ dimensional generic action 
\begin{equation}
I=\frac{2}{\kappa\gamma}\int d^{n}x\,\left[\sqrt{-\det\left(g_{\mu\nu}+\gamma A_{\mu\nu}\right)}-\left(\gamma\Lambda_{0}+1\right)\sqrt{-\det g}\right],\label{eq:Generic_BI_intro}
\end{equation}
where $\kappa$ is the modified Newton's constant which in four dimensions
reads $\kappa=16\pi G$ and $G$ is the Newton's constant and $\gamma$
is a dimensionful BI parameter with mass dimension $-2$ in four dimensions.
To stick to the idea of obtaining \emph{minimal} theories, we will
find the simplest two tensor $A_{\mu\nu}$ which does not have derivatives
of the Riemann tensor and which has as small powers and contractions
of the Riemann tensor as possible. The most naive approach would be
to take $A_{\mu\nu}=G_{\mu\nu}+\beta g_{\mu\nu}R$, similar to the
$2+1$ dimensional case. As shown in Appendix-\ref{sec:Naive-BI-Gravity},
upon small curvature expansion, this theory will generate quadratic
terms which have massless spin-2, massive spin-0 and massive spin-2
modes; the last one being a ghost even around flat spacetime.%
\footnote{Nevertheless, it is remarkable that instead of the square root Lagrangian
if one considers a different power in $n$ dimensions $\left[\det\left(g+\gamma G\right)\right]^{1/\left(n-1\right)}$
then one has a massive gravity theory without the nonlinear Boulware-Deser
ghost \cite{Yi,Paulos_Tolley}.%
} Hence, this too optimistic guess does not lead to a perturbatively
viable theory. In the small curvature expansion ($|\gamma A_{\mu\nu}|\ll1$)
of the action (\ref{eq:Generic_BI_intro}), either quadratic terms
must be eliminated or they must appear in the benign Gauss-Bonnet
combination to get rid off the massive modes; therefore, in four dimensions
and beyond, to built viable BI-gravity theories, one has to take $A_{\mu\nu}$
to be up to \emph{at least quadratic} order in the curvature, which
of course leads to an eight order theory in the curvature under the
square root when the determinant is explicitly written in terms of
the traces.\textcolor{red}{{} }

Upon inspection, one can see that the most general two-tensor up to
and including quadratic order can be written as 
\begin{align}
A_{\mu\nu}= & R_{\mu\nu}+\beta S_{\mu\nu}\nonumber \\
 & +\gamma\left(a_{1}C_{\mu\rho\sigma\lambda}C_{\nu}^{\phantom{\nu}\rho\sigma\lambda}+a_{2}C_{\mu\rho\nu\sigma}R^{\rho\sigma}+a_{3}R_{\mu\rho}R_{\nu}^{\rho}+a_{4}S_{\mu\rho}S_{\nu}^{\rho}\right)\nonumber \\
 & +\frac{\gamma}{n}g_{\mu\nu}\left(b_{1}C_{\rho\sigma\lambda\gamma}C^{\rho\sigma\lambda\gamma}+b_{2}R_{\rho\sigma}R^{\rho\sigma}+b_{3}S_{\rho\sigma}S^{\rho\sigma}\right),\label{eq:Generic_Amn}
\end{align}
where $S_{\mu\nu}\equiv R_{\mu\nu}-\frac{1}{n}g_{\mu\nu}R$ is the
traceless-Ricci tensor and $C_{\mu\alpha\nu\beta}$ is the Weyl tensor,
and $\beta$, $a_{i}$, and $b_{i}$ are dimensionless constants.
Observe that there is no $R_{\mu\nu}S^{\mu\nu}$ cross term because
of the following relation 
\begin{equation}
R_{\mu\rho}S_{\nu}^{\rho}=\frac{1}{2}R_{\mu\rho}R_{\nu}^{\rho}+\frac{1}{2}S_{\mu\rho}S_{\nu}^{\rho}-\frac{1}{2n}g_{\mu\nu}\left(R_{\rho\sigma}R^{\rho\sigma}-S_{\rho\sigma}S^{\rho\sigma}\right).
\end{equation}
Suppose $\bar{g}_{\mu\nu}$ is a maximally symmetric vacuum of the
theory and we would like to study excitations ($h_{\mu\nu}$) about
this vacuum. If this vacuum is flat, then our task is easy since all
we need is to expand the action up to quadratic order in the curvature,
then expand the resultant action up to $O\left(h_{\mu\nu}^{2}\right)$
and check the propagating modes in the theory. But, if this vacuum
is an (A)dS space, then in principle all the terms in the curvature
expansion contribute to the free theory (that is the vacuum and the
$O\left(h_{\mu\nu}^{2}\right)$ theory), and hence one has a highly
complicated task. Therefore, in building viable BI actions for gravity
the main hurdle is to satisfy the tree-level unitarity around non-flat
maximally symmetric backgrounds; but, fortunately we have built the
necessary formalism to carry out this task in \cite{Gullu-UniBI}.
{[}PhD thesis \cite{Sisman-Thesis} of one of the authors is devoted
to these issues and expounds upon many of the discussions in the published
papers.{]}

It was shown in these works (and we shall give another argument in
this paper) that in \emph{four dimensions} no terms beyond $O\left(A_{\mu\nu}^{2}\right)$
expansion around $A_{\mu\nu}=0$ contribute to the free theory, namely
the vacuum and the excitations. Therefore, to study the excitations
of (\ref{eq:Generic_BI_intro}) about its maximally symmetric vacua
in four dimensions, all one needs to study is the following theory\begin{empheq}[box=\widefbox]{equation}
I = \frac{1}{\kappa\gamma} \int d^{4}x \, \sqrt{-g} \left[ A - 2 \gamma \Lambda_{0} + \frac{1}{4} A^{2} - \frac{1}{2} A_{\mu\nu} A^{\mu\nu} \right], \label{eq:OA2_expansion}
\end{empheq}with $A\equiv A_{\mu}^{\mu}$ and note that this is a fourth order
theory in the curvature. For generic even $n$ dimensions, one needs
to expand up to $O\left(A_{\mu\nu}^{n/2}\right)$, and for odd dimensions,
all the powers contribute. In four dimensions, because of the identity
\begin{equation}
C_{\mu\rho\sigma\lambda}C_{\nu}^{\phantom{\nu}\rho\sigma\lambda}=\frac{1}{4}g_{\mu\nu}C_{\alpha\rho\sigma\lambda}C^{\alpha\rho\sigma\lambda},
\end{equation}
we can also eliminate $a_{1}$ or $b_{1}$, without loss of generality,
we choose $a_{1}=0$. Note that instead of this basis (namely the
Weyl, Ricci and traceless-Ricci tensors), one can use the Riemann,
Ricci tensors and the scalar curvature which we do in Appendix-\ref{sec:Conversions-Between-Basis}
for the purpose of comparison. There, we also give formulas relating
one basis to the other. The $A_{\mu\nu}$ tensor with these seven
dimensionless parameters looks cumbersome; but, in what follows unitarity
of theory with only massless spin-2 excitations about the (A)dS vacua
will eliminate three (or four depending on the theory) of these parameters,
and in addition, conforming to the notion of minimality will lead
to a theory without free dimensionless parameters. Note that, we do
not count the dimensionful BI parameter $\gamma$ which can be constrained
by experiments: As long as $\gamma R$ is small, any $\gamma$ is
viable in our analysis. Hence, it should be considered as a new dimensionful
parameter. Of course, not to introduce a new dimensionful parameter,
one can choose $\gamma=\kappa$ since they are of the same dimensions.%
\footnote{In Born-Infeld electrodynamics, one necessarily introduces a dimensionful
BI parameter; but, in BI gravity one can simply recycle Newton's constant
and no new parameter is introduced.%
} {[}Observe that, since $\kappa=4\pi\ell_{p}^{2}$ with $\ell_{p}$
being the Planck length, the condition $\kappa R\ll1$ is satisfied
as long as we are far away from the Planck regime: $R\ll\frac{1}{\ell_{p}^{2}}$.{]}

The layout of the paper is as follows: In Section II, we recall that
the ``free theory'' of BI gravity should be the same as the free
theory of Einstein--Gauss-Bonnet theory that describes unitary massless
spin-2 excitations around flat and (A)dS spaces. In Section III, we
give details of finding the maximally symmetric vacua of generic gravity
theories including the BI gravity with the help of equivalent linear
actions which circumvent the complicated task of deriving the field
equations. In that section, we also derive the equivalent quadratic
curvature action that has the same free theory including the vacuum
of the original generic gravity, specifically the BI theory. In Section
IV, we determine the vacua of the BI gravity. In Section V, we impose
that the BI gravity describes unitary massless spin-2 gravitons around
its flat background. In Section VI, we study the unitarity of the
BI theory around its unique viable (A)dS vacuum and impose the condition
that only massless spin-2 particle is allowed. In the Appendices,
we give details of the computations relevant to the results in the
text.

\section{Constructing the Born-Infeld Action}

The most general quadratic theory in $n$ dimensions that describes
\emph{only} massless spin two excitations around its flat or (A)dS
vacuum is the Einstein--Gauss-Bonnet (EGB) theory with the Lagrangian
\begin{align}
\mathcal{L} & =\frac{1}{\kappa}\left(R-2\Lambda_{0}+\gamma\chi_{\text{GB}}\right),
\end{align}
where the GB combination is given as 
\begin{equation}
\chi_{\text{GB}}\equiv R^{\mu\nu\rho\sigma}R_{\mu\nu\rho\sigma}-4R^{\mu\nu}R_{\mu\nu}+R^{2}.
\end{equation}
In four dimensions, the GB part is a total derivative, and hence does
not contribute to the field equations and plays no role in the particle
spectrum or the vacuum of the theory. But as we shall see here, it
plays a major role in constructing BI type actions: Namely, we will
see that at the quadratic level BI gravity reduces to the EGB theory
instead of the Einstein's theory even though classically they are
equivalent. In some sense, the dimensionful parameter $\gamma$ in
front of the GB term, plays the role of the BI parameter.

It is clear that flat space is a vacuum for $\Lambda_{0}=0$ and if
$\Lambda_{0}\ne0$ (A)dS is the vacuum with $\Lambda=\Lambda_{0}$.
In the basis discussed in the Introduction, we can recast the EGB
action as 
\begin{align}
\mathcal{L} & =\frac{1}{\kappa}\left[R-2\Lambda_{0}+\gamma\left(C^{\mu\nu\rho\sigma}C_{\mu\nu\rho\sigma}+\frac{2}{3}R^{\mu\nu}R_{\mu\nu}-\frac{8}{3}S^{\mu\nu}S_{\mu\nu}\right)\right],\label{GB_combination}
\end{align}
where we have used the \emph{four} dimensional identity 

\begin{equation}
C^{\mu\nu\rho\sigma}C_{\mu\nu\rho\sigma}=R^{\mu\nu\rho\sigma}R_{\mu\nu\rho\sigma}-2R^{\mu\nu}R_{\mu\nu}+\frac{1}{3}R^{2}.
\end{equation}
It is not difficult to see that (\ref{GB_combination}) describes
only massless spin two excitations in flat and AdS vacua. There are
many ways to show this but because this is almost common knowledge
let us briefly sketch the proof without going into further details:
Linearization of the field equations derived from (\ref{GB_combination})
about its (A)dS vacuum yields 
\begin{align}
\frac{1}{\kappa}\mathcal{G}_{\mu\nu}= & 0,\label{linearized_EGB}
\end{align}
where $\mathcal{G_{\mu\nu}}$ is the linearized Einstein tensor, which
in the transverse-traceless gauge for perturbations $h_{\mu\nu}=g_{\mu\nu}-\bar{g}_{\mu\nu}$,
reads 
\begin{align}
\frac{1}{\kappa}\mathcal{G_{\mu\nu}}= & -\frac{1}{2\kappa}\left(\bar{\square}-\frac{2\Lambda_{0}}{3}\right)h_{\mu\nu}=0.\label{linearized_EGB_2}
\end{align}
Despite the appearance of a mass-like term, (\ref{linearized_EGB_2})
together with the transversality and the tracelessness conditions,
describe a massless spin two excitation. This can be easily seen if
one writes the AdS metric in its conformal to flat coordinates $\bar{g}_{\mu\nu}=\Omega^{2}\eta_{\mu\nu}$
with $\Omega=\left(1-\frac{3}{4}\Lambda_{0}x^{2}\right)^{-1}$ where
$x^{2}=\eta_{\mu\nu}x^{\mu}x^{\nu}$ which reduces (\ref{linearized_EGB_2})
to the massless wave equation in flat space $\partial^{2}h_{\mu\nu}=0$
\cite{Deser-Nepo}. For this massless spin-2 excitation to be unitary,
there is only one condition that is the positivity of the Newton's
constant, $\kappa>0$, namely gravity is attractive or similarly massless
spin-2 field has a positive kinetic energy as seen from the action
{[}in the mostly positive signature convention{]} 
\begin{equation}
I=\int d^{4}x\sqrt{-\bar{g}}\left\{ \frac{1}{4\kappa}h^{\mu\nu}\left(\bar{\square}-\frac{2\Lambda_{0}}{3}\right)h_{\mu\nu}+h^{\mu\nu}T_{\mu\nu}\right\} .
\end{equation}
If one requires a generic gravity theory of the form $\mathcal{L}\equiv\sqrt{-g}\, f\left(R_{\rho\sigma}^{\mu\nu}\right)$,
namely with a Lagrangian density built from arbitrary powers of the
Riemann tensor and its contractions but not its derivatives to propagate
only unitary massless spin two excitations, it should have the same
propagator structure as the EGB theory (or equivalently the cosmological
Einstein's theory). It appears to be a highly cumbersome task to find
the propagator of a generic gravity theory or a BI type gravity theory
in constant curvature backgrounds because in principle infinitely
many terms contribute to the propagator. Fortunately, there is a highly
useful shortcut which works by constructing an \emph{equivalent quadratic
curvature action} that has the same propagator structure and the vacua
as the generic theory under study. We work this out in the next section.

\section{Equivalent Linear Action And Equivalent Quadratic Curvature Action\label{sec:Equivalent-Linear-Action}}

The first step in finding the particle spectrum of a given gravity
theory about its maximally symmetric vacuum is to show that the theory
in fact admits such a vacuum and if it does admit such a solution,
one must find the effective cosmological constant of the vacuum. The
most direct way to find the maximally symmetric vacuum is to derive
the field equations first and then solve these equations. But, for
the determinantal actions of the form that we study in this work or
for higher derivative theories with many powers of curvature, finding
the field equations is by itself a difficult task. As a demonstration
of the complication, the reader could check the field equations of
the action we study in this work in Appendix-\ref{sec:Field-Equations}.

In this section, we shall give a method to find the maximally symmetric
vacuum or vacua of a given theory which circumvents the procedure
of deriving the field equations. The method involves constructing
an equivalent linear action (ELA) that has the same vacuum or vacua
as the original action and it is so powerful that it pays to lay out
some details here.

Consider a generic action of the form%
\footnote{Of course one can work in the basis introduced in the Introduction,
but here we shall work with the Riemann tensor, as this basis is more
common in many other applications.%
} 
\begin{equation}
I=\int d^{n}x\,\sqrt{-g}f\left(g^{\alpha\beta},\, R_{\phantom{\mu}\nu\rho\sigma}^{\mu},\,\nabla_{\rho}R_{\phantom{\mu}\nu\rho\sigma}^{\mu},\dots,\,\nabla_{\rho_{1}}\nabla_{\rho_{2}}\dots\nabla_{\rho_{m}}R_{\phantom{\mu}\nu\rho\sigma}^{\mu}\right),\label{general_theory}
\end{equation}
for which we ask if it admits a maximally symmetric vacuum and if
it does so what is the effective cosmological constant? Here we shall
work in generic $n$ dimensions. At this stage, it is clear that the
derivative terms will not contribute to the maximally symmetric vacuum
since they will yield covariant derivatives of the metric tensor at
the level of the field equations which vanish by metric compatibility.
Therefore, for notational simplicity let us denote the action as 
\begin{equation}
I=\int d^{n}x\,\sqrt{-g}f\left(R_{\alpha\beta}^{\mu\nu}\right),\label{f_gen}
\end{equation}
where we have also gotten rid off the inverse metric without loss
of generality and taken the independent variable to be $R_{\alpha\beta}^{\mu\nu}$
that could stand for the Riemann tensor, or if once contracted to
the Ricci tensor, and if twice contracted to the scalar curvature.
For example, the Einstein-Hilbert action in this language reads $\int d^{n}x\,\sqrt{-g}\delta_{\mu}^{\alpha}\delta_{\nu}^{\beta}R_{\alpha\beta}^{\mu\nu}$.
To find the field equations for the maximally symmetric spacetime,
one varies the action as
\begin{equation}
\delta I=\int d^{n}x\,\left(\delta\sqrt{-g}f\left(R_{\alpha\beta}^{\mu\nu}\right)+\sqrt{-g}\frac{\partial f}{\partial R_{\rho\sigma}^{\mu\nu}}\delta R_{\rho\sigma}^{\mu\nu}\right).\label{var_action}
\end{equation}
Needless to say that, this procedure will not yield the full equations
of the most general theory (\ref{general_theory}) but only the part
relevant for the maximally symmetric spacetime. On the other hand,
if $f$ does not depend on the derivatives of the Riemann tensor,
as will be the case in this work, it will yield the full equations.
We can write the variation of the Riemann tensor as 
\begin{align}
\delta R_{\rho\sigma}^{\mu\nu}= & \frac{1}{2}\left(g_{\alpha\rho}\nabla_{\sigma}\nabla^{\nu}-g_{\alpha\sigma}\nabla_{\rho}\nabla^{\nu}\right)\delta g^{\mu\alpha}-\frac{1}{2}\left(g_{\alpha\rho}\nabla_{\sigma}\nabla^{\mu}-g_{\alpha\sigma}\nabla_{\rho}\nabla^{\mu}\right)\delta g^{\alpha\nu}\nonumber \\
 & -\frac{1}{2}R_{\rho\sigma\phantom{\nu}\alpha}^{\phantom{\rho\sigma}\nu}\delta g^{\mu\alpha}+\frac{1}{2}R_{\rho\sigma\phantom{\mu}\alpha}^{\phantom{\rho\sigma}\mu}\delta g^{\alpha\nu},\label{var_riem}
\end{align}
which was obtained from 
\begin{equation}
\delta R_{\phantom{\mu}\nu\rho\sigma}^{\mu}=\nabla_{\rho}\delta\Gamma_{\nu\sigma}^{\mu}-\nabla_{\sigma}\delta\Gamma_{\nu\rho}^{\mu}.
\end{equation}
In calculating the derivative $\frac{\partial f}{\partial R_{\alpha\beta}^{\mu\nu}}$
in (\ref{var_action}), one may try to symmetrize it in such a way
that it satisfies the symmetries of the Riemann tensor. However, this
is not required since at the end, it is multiplied with $\delta R_{\rho\sigma}^{\mu\nu}$
which kills the parts of $\frac{\partial f}{\partial R_{\alpha\beta}^{\mu\nu}}$
that do not obey the symmetries of the Riemann tensor. Then, inserting
the variation of the Riemann tensor (\ref{var_riem}) into the varied
action (\ref{var_action}) leads to a bunch of terms 
\begin{align}
\delta I= & \int d^{n}x\,\left(-\frac{1}{2}g_{\mu\nu}\sqrt{-g}f\left(R_{\rho\sigma}^{\alpha\beta}\right)\delta g^{\mu\nu}\right)\nonumber \\
 & +\frac{1}{2}\int d^{n}x\,\sqrt{-g}\frac{\partial f}{\partial R_{\rho\sigma}^{\mu\nu}}\left(g_{\alpha\rho}\nabla_{\sigma}\nabla^{\nu}-g_{\alpha\sigma}\nabla_{\rho}\nabla^{\nu}\right)\delta g^{\mu\alpha}\nonumber \\
 & -\frac{1}{2}\int d^{n}x\,\sqrt{-g}\frac{\partial f}{\partial R_{\rho\sigma}^{\mu\nu}}\left(g_{\alpha\rho}\nabla_{\sigma}\nabla^{\mu}-g_{\alpha\sigma}\nabla_{\rho}\nabla^{\mu}\right)\delta g^{\alpha\nu}\nonumber \\
 & -\frac{1}{2}\int d^{n}x\,\sqrt{-g}\frac{\partial f}{\partial R_{\rho\sigma}^{\mu\nu}}\left(R_{\rho\sigma\phantom{\nu}\alpha}^{\phantom{\rho\sigma}\nu}\delta g^{\mu\alpha}-R_{\rho\sigma\phantom{\mu}\alpha}^{\phantom{\rho\sigma}\mu}\delta g^{\alpha\nu}\right).
\end{align}
After integration by parts and dropping the boundary terms, one arrives
at the field equations 
\begin{align}
\frac{1}{2}\left(g_{\nu\rho}\nabla^{\lambda}\nabla_{\sigma}-g_{\nu\sigma}\nabla^{\lambda}\nabla_{\rho}\right)\frac{\partial f}{\partial R_{\rho\sigma}^{\mu\lambda}}-\frac{1}{2}\left(g_{\mu\rho}\nabla^{\lambda}\nabla_{\sigma}-g_{\mu\sigma}\nabla^{\lambda}\nabla_{\rho}\right)\frac{\partial f}{\partial R_{\rho\sigma}^{\lambda\nu}}\nonumber \\
-\frac{1}{2}\left(\frac{\partial f}{\partial R_{\rho\sigma}^{\mu\lambda}}R_{\rho\sigma\phantom{\lambda}\nu}^{\phantom{\rho\sigma}\lambda}-\frac{\partial f}{\partial R_{\rho\sigma}^{\lambda\nu}}R_{\rho\sigma\phantom{\lambda}\mu}^{\phantom{\rho\sigma}\lambda}\right)-\frac{1}{2}g_{\mu\nu}f\left(R_{\rho\sigma}^{\alpha\beta}\right) & =0.\label{eq:Field_eqns_of_fRiem}
\end{align}
For the maximally symmetric spacetimes, the first line of the field
equations just yields zero. Therefore, the relevant part of the field
equations that determines the effective cosmological constant is the
second line 
\begin{equation}
\left[\frac{\partial f}{\partial R_{\rho\sigma}^{\mu\lambda}}\right]_{\bar{R}_{\rho\sigma}^{\mu\lambda}}\bar{R}_{\rho\sigma\phantom{\lambda}\nu}^{\phantom{\rho\sigma}\lambda}-\left[\frac{\partial f}{\partial R_{\rho\sigma}^{\lambda\nu}}\right]_{\bar{R}_{\rho\sigma}^{\mu\lambda}}\bar{R}_{\rho\sigma\phantom{\lambda}\mu}^{\phantom{\rho\sigma}\lambda}+g_{\mu\nu}f\left(\bar{R}_{\rho\sigma}^{\alpha\beta}\right)=0,\label{back_eq}
\end{equation}
where the barred quantities are evaluated at the maximally symmetric
value of the Riemann tensor given as 
\begin{equation}
\bar{R}_{\rho\sigma}^{\mu\lambda}=\frac{2\Lambda}{\left(n-1\right)\left(n-2\right)}\left(\delta_{\rho}^{\mu}\delta_{\sigma}^{\lambda}-\delta_{\sigma}^{\mu}\delta_{\rho}^{\lambda}\right).\label{AdS_background}
\end{equation}
Equation (\ref{back_eq}) is the vacuum field equation and the information
on the functional form of the Lagrangian enters the field equation
through \emph{only two} background-evaluated quantities 
\begin{equation}
\left[\frac{\partial f}{\partial R_{\rho\sigma}^{\mu\lambda}}\right]_{\bar{R}_{\rho\sigma}^{\mu\lambda}},\qquad f\left(\bar{R}_{\rho\sigma}^{\alpha\beta}\right).
\end{equation}
Therefore, (\ref{back_eq}) tells us that if these two quantities
are the same for any given two gravity theories, then those two gravity
theories have the same maximally symmetric vacua, namely their effective
cosmological constants are equal. Then, for a theory defined by a
given $f\left(R_{\alpha\beta}^{\mu\nu}\right)$, one can determine
the vacua of the theory by performing a first order Taylor series
expansion around yet to be determined maximally symmetric background
as 
\begin{equation}
I=\int d^{n}x\,\sqrt{-g}\left\{ f\left(\bar{R}_{\alpha\beta}^{\mu\nu}\right)+\left[\frac{\partial f}{\partial R_{\rho\sigma}^{\lambda\nu}}\right]_{\bar{R}_{\rho\sigma}^{\mu\lambda}}\left(R_{\rho\sigma}^{\lambda\nu}-\bar{R}_{\rho\sigma}^{\lambda\nu}\right)\right\} ,\label{Taylor_expanded}
\end{equation}
which from now on will be called as equivalent linearized action (ELA).
Risking to be a little pedantic, let us reiterate the above observation:
Considered as another generic gravity theory, (\ref{Taylor_expanded})
has the same vacua as (\ref{f_gen}). As a result,\emph{ to get the
effective cosmological constant of the vacuum (or vacua) of the most
general gravity theory, all one needs to do is a first order Taylor
series expansion of the generic theory in the Riemann tensor and construct
the equivalent linear action.}

Furthermore, let us show that (\ref{Taylor_expanded}) reduces to
a cosmological Einstein-Hilbert action. Let us define $\zeta$ which
satisfies 
\begin{equation}
\left[\frac{\partial f}{\partial R_{\rho\sigma}^{\mu\nu}}\right]_{\bar{R}_{\rho\sigma}^{\mu\lambda}}R_{\rho\sigma}^{\mu\nu}\equiv\zeta R.\label{zeta_defn}
\end{equation}
Here, the term $\left[\partial f/\partial R_{\rho\sigma}^{\mu\nu}\right]_{\bar{R}_{\rho\sigma}^{\mu\nu}}$
is made up of the Kronecker-deltas such as $\delta_{\mu}^{\rho}\delta_{\nu}^{\sigma}$,
and it should satisfy the symmetries of the Riemann tensor, so antisymmetrizing
$\delta_{\mu}^{\rho}\delta_{\nu}^{\sigma}$ yields $\delta_{\mu}^{[\rho}\delta_{\nu}^{\sigma]}$.%
\footnote{Note that the total antisymmetrization in the up indices implies the
total antisymmetrization in the down indices, that is $\delta_{\mu}^{[\rho}\delta_{\lambda}^{\sigma]}=\delta_{[\mu}^{[\rho}\delta_{\lambda]}^{\sigma]}$.%
} Considering this together with (\ref{zeta_defn}) yields the background
evaluated first derivative of $f\left(R_{\alpha\beta}^{\mu\nu}\right)$
as 
\begin{equation}
\left[\frac{\partial f}{\partial R_{\rho\sigma}^{\mu\lambda}}\right]_{\bar{R}_{\rho\sigma}^{\mu\lambda}}=\zeta\delta_{\mu}^{[\rho}\delta_{\nu}^{\sigma]}.\label{eq:Background_value_of_df/dR}
\end{equation}
Using these, one can recast the equivalent linear action (\ref{Taylor_expanded})
in a more explicit form such that it becomes the usual cosmological
Einstein-Hilbert action as
\begin{equation}
I_{\text{ELA}}=\frac{1}{\kappa_{l}}\int d^{n}x\,\sqrt{-g}\left(R-2\Lambda_{0,l}\right),\label{eq:ELA}
\end{equation}
where the subindex $l$ denotes the equivalent linear action values.
The effective Newton's constant and the effective ``bare'' cosmological
constant are 
\begin{equation}
\frac{1}{\kappa_{l}}=\zeta,\qquad\frac{\Lambda_{0,l}}{\kappa_{l}}=-\frac{1}{2}\bar{f}+\frac{n\Lambda}{n-2}\zeta,
\end{equation}
where we have used $\bar{R}=\frac{2n\Lambda}{n-2}$ and defined $\bar{f}\equiv f\left(\bar{R}_{\rho\sigma}^{\alpha\beta}\right)$.
Then, the field equation for the maximally symmetric background is
simply $\Lambda=\Lambda_{0,l}$ which yields $\Lambda=\frac{n-2}{4\zeta}\bar{f}$.
Note that this is definitely the field equation that one gets after
putting (\ref{AdS_background}) and (\ref{eq:Background_value_of_df/dR})
in (\ref{back_eq}). 

This construction implies that the maximally symmetric vacua of a
generic gravity theory can also be found by expanding the original
action in the metric perturbation $h_{\mu\nu}$ up to the first order
$O\left(h_{\mu\nu}\right)$ and taking the variation with respect
to $h_{\mu\nu}$. 

Once the vacuum of the theory is established, one can move on to discuss
the particle spectrum around this vacuum by expanding the action up
to $O\left(h_{\mu\nu}^{2}\right)$ in the metric perturbation. Directly
expanding the action in powers of $h_{\mu\nu}$ is a highly complicated
task but again fortunately a similar method to the one described above
exists \cite{Hindawi}. The method amounts to finding an equivalent
quadratic curvature action (EQCA) that has the same degrees of freedom
around the same vacua as the original theory. EQCA can be found by
expanding the action in Taylor series up to quadratic order in the
Riemann tensor as we show below. Here we shall assume that the action
does not depend on the derivatives of the Riemann tensor.

For an action that does not depend on the derivatives of the Riemann
tensor, the field equations are (\ref{eq:Field_eqns_of_fRiem}). To
analyze the spectrum of the $f\left(R_{\rho\sigma}^{\alpha\beta}\right)$
theory, that is the excitations around a given background, one needs
the linearized field equations from which one can identify the excitations
by decoupling the linearized field equations into a set of individual
wave equations for each excitation. As we discussed above, one way
to obtain this linearized field equation is to expand the action in
$h_{\mu\nu}$ up to second order and perform variation with respect
to $h_{\mu\nu}$. On the other hand, naturally, one can also directly
linearize the field equations (\ref{eq:Field_eqns_of_fRiem}). To
obtain the linearized field equations, one needs the following linearized
two tensors
\begin{align}
\left[g_{\mu\nu}f\left(R_{\alpha\beta}^{\mu\nu}\right)\right]_{L}= & h_{\mu\nu}f\left(\bar{R}_{\alpha\beta}^{\mu\nu}\right)+\bar{g}_{\mu\nu}\left[\frac{\partial f}{\partial R_{\rho\sigma}^{\alpha\beta}}\right]_{\bar{R}_{\rho\sigma}^{\alpha\beta}}\left(R_{\rho\sigma}^{\alpha\beta}\right)_{L},
\end{align}
and
\begin{align}
\left(\frac{\partial f}{\partial R_{\rho\sigma}^{\mu\lambda}}R_{\rho\sigma\phantom{\lambda}\nu}^{\phantom{\rho\sigma}\lambda}\right)_{L}= & \left[\frac{\partial^{2}f}{\partial R_{\alpha\tau}^{\eta\theta}\partial R_{\rho\sigma}^{\mu\lambda}}\right]_{\bar{R}_{\rho\sigma}^{\mu\lambda}}\left(R_{\alpha\tau}^{\eta\theta}\right)_{L}\bar{R}_{\rho\sigma\phantom{\lambda}\nu}^{\phantom{\rho\sigma}\lambda}+\left[\frac{\partial f}{\partial R_{\rho\sigma}^{\mu\lambda}}\right]_{\bar{R}_{\rho\sigma}^{\mu\lambda}}\left(R_{\rho\sigma\phantom{\lambda}\nu}^{\phantom{\rho\sigma}\lambda}\right)_{L},
\end{align}
and
\begin{align}
\left(g_{\nu\rho}\nabla^{\lambda}\nabla_{\sigma}\frac{\partial f}{\partial R_{\rho\sigma}^{\mu\lambda}}\right)_{L}= & \bar{g}_{\nu\rho}\left[\frac{\partial^{2}f}{\partial R_{\alpha\tau}^{\eta\theta}\partial R_{\rho\sigma}^{\mu\lambda}}\right]_{\bar{R}_{\rho\sigma}^{\mu\lambda}}\bar{\nabla}^{\lambda}\bar{\nabla}_{\sigma}\left(R_{\alpha\tau}^{\eta\theta}\right)_{L}+\bar{g}_{\nu\rho}\left[\frac{\partial f}{\partial R_{\rho\alpha}^{\mu\lambda}}\right]_{\bar{R}_{\rho\sigma}^{\mu\lambda}}\bar{\nabla}^{\lambda}\left(\Gamma_{\sigma\alpha}^{\sigma}\right)_{L}\nonumber \\
 & -\bar{g}_{\nu\rho}\left[\frac{\partial f}{\partial R_{\rho\sigma}^{\alpha\lambda}}\right]_{\bar{R}_{\rho\sigma}^{\mu\lambda}}\bar{\nabla}^{\lambda}\left(\Gamma_{\sigma\mu}^{\alpha}\right)_{L}-\bar{g}_{\nu\rho}\left[\frac{\partial f}{\partial R_{\rho\sigma}^{\mu\alpha}}\right]_{\bar{R}_{\rho\sigma}^{\mu\lambda}}\bar{\nabla}^{\lambda}\left(\Gamma_{\sigma\lambda}^{\alpha}\right)_{L},
\end{align}
where the subindex $L$ means that the quantity is expanded up to
$O\left(h_{\mu\nu}\right)$. The linearization of the other terms
in (\ref{eq:Field_eqns_of_fRiem}) follow from these terms upon symmetrization
and antisymmetrization. Notice that the information on the functional
form of the Lagrangian enters the linearized field equations through
the following three background-evaluated quantities:
\begin{equation}
\left[\frac{\partial^{2}f}{\partial R_{\alpha\tau}^{\eta\theta}\partial R_{\rho\sigma}^{\mu\lambda}}\right]_{\bar{R}_{\rho\sigma}^{\mu\lambda}},\qquad\left[\frac{\partial f}{\partial R_{\rho\sigma}^{\mu\lambda}}\right]_{\bar{R}_{\rho\sigma}^{\mu\lambda}},\qquad f\left(\bar{R}_{\rho\sigma}^{\alpha\beta}\right).
\end{equation}
Therefore, if these three quantities are the same for any given two
gravity theories, then those two theories have the same spectrum around
the same vacua. Then, for a theory defined by a given $f\left(R_{\alpha\beta}^{\mu\nu}\right)$,
one can determine the spectrum of the theory through the quadratic
gravity defined by the up to second order Taylor series expansion
of $f\left(R_{\alpha\beta}^{\mu\nu}\right)$ around the maximally
symmetric background as 
\begin{align}
I=\int d^{n}x\,\sqrt{-g} & \left\{ f\left(\bar{R}_{\alpha\beta}^{\mu\nu}\right)+\left[\frac{\partial f}{\partial R_{\rho\sigma}^{\lambda\nu}}\right]_{\bar{R}_{\rho\sigma}^{\mu\nu}}\left(R_{\rho\sigma}^{\lambda\nu}-\bar{R}_{\rho\sigma}^{\lambda\nu}\right)\right.\nonumber \\
 & \left.+\frac{1}{2}\left[\frac{\partial^{2}f}{\partial R_{\alpha\tau}^{\eta\theta}\partial R_{\rho\sigma}^{\mu\lambda}}\right]_{\bar{R}_{\rho\sigma}^{\mu\nu}}\left(R_{\alpha\tau}^{\eta\theta}-\bar{R}_{\alpha\tau}^{\eta\theta}\right)\left(R_{\rho\sigma}^{\mu\lambda}-\bar{R}_{\rho\sigma}^{\mu\lambda}\right)\right\} ,\label{eq:EQCA_as_Taylor_epansion}
\end{align}
which from now on will be called as the equivalent quadratic curvature
action (EQCA). Note that this action not only has the same spectrum
but also has the same vacua as the original $f\left(R_{\alpha\beta}^{\mu\nu}\right)$
theory.

Now, let us further recast (\ref{eq:EQCA_as_Taylor_epansion}) in
the form of a quadratic gravity theory. To do this, first let us define
the quadratic curvature parameters $\alpha$, $\beta$, and $\gamma$
as 
\begin{equation}
\frac{1}{2}\left[\frac{\partial^{2}f}{\partial R_{\alpha\tau}^{\eta\theta}\partial R_{\rho\sigma}^{\mu\lambda}}\right]_{\bar{R}_{\rho\sigma}^{\mu\nu}}R_{\alpha\tau}^{\eta\theta}R_{\rho\sigma}^{\mu\lambda}\equiv\alpha R^{2}+\beta R_{\sigma}^{\lambda}R_{\lambda}^{\sigma}+\gamma\left(R_{\rho\sigma}^{\eta\lambda}R_{\eta\lambda}^{\rho\sigma}-4R_{\sigma}^{\lambda}R_{\lambda}^{\sigma}+R^{2}\right).\label{eq:a_b_g_defns}
\end{equation}
Since the background evaluated second order derivative of $f\left(R_{\alpha\beta}^{\mu\nu}\right)$
just involves Kronecker deltas and obeys the symmetries of the Riemann
tensors $R_{\alpha\tau}^{\eta\theta}$ and $R_{\rho\sigma}^{\mu\lambda}$,
one has
\begin{equation}
\left[\frac{\partial^{2}f}{\partial R_{\alpha\tau}^{\eta\theta}\partial R_{\rho\sigma}^{\mu\lambda}}\right]_{\bar{R}_{\rho\sigma}^{\mu\nu}}=2\alpha\delta_{\eta}^{[\alpha}\delta_{\theta}^{\tau]}\delta_{\mu}^{[\rho}\delta_{\lambda}^{\sigma]}+\beta\left(\delta_{[\eta}^{\alpha}\delta_{\theta]}^{[\rho}\delta_{[\mu}^{\left|\tau\right|}\delta_{\lambda]}^{\sigma]}-\delta_{[\eta}^{\tau}\delta_{\theta]}^{[\rho}\delta_{[\mu}^{\left|\alpha\right|}\delta_{\lambda]}^{\sigma]}\right)+12\gamma\delta_{\eta}^{[\alpha}\delta_{\theta}^{\tau}\delta_{\mu}^{\rho}\delta_{\lambda}^{\sigma]},
\end{equation}
where the last term, clearly, should have the totally antisymmetric
form since the Gauss-Bonnet combination is the quadratic Lovelock
term%
\footnote{More explicitly, the Lovelock Lagrangian density can be written as
$\delta_{\nu_{1}\nu_{2}\nu_{3}\nu_{4}}^{\mu_{1}\mu_{2}\mu_{3}\mu_{4}}R_{\mu_{1}\mu_{2}}^{\nu_{1}\nu_{2}}R_{\mu_{3}\mu_{4}}^{\nu_{3}\nu_{4}}=4\chi_{\text{GB}}$
where $\delta_{\nu_{1}\nu_{2}\nu_{3}\nu_{4}}^{\mu_{1}\mu_{2}\mu_{3}\mu_{4}}=\epsilon_{abcd}\delta_{\nu_{a}}^{\mu_{1}}\delta_{\nu_{b}}^{\mu_{2}}\delta_{\nu_{c}}^{\mu_{3}}\delta_{\nu_{d}}^{\mu_{4}}=4!\delta_{\nu_{[a}}^{\mu_{1}}\delta_{\nu_{b}}^{\mu_{2}}\delta_{\nu_{c}}^{\mu_{3}}\delta_{\nu_{d]}}^{\mu_{4}}$.%
}. Using these together with (\ref{zeta_defn}), one can put (\ref{eq:EQCA_as_Taylor_epansion})
in a more explicit form as a quadratic gravity theory \cite{Senturk};
\begin{equation}
I_{\text{EQCA}}=\int d^{n}x\,\sqrt{-g}\left[\frac{1}{\tilde{\kappa}}\left(R-2\tilde{\Lambda}_{0}\right)+\alpha R^{2}+\beta R_{\sigma}^{\lambda}R_{\lambda}^{\sigma}+\gamma\chi_{\text{GB}}\right],\label{eq:EQCA}
\end{equation}
where the effective Newton's constant is given as 
\begin{align}
\frac{1}{\tilde{\kappa}} & =\zeta-\frac{4\Lambda}{n-2}\left[n\alpha+\beta+\gamma\frac{\left(n-2\right)\left(n-3\right)}{\left(n-1\right)}\right],\label{eq:kappa_eff}
\end{align}
and the effective ``bare'' cosmological constant reads
\begin{align}
\frac{\tilde{\Lambda}_{0}}{\tilde{\kappa}} & =-\frac{1}{2}f\left(\bar{R}_{\rho\sigma}^{\alpha\beta}\right)+\frac{n\Lambda}{n-2}\zeta-\frac{2\Lambda^{2}n}{\left(n-2\right)^{2}}\left[n\alpha+\beta+\gamma\frac{\left(n-2\right)\left(n-3\right)}{\left(n-1\right)}\right].\label{eq:Lambda_0_eff}
\end{align}
The maximally symmetric solution of (\ref{eq:EQCA}) satisfies \cite{Deser_Tekin}
\begin{equation}
\frac{\Lambda-\tilde{\Lambda}_{0}}{2\tilde{\kappa}}+\left[\left(n\alpha+\beta\right)\frac{\left(n-4\right)}{\left(n-2\right)^{2}}+\gamma\frac{\left(n-3\right)\left(n-4\right)}{\left(n-1\right)\left(n-2\right)}\right]\Lambda^{2}=0,\label{quadratic}
\end{equation}
which certainly is the same vacuum equation as that of the $f\left(R_{\alpha\beta}^{\mu\nu}\right)$
theory and its equivalent linearized version (\ref{eq:ELA}). We made
the equivalence between the linearized field equations of the $f\left(R_{\alpha\beta}^{\mu\nu}\right)$
theory and (\ref{eq:EQCA}) more explicit in the Appendix-\ref{sec:Linearization-of-the}.

\subsection*{ELA and EQCA construction for Born-Infeld gravity}

The above discussion was for generic $f\left(R_{\alpha\beta}^{\mu\nu}\right)$
theories, let us now focus on the BI-type theories. To calculate the
EQCA, one basically needs (A)dS background calculated values for the
matrix function $\sqrt{\det\left(\delta_{\nu}^{\rho}+\gamma A_{\nu}^{\rho}\right)}$
and its first and second derivatives. {[}Note that, we are pulling
out a factor of $\sqrt{-\text{det\,\ }g}$ so that we can work with
the Kronecker delta $\delta_{\mu}^{\nu}$ whose variation is zero.{]}
First, the background value of $\sqrt{\det\left(\delta_{\nu}^{\rho}+\gamma A_{\nu}^{\rho}\right)}$
is given as 
\begin{equation}
\sqrt{\det\left(\delta_{\nu}^{\rho}+\gamma\bar{A}_{\nu}^{\rho}\right)}=\left(1+\bar{a}\right)^{\frac{n}{2}},\label{eq:Back_val_of_I-plus-A}
\end{equation}
where $\bar{a}$ is defined via $\gamma\bar{A}_{\nu}^{\rho}=\bar{a}\delta_{\nu}^{\rho}$.
Then, by using $\det N=\exp\left(\text{Tr}\left(\text{ln}N\right)\right)$,
the first and second order differentials of $\sqrt{\det\left(\delta_{\nu}^{\rho}+\gamma A_{\nu}^{\rho}\right)}$
can be, respectively expressed as 
\begin{equation}
\partial\sqrt{\det\left(\delta_{\nu}^{\rho}+\gamma A_{\nu}^{\rho}\right)}=\frac{\gamma}{2}\sqrt{\det\left(\delta_{\nu}^{\rho}+\gamma A_{\nu}^{\rho}\right)}B_{\gamma}^{\lambda}\partial A_{\lambda}^{\gamma},\label{eq:First_diff_of_I-plus-A}
\end{equation}
and
\begin{equation}
\partial^{2}\sqrt{\det\left(\delta_{\nu}^{\rho}+\gamma A_{\nu}^{\rho}\right)}=\frac{\gamma}{2}\sqrt{\det\left(\delta_{\nu}^{\rho}+\gamma A_{\nu}^{\rho}\right)}\left[B_{\gamma}^{\lambda}\partial^{2}A_{\lambda}^{\gamma}-\gamma B_{\theta}^{\lambda}B_{\gamma}^{\tau}\left(\partial A_{\tau}^{\theta}\right)\partial A_{\lambda}^{\gamma}+\frac{\gamma}{2}\left(B_{\gamma}^{\lambda}\partial A_{\lambda}^{\gamma}\right)^{2}\right],\label{eq:Sec_diff_of_I-plus-A}
\end{equation}
where $B_{\gamma}^{\lambda}$ represents the inverse of the matrix
$\left(\delta_{\gamma}^{\lambda}+\gamma A_{\gamma}^{\lambda}\right)$
and for the differential of $B$ we use $\partial B=-\gamma B\left(\partial A\right)B$.
Note that one may not be able to find the explicit form of the $B$
matrix for a given $A$ matrix, and in fact, even for the simple case
of $A_{\mu\nu}=R_{\mu\nu}$ it is not possible to find the explicit
form of $B$. However, this is not needed since just the (A)dS background
value of the $B$ matrix is required to calculate the background values
for the first and second derivatives of $\sqrt{\det\left(\delta_{\nu}^{\rho}+\gamma A_{\nu}^{\rho}\right)}$.
One can calculate it as
\begin{equation}
\bar{B}_{\gamma}^{\lambda}=\left(1+\bar{a}\right)^{-1}\delta_{\gamma}^{\lambda}.\label{eq:AdS_val_B}
\end{equation}
Note that the matrix $\left(I+A\right)$ becomes singular for $\bar{a}=-1$,
so we assume that $\bar{a}\ne-1$. In the absence of the specific
definition for the $A$ tensor, there is no need to further study
the background values of (\ref{eq:First_diff_of_I-plus-A}) and (\ref{eq:Sec_diff_of_I-plus-A})
by employing (\ref{eq:Back_val_of_I-plus-A}) and (\ref{eq:AdS_val_B}). 

To find the EQCA for a specific BI gravity theory, one needs to find
$\overline{a}$ and needs to calculate the first and second derivatives
of the $A$ tensor with respect to the Riemann tensor, $R_{\rho\sigma}^{\mu\nu}$.
Then, the formulas (\ref{eq:Back_val_of_I-plus-A})--(\ref{eq:AdS_val_B})
are enough to work out the EQCA for the BI gravity theory.

\subsubsection*{Even-dimensional EQCA and ELA\label{sub:4D-EQCA-and-ELA}}

In order to calculate the EQCA and ELA of BI gravity, one needs to
calculate the following three (A)dS background evaluated quantities
as explicitly seen from (\ref{eq:EQCA_as_Taylor_epansion}); 
\begin{equation}
\sqrt{\det\left(\delta_{\nu}^{\beta}+\gamma\bar{A}_{\nu}^{\beta}\right)},\qquad\left[\frac{\partial}{\partial R_{\rho\sigma}^{\lambda\nu}}\sqrt{\det\left(\delta_{\nu}^{\beta}+\gamma A_{\nu}^{\beta}\right)}\right]_{\bar{R}_{\rho\sigma}^{\mu\nu}},\qquad\left[\frac{\partial}{\partial R_{\alpha\tau}^{\eta\theta}\partial R_{\rho\sigma}^{\mu\lambda}}\sqrt{\det\left(\delta_{\nu}^{\beta}+\gamma A_{\nu}^{\beta}\right)}\right]_{\bar{R}_{\rho\sigma}^{\mu\nu}}.
\end{equation}
Using (\ref{eq:Back_val_of_I-plus-A}--\ref{eq:AdS_val_B}) and $\gamma\bar{A}_{\nu}^{\rho}=\bar{a}\delta_{\nu}^{\rho}$,
the (A)dS evaluated value of the Lagrangian can be calculated as 
\begin{equation}
\sqrt{\det\left(\delta_{\nu}^{\beta}+\gamma\bar{A}_{\nu}^{\beta}\right)}=\left(1+\bar{a}\right)^{\frac{n}{2}},\label{eq:AdS_val_of_Lag}
\end{equation}
and its first derivative reads as,
\begin{align}
\left[\frac{\partial}{\partial R_{\rho\sigma}^{\lambda\nu}}\sqrt{\det\left(\delta_{\nu}^{\beta}+\gamma A_{\nu}^{\beta}\right)}\right]_{\bar{R}_{\rho\sigma}^{\mu\lambda}} & =\frac{\gamma}{2}\sqrt{\det\left(\delta_{\nu}^{\beta}+\gamma\bar{A}_{\nu}^{\beta}\right)}\bar{B}_{\gamma}^{\kappa}\left[\frac{\partial A_{\kappa}^{\gamma}}{\partial R_{\rho\sigma}^{\lambda\nu}}\right]_{\bar{R}_{\rho\sigma}^{\mu\lambda}}\nonumber \\
 & =\frac{\gamma}{2}\left(1+\bar{a}\right)^{\frac{\left(n-2\right)}{2}}\delta_{\gamma}^{\kappa}\left[\frac{\partial A_{\kappa}^{\gamma}}{\partial R_{\rho\sigma}^{\lambda\nu}}\right]_{\bar{R}_{\rho\sigma}^{\mu\lambda}},\label{eq:AdS_val_of_1st_derv_BI}
\end{align}
and finally its second derivative boils down to
\begin{align}
\left[\frac{\partial^{2}}{\partial R_{\alpha\tau}^{\eta\theta}\partial R_{\rho\sigma}^{\mu\lambda}}\sqrt{\det\left(\delta_{\nu}^{\beta}+\gamma A_{\nu}^{\beta}\right)}\right]_{\bar{R}_{\rho\sigma}^{\mu\nu}}= & \frac{\gamma}{2}\left(1+\bar{a}\right)^{\frac{\left(n-2\right)}{2}}\delta_{\gamma}^{\kappa}\left[\frac{\partial^{2}A_{\kappa}^{\gamma}}{\partial R_{\alpha\tau}^{\eta\theta}\partial R_{\rho\sigma}^{\mu\lambda}}\right]_{\bar{R}_{\rho\sigma}^{\mu\nu}}\nonumber \\
 & -\frac{\gamma^{2}}{2}\left(1+\bar{a}\right)^{\frac{\left(n-4\right)}{2}}\delta_{\xi}^{\kappa}\delta_{\gamma}^{\zeta}\left[\frac{\partial A_{\zeta}^{\xi}}{\partial R_{\alpha\tau}^{\eta\theta}}\right]_{\bar{R}_{\rho\sigma}^{\mu\nu}}\left[\frac{\partial A_{\kappa}^{\gamma}}{\partial R_{\rho\sigma}^{\mu\lambda}}\right]_{\bar{R}_{\rho\sigma}^{\mu\nu}}\nonumber \\
 & +\frac{\gamma^{2}}{4}\left(1+\bar{a}\right)^{\frac{\left(n-4\right)}{2}}\delta_{\gamma}^{\kappa}\delta_{\xi}^{\zeta}\left[\frac{\partial A_{\kappa}^{\gamma}}{\partial R_{\rho\sigma}^{\mu\lambda}}\right]_{\bar{R}_{\rho\sigma}^{\mu\nu}}\left[\frac{\partial A_{\zeta}^{\xi}}{\partial R_{\alpha\tau}^{\eta\theta}}\right]_{\bar{R}_{\rho\sigma}^{\mu\nu}}.\label{eq:AdS_val_of_2nd_derv_BI}
\end{align}
Here, a difference between the odd and even dimensional cases arises:
it is important to notice that only finite integer powers of $\bar{a}$
appears in these expansions for even dimensions. But, for odd dimensions,
infinite powers of $\bar{a}$ appears. This observation is crucial,
because the same second order expansion in curvature around an (A)dS
background can be obtained by first performing an infinite order expansion
in $A_{\mu\nu}$ ``around $A_{\mu\nu}=0$'', then carrying out the
second order expansion in curvature by using this infinite order series
in $A_{\mu\nu}$. \emph{A priori}, all orders in $A_{\mu\nu}$ contribute
to the second order expansion in curvature in (A)dS backgrounds; however,
as we just observed for even dimensions only a finite number of terms
in the $A_{\mu\nu}$ expansion contribute to the EQCA: More specifically
for even $n$-dimensions expansion up to $O\left(A_{\mu\nu}^{\frac{n}{2}}\right)$
is necessary and sufficient. In four dimensions, we need to expand
up to $O\left(A_{\mu\nu}^{2}\right)$ as given in (\ref{eq:OA2_expansion}).
On the other hand, one needs all the powers of $A_{\mu\nu}^{i}$ for
odd dimensions.

Since there are nontrivial cancellations, it pays to make this observation
more explicit in four dimensions. To construct the EQCA of the $O\left(A_{\mu\nu}^{2}\right)$
which is 
\begin{equation}
\left[\sqrt{\det\left(\delta_{\nu}^{\beta}+\gamma A_{\nu}^{\beta}\right)}\right]_{O\left(A^{2}\right)}=1+\frac{\gamma}{2}A_{\mu}^{\mu}+\frac{\gamma^{2}}{8}A_{\mu}^{\mu}A_{\nu}^{\nu}-\frac{\gamma^{2}}{4}A_{\mu}^{\nu}A_{\nu}^{\mu},\label{eq:OA2_lag}
\end{equation}
one first needs the (A)dS background evaluated value of the $O\left(A_{\mu\nu}^{2}\right)$
Lagrangian which can be calculated by putting $\gamma\bar{A}_{\nu}^{\rho}=\bar{a}\delta_{\nu}^{\rho}$
in (\ref{eq:OA2_lag}) as 
\begin{equation}
\left[\sqrt{\det\left(\delta_{\nu}^{\beta}+\gamma\bar{A}_{\nu}^{\beta}\right)}\right]_{O\left(A^{2}\right)}=1+2\bar{a}+2\bar{a}^{2}-\bar{a}^{2}=\left(1+\bar{a}\right)^{2},
\end{equation}
which is an exact expression represented by finite number of terms
in the $A_{\mu\nu}$ expansion and matches (\ref{eq:AdS_val_of_Lag})
when $n=4$. Moving on to the first and second derivatives (\ref{eq:OA2_lag}),
one respectively gets 
\begin{equation}
\frac{\partial}{\partial R_{\rho\sigma}^{\lambda\nu}}\left[\sqrt{\det\left(\delta_{\nu}^{\beta}+\gamma A_{\nu}^{\beta}\right)}\right]_{O\left(A^{2}\right)}=\frac{\gamma}{2}\frac{\partial A_{\kappa}^{\kappa}}{\partial R_{\rho\sigma}^{\lambda\nu}}+\frac{\gamma^{2}}{4}A_{\beta}^{\beta}\frac{\partial A_{\kappa}^{\kappa}}{\partial R_{\rho\sigma}^{\lambda\nu}}-\frac{\gamma^{2}}{2}A_{\kappa}^{\beta}\frac{\partial A_{\beta}^{\kappa}}{\partial R_{\rho\sigma}^{\lambda\nu}},
\end{equation}
and
\begin{align}
\frac{\partial^{2}}{\partial R_{\alpha\tau}^{\eta\theta}\partial R_{\rho\sigma}^{\mu\lambda}}\left[\sqrt{\det\left(\delta_{\nu}^{\beta}+\gamma A_{\nu}^{\beta}\right)}\right]_{O\left(A^{2}\right)}= & \frac{\gamma}{2}\frac{\partial^{2}A_{\kappa}^{\kappa}}{\partial R_{\alpha\tau}^{\eta\theta}\partial R_{\rho\sigma}^{\lambda\nu}}+\frac{\gamma^{2}}{4}A_{\beta}^{\beta}\frac{\partial^{2}A_{\kappa}^{\kappa}}{\partial R_{\alpha\tau}^{\eta\theta}\partial R_{\rho\sigma}^{\lambda\nu}}-\frac{\gamma^{2}}{2}A_{\kappa}^{\beta}\frac{\partial^{2}A_{\beta}^{\kappa}}{\partial R_{\alpha\tau}^{\eta\theta}\partial R_{\rho\sigma}^{\lambda\nu}}\nonumber \\
 & +\frac{\gamma^{2}}{4}\frac{\partial A_{\beta}^{\beta}}{\partial R_{\alpha\tau}^{\eta\theta}}\frac{\partial A_{\kappa}^{\kappa}}{\partial R_{\rho\sigma}^{\lambda\nu}}-\frac{\gamma^{2}}{2}\frac{\partial A_{\kappa}^{\beta}}{\partial R_{\alpha\tau}^{\eta\theta}}\frac{\partial A_{\beta}^{\kappa}}{\partial R_{\rho\sigma}^{\lambda\nu}}.
\end{align}
These derivatives can be evaluated for the (A)dS background respectively
as
\begin{align}
\left[\frac{\partial}{\partial R_{\rho\sigma}^{\lambda\nu}}\left[\sqrt{\det\left(\delta_{\nu}^{\beta}+\gamma A_{\nu}^{\beta}\right)}\right]_{O\left(A^{2}\right)}\right]_{\bar{R}_{\rho\sigma}^{\mu\nu}} & =\frac{\gamma}{2}\left[\frac{\partial A_{\kappa}^{\kappa}}{\partial R_{\rho\sigma}^{\lambda\nu}}\right]_{\bar{R}_{\rho\sigma}^{\mu\nu}}+\gamma\bar{a}\left[\frac{\partial A_{\kappa}^{\kappa}}{\partial R_{\rho\sigma}^{\lambda\nu}}\right]_{\bar{R}_{\rho\sigma}^{\mu\nu}}-\frac{\gamma}{2}\bar{a}\delta_{\kappa}^{\beta}\left[\frac{\partial A_{\beta}^{\kappa}}{\partial R_{\rho\sigma}^{\lambda\nu}}\right]_{\bar{R}_{\rho\sigma}^{\mu\nu}}\nonumber \\
 & =\frac{\gamma}{2}\left(1+\bar{a}\right)\delta_{\kappa}^{\beta}\left[\frac{\partial A_{\beta}^{\kappa}}{\partial R_{\rho\sigma}^{\lambda\nu}}\right]_{\bar{R}_{\rho\sigma}^{\mu\nu}},
\end{align}
\begin{align}
\left[\frac{\partial^{2}}{\partial R_{\alpha\tau}^{\eta\theta}\partial R_{\rho\sigma}^{\mu\lambda}}\left[\sqrt{\det\left(\delta_{\nu}^{\beta}+\gamma A_{\nu}^{\beta}\right)}\right]_{O\left(A^{2}\right)}\right]_{\bar{R}_{\rho\sigma}^{\mu\nu}}= & \frac{\gamma}{2}\left(1+\overline{a}\right)\left[\frac{\partial^{2}A_{\kappa}^{\kappa}}{\partial R_{\alpha\tau}^{\eta\theta}\partial R_{\rho\sigma}^{\lambda\nu}}\right]_{\bar{R}_{\rho\sigma}^{\mu\nu}}\nonumber \\
 & +\frac{\gamma^{2}}{4}\left[\frac{\partial A_{\beta}^{\beta}}{\partial R_{\alpha\tau}^{\eta\theta}}\right]_{\bar{R}_{\rho\sigma}^{\mu\nu}}\left[\frac{\partial A_{\kappa}^{\kappa}}{\partial R_{\rho\sigma}^{\lambda\nu}}\right]_{\bar{R}_{\rho\sigma}^{\mu\nu}}\nonumber \\
 & -\frac{\gamma^{2}}{2}\left[\frac{\partial A_{\kappa}^{\beta}}{\partial R_{\alpha\tau}^{\eta\theta}}\right]_{\bar{R}_{\rho\sigma}^{\mu\nu}}\left[\frac{\partial A_{\beta}^{\kappa}}{\partial R_{\rho\sigma}^{\lambda\nu}}\right]_{\bar{R}_{\rho\sigma}^{\mu\nu}}
\end{align}
which respectively matches (\ref{eq:AdS_val_of_1st_derv_BI}) and
(\ref{eq:AdS_val_of_2nd_derv_BI}) when $n=4$ . Hence, we have achieved
our goal of showing that in four dimensions to get the vacuum and
the spectrum of a determinantal BI gravity theory one needs to expand
only up to $O\left(A_{\mu\nu}^{2}\right)$ around $A_{\mu\nu}=0$.
It is a remarkable fact about this determinantal actions that at every
order $O\left(A_{\mu\nu}^{2+i}\right)$, contributions of the various
terms to the EQCA cancel among each other, that is they do not contribute
to the free theory around the maximally symmetric background. Moreover,
this cancellation works in such a way that $O\left(\gamma^{0}\right)$,
$O\left(\gamma^{1}\right)$ and $O\left(\beta\right)$ terms cancel
among each other. This fact also means that when $A_{\mu\nu}$ is
at most quadratic in curvature as in (\ref{eq:Generic_Amn}), one
can also get the same result by performing a Taylor series expansion
in small curvature, that is around $R_{\rho\sigma}^{\mu\nu}=0$, up
to $O\left(R^{4}\right)$ in four dimension; hence, end up with a
quartic gravity theory. This also immediately leads to the fact that
\emph{a priori} there will be four possible maximally symmetric vacua
of the theory which we next study.

\section{Determining the Vacua of the BI Theory}

Let us find the maximally symmetric vacua (generically there will
be four different vacua in four dimensions as noted above) of our
theory 
\begin{equation}
\kappa\mathcal{L}=\frac{2}{\gamma}\left[\sqrt{\det\left(\delta_{\sigma}^{\rho}+\gamma A_{\sigma}^{\rho}\right)}-\left(\lambda_{0}+1\right)\right],\label{action_lambda_0}
\end{equation}
with $A_{\mu\nu}$ as (\ref{eq:Generic_Amn}) and we have defined
a dimensionless cosmological parameter $\lambda_{0}\equiv\gamma\Lambda_{0}$
which we shall use from now on. We resort to the equivalent linear
action formalism described above which in this basis follows from
the zeroth and the first order Taylor series expansion of the action
(\ref{action_lambda_0}) 
\begin{align}
\kappa\mathcal{L}_{\text{ELA}}= & \frac{2}{\gamma}\left[\sqrt{\det\left(\delta_{\sigma}^{\rho}+\gamma\bar{A}_{\sigma}^{\rho}\right)}-\left(\lambda_{0}+1\right)\right]\nonumber \\
 & +\left[\frac{\partial\mathcal{L}}{\partial C_{\alpha\beta}^{\mu\nu}}\right]_{\bar{R}_{\rho\sigma}^{\mu\nu}}C_{\alpha\beta}^{\mu\nu}+\left[\frac{\partial\mathcal{L}}{\partial S_{\nu}^{\mu}}\right]_{\bar{R}_{\rho\sigma}^{\mu\nu}}S_{\nu}^{\mu}+\left[\frac{\partial\mathcal{L}}{\partial R_{\nu}^{\mu}}\right]_{\bar{R}_{\rho\sigma}^{\mu\nu}}\left(R_{\nu}^{\mu}-\bar{R}_{\nu}^{\mu}\right).
\end{align}
Note that the background values $\bar{C}_{\alpha\beta}^{\mu\nu}$
and $\bar{S}_{\nu}^{\mu}$ are zero, and $\bar{R}_{\rho\sigma}^{\mu\nu}$
is given in (\ref{AdS_background}) with $n=4$ and $\bar{R}_{\nu}^{\mu}=\Lambda\delta_{\nu}^{\mu}$.
The background values of the Lagrangian density and its first order
derivatives are calculated in the Appendix-\ref{sec:ELA_and_EQCA_terms},
and using these results, the equivalent linearized action of the BI
theory given in (\ref{eq:Generic_BI_intro}) becomes
\begin{equation}
\mathcal{L}_{\text{ELA}}=\frac{1}{\kappa_{l}}\left(R-\frac{2}{\gamma}\lambda_{0,l}\right),\label{eq:ELA-BI}
\end{equation}
the Newton's constant and the cosmological constant, upon using the
computations in Appendix-\ref{sec:ELA_and_EQCA_terms}, turn out to
be 
\begin{align}
\frac{1}{\kappa_{l}}=\left(1+\bar{a}\right)\left[1+2\lambda\left(a_{3}+b_{2}\right)\right],\qquad\bar{a} & =\lambda+\lambda^{2}\left(a_{3}+b_{2}\right),\nonumber \\
\lambda_{0,l}=\kappa_{l}\left(1+\lambda_{0}-\left(1+\bar{a}\right)^{2}\right) & +2\lambda,\label{a_bar}
\end{align}
where we assumed $1/\kappa_{l}$ is not zero, otherwise the theory
would not reproduce Einstein's gravity.\textcolor{red}{{} }Then, since
the vacua of the equivalent linear theory is determined by 
\begin{equation}
\lambda=\lambda_{0,l},
\end{equation}
one arrives at the quartic equation that gives the four possible maximally
symmetric vacua
\begin{equation}
c^{2}\lambda^{4}+c\lambda^{3}-\lambda+\lambda_{0}=0,\label{vacuum_eq}
\end{equation}
where we have defined $c\equiv a_{3}+b_{2}$. Let us not depict the
solutions in their general form since they are not needed, but let
us note some specific points. If $c=0$ then we have a unique vacuum
with $\lambda=\lambda_{0}$. If $c\ne0$, then there are real and
complex solutions depending on the values of $c$ and $\lambda_{0}$.
It is possible to have four real solutions if $0<c<\frac{1}{4}$ and
if a complicated condition on $\lambda_{0}$ is satisfied. As an example,
let us take $c=\frac{1}{8}$ and $\lambda_{0}=\frac{3}{4}$, then
we have $\lambda=\left(-6,\,-2\left(1\pm\sqrt{2}\right),\,2\right)$.
But, as we shall see later on, unitarity will demand that $c=\frac{1}{4}$
and $\lambda<1$, and the nonvanishing of the effective Newton's constant
demands that $\lambda\ne-2$. These conditions are satisfied if: $\lambda_{0}<\frac{11}{16}$
and $\lambda_{0}\ne-1$. One can compute the discriminant to be $\Delta=\frac{1}{256}\left(1+\lambda_{0}\right)^{2}\left(-11+16\lambda_{0}\right)$
which is always negative in the allowed region. This says that there
are two real and two complex-conjugate roots. One of the real roots
does not satisfy the $\lambda<1$ condition; but, the other one always
satisfies this condition. Therefore, we have a unique \emph{viable}
vacuum.

\section{Unitarity Around Flat Backgrounds}

First, let us start with the flat space $\lambda=0$ for which we
must take the bare cosmological parameter to be $\lambda_{0}=0$.
From (\ref{vacuum_eq}) we see that flat space is the vacuum of the
theory but it is not the only vacuum since the equation reduces to
\begin{equation}
\lambda\left(c^{2}\lambda^{3}+c\lambda^{2}-1\right)=0,
\end{equation}
with at least one more real solution with a non-zero $\lambda$ with
the exception that $c=0$. For example, for $c=\frac{1}{4}$, whose
relevance will appear below, one has $\lambda\approx1.679$.

Let us focus on the flat vacuum. In this case our job is not complicated
at all: All we need to do is to expand (\ref{eq:Generic_BI_intro})
up to $O\left(R^{2}\right)$ and demand that it matches either with
the Einstein-Hilbert action or with the Einstein-Gauss-Bonnet action.

The determinant $\left[\det\left(1+M\right)\right]^{1/2}$ can be
expanded up to $O\left(M^{4}\right)$ as 
\begin{equation}
\left[\det\left(1+M\right)\right]^{1/2}=1+\frac{1}{2}\text{Tr}M+\frac{1}{8}\left(\text{Tr}M\right)^{2}-\frac{1}{4}\text{Tr}\left(M^{2}\right)+\frac{1}{6}\text{Tr}\left(M^{3}\right)-\frac{1}{8}\text{Tr}M\text{Tr}\left(M^{2}\right)+\frac{1}{48}\left(\text{Tr}M\right)^{3},\label{eq:OM3_expansion}
\end{equation}
where $\text{Tr}M=g^{\mu\nu}M_{\mu\nu}$. Using this expression, the
$O\left(R^{2}\right)$ expansion of (\ref{eq:Generic_BI_intro}) yields
the quadratic Lagrangian
\begin{align}
\kappa\mathcal{L}_{O\left(R^{2}\right)}= & R+\gamma b_{1}C_{\mu\rho\sigma\lambda}C^{\mu\rho\sigma\lambda}\nonumber \\
 & +\gamma\left(a_{3}+b_{2}+\frac{1}{2}\right)R_{\mu\rho}R^{\mu\rho}+\gamma\left(a_{4}+b_{3}-1-\frac{\beta\left(\beta+2\right)}{2}\right)S_{\mu\rho}S^{\mu\rho}.\label{flat_space_o(R^2)}
\end{align}
Here, note that $a_{2}$ does not appear in the $O\left(R^{2}\right)$
expansion, so unitarity constraints around the flat background do
not put any condition on the $C_{\mu\rho\nu\sigma}R^{\rho\sigma}$
term. As already noted, there are two unitary theories that (\ref{flat_space_o(R^2)})
can reduce to: The Einstein theory and the EGB theory which need separate
attention even though they are classically equivalent in four dimensions.
Let us start with the reduction to Einstein's theory.

\subsection{Reduction to the Einstein theory}

We will compare (\ref{flat_space_o(R^2)}) with 
\begin{align}
\kappa\mathcal{L} & =R,
\end{align}
which yields the elimination of three parameters 
\begin{equation}
b_{1}=0,\qquad a_{3}=-\frac{1}{2}-b_{2},\qquad a_{4}=\frac{\beta\left(\beta+2\right)}{2}+1-b_{3},\label{flat_cond}
\end{equation}
leaving a theory with four dimensionless parameters that can be built
from 
\begin{align}
A_{\mu\nu}= & R_{\mu\nu}+\beta S_{\mu\nu}+\frac{\gamma}{4}g_{\mu\nu}\left(b_{2}R_{\rho\sigma}R^{\rho\sigma}+b_{3}S_{\rho\sigma}S^{\rho\sigma}\right)\nonumber \\
 & +\gamma\left[a_{2}C_{\mu\rho\nu\sigma}R^{\rho\sigma}-\left(\frac{1}{2}+b_{2}\right)R_{\mu\rho}R_{\nu}^{\rho}+\left(\frac{\beta\left(\beta+2\right)}{2}+1-b_{3}\right)S_{\mu\rho}S_{\nu}^{\rho}\right].\label{Generic_Amn_flat}
\end{align}
It is important to understand that the BI gravity defined with this
$A_{\mu\nu}$ describes a massless, unitary spin-2 graviton about
its flat vacuum in all finite orders in the curvature expansion and
as well as the full theory, namely in the infinite order in the curvature
expansion. This is because in flat backgrounds, only terms up to quadratic
in curvature contribute to the propagator of the theory but this theory
does not have quadratic terms and when expanded in curvature it symbolically
reads $\mathcal{L}=R+R^{3}+\dots$. 

Let us also note that if we require the uniqueness of the vacuum namely
that the flat space is the unique vacuum then one cannot reduce the
theory to Einstein's gravity since unique vacuum condition is achieved
with $c=0$ but Einsteinian reduction is achieved with $c=-\frac{1}{2}$
as seen from the second equation of (\ref{flat_cond}), hence the
contradiction. Therefore, the theory (\ref{Generic_Amn_flat}) has
two vacua one with $\lambda=0$ and the other with $\lambda\approx2.594$.
Of course, to have a consistent theory we must check its unitarity
about the second vacuum. As we shall see in the next section, $c=-\frac{1}{2}$
is excluded. Before that discussion, let us consider some specific
theories by taking the undetermined dimensionless parameters to be
zero. 

Unitarity about its flat vacuum does not constrain this theory any
further: Let us use the notion of minimality and fix the undetermined
parameters. There could be many ways to define minimal theories here:
For example, if we set $\beta=a_{2}=a_{3}=a_{4}=0$ we arrive at a
unitary theory around its flat background with the action 
\begin{equation}
I=\frac{2}{\kappa\gamma}\int d^{4}x\left\{ \sqrt{-\text{det}\left[g_{\mu\nu}+\gamma R_{\mu\nu}+\frac{\gamma^{2}}{8}g_{\mu\nu}R_{\sigma\rho}G^{\sigma\rho}\right]}-\sqrt{-\text{det}\, g}\right\} ,
\end{equation}
which was already given in \cite{Deser_Gibbons}. Another option is
choosing $\beta=b_{2}=b_{3}=a_{2}=0$ which yields another theory

\begin{align}
I & =\frac{2}{\kappa\gamma}\int d^{4}x\left\{ \sqrt{-\text{det}\left[g_{\mu\nu}+\gamma R_{\mu\nu}+\frac{\gamma^{2}}{2}\left(G_{\mu\rho}G_{\nu}^{\rho}-\frac{1}{8}g_{\mu\nu}R^{2}\right)\right]}-\sqrt{-\text{det}\, g}\right\} .\label{flat_second_option}
\end{align}
Of course, with four free parameters there are many other options,
but in any case the most general theory that has a unitary massless
spin-2 excitation around its flat vacuum is constructed with (\ref{Generic_Amn_flat}).

\subsection{Reduction to the Einstein--Gauss-Bonnet theory}

The next possible option is to try to reduce (\ref{flat_space_o(R^2)})
to the EGB theory which has the same spectrum, field equations, \emph{etc
}in four dimensions with Einstein's gravity. We will compare (\ref{flat_space_o(R^2)})
with (\ref{GB_combination}) for $\lambda_{0}=0$ which yields the
following relations between the parameters: 
\begin{equation}
a_{3}=\frac{2}{3}b_{1}-b_{2}-\frac{1}{2},\qquad a_{4}=\frac{\beta\left(\beta+2\right)}{2}-\frac{8}{3}b_{1}-b_{3}+1,\label{EGB_reduction_conditions}
\end{equation}
eliminating two of them and yielding the following $A_{\mu\nu}$ tensor
with five parameters :
\begin{align}
A_{\mu\nu}= & R_{\mu\nu}+\beta S_{\mu\nu}\nonumber \\
 & +\gamma\left(a_{2}C_{\mu\rho\nu\sigma}R^{\rho\sigma}+\left(\frac{2}{3}b_{1}-b_{2}-\frac{1}{2}\right)R_{\mu\rho}R_{\nu}^{\rho}+\left(\frac{\beta\left(\beta+2\right)}{2}-\frac{8}{3}b_{1}-b_{3}+1\right)S_{\mu\rho}S_{\nu}^{\rho}\right)\nonumber \\
 & +\frac{\gamma}{4}g_{\mu\nu}\left(b_{1}C_{\rho\sigma\lambda\gamma}C^{\rho\sigma\lambda\gamma}+b_{2}R_{\rho\sigma}R^{\rho\sigma}+b_{3}S_{\rho\sigma}S^{\rho\sigma}\right).
\end{align}
This defines the most general theory that has a massless unitary graviton
about its flat vacuum. Unlike the previous case, we can further require
that the flat vacuum is the \emph{unique vacuum}, that is $c=0$,
then we arrive at the relations 

\begin{equation}
a_{3}=-b_{2},\qquad b_{1}=\frac{3}{4},\qquad a_{4}=\frac{\beta\left(\beta+2\right)}{2}-b_{3}-1,
\end{equation}
which reduces the general $A_{\mu\nu}$ tensor to 
\begin{align}
A_{\mu\nu}= & R_{\mu\nu}+\beta S_{\mu\nu}\nonumber \\
 & +\gamma\left(a_{2}C_{\mu\rho\nu\sigma}R^{\rho\sigma}+a_{3}R_{\mu\rho}R_{\nu}^{\rho}+a_{4}S_{\mu\rho}S_{\nu}^{\rho}\right)\nonumber \\
 & +\frac{\gamma}{4}g_{\mu\nu}\left[\frac{3}{4}C_{\rho\sigma\lambda\gamma}C^{\rho\sigma\lambda\gamma}-a_{3}R_{\rho\sigma}R^{\rho\sigma}+\left(\frac{\beta\left(\beta+2\right)}{2}-a_{4}-1\right)S_{\rho\sigma}S^{\rho\sigma}\right],
\end{align}
Hence, the unitarity and the unique vacuum conditions gave us a four
parameter theory. By judiciously choosing some of these parameters
to vanish we can define various minimal theories. First choice can
be to set $\beta=a_{2}=b_{2}=b_{3}=0$ yielding 
\begin{align}
A_{\mu\nu} & =R_{\mu\nu}+\frac{3\gamma}{16}g_{\mu\nu}\chi_{\text{GB}}+\frac{3\gamma}{8}g_{\mu\nu}R_{\sigma\rho}G^{\sigma\rho}-\gamma R_{\mu\rho}G_{\nu}^{\rho}.
\end{align}
Another minimal theory option is obtained after setting $\beta=a_{2}=a_{3}=a_{4}=0$
yielding an $A_{\mu\nu}$ in terms of the Ricci tensor and the metric
tensor multiplied with specific quadratic terms as 
\begin{align}
A_{\mu\nu} & =R_{\mu\nu}+\frac{\gamma}{8}g_{\mu\nu}\left(\frac{3}{2}\chi_{\text{GB}}+R_{\rho\sigma}G^{\rho\sigma}\right),
\end{align}
where we have made use of the GB identity. The second option leads
to the action 
\begin{equation}
I=\frac{2}{\kappa\gamma}\int d^{4}x\left\{ \sqrt{\text{-det}\left[g_{\mu\nu}+\gamma R_{\mu\nu}+\frac{\gamma^{2}}{8}g_{\mu\nu}\left(\frac{3}{2}\chi_{\text{GB}}+R_{\sigma\rho}G^{\sigma\rho}\right)\right]}-\sqrt{-\text{det}\, g}\right\} ,\label{flatspace_minimal}
\end{equation}
which should be considered as an exact theory for all values of the
curvature: At any order in the curvature expansion the flat vacuum
is the unique vacuum solution and the theory describes a unitary massless
graviton. Now, let us see in small curvature expansion what kind of
a theory we get up to $O\left(R^{3}\right)$. For this purpose we
use (\ref{eq:OM3_expansion}) and get from (\ref{flatspace_minimal})
the following effective theory
\begin{equation}
I=\frac{1}{\kappa}\int d^{4}x\sqrt{-\text{det}\, g}\left\{ R+\frac{3\gamma}{4}\chi_{\text{GB}}+\frac{\gamma^{2}}{48}\left(9RR_{\mu\sigma\nu\rho}R^{\mu\sigma\nu\rho}+16R_{\mu}^{\alpha}R_{\alpha}^{\beta}R_{\beta}^{\mu}-42RR_{\mu\nu}R^{\mu\nu}+8R^{3}\right)\right\} .\label{cubic_b=000131}
\end{equation}
The Gauss-Bonnet term does not contribute to the field equations.
In case it is not apparent that this theory has a unique vacuum and
a unitary massless spin-2 excitation from our construction above,
let us show this here in a different way. In fact these can be seen
either from the field equations or from the equivalent quadratic curvature
action that the flat space is the unique vacuum. Let us follow the
second path and find the vacuum and the excitations for this BI-generated
cubic curvature modification of Einstein's theory. Unitarity and the
particle spectrum of all cubic curvature gravity theories based on
the Riemann tensor and its contractions were studied in \cite{Sisman-AllUniD}.
The most general cubic curvature gravity is defined with the action
as 
\begin{equation}
I=\frac{1}{\kappa}\int d^{4}x\sqrt{-g}\left[R-2\Lambda_{0}+\alpha R^{2}+\beta R^{\mu\nu}R_{\mu\nu}+\gamma\chi_{\text{GB}}+F\left(R_{\rho\sigma}^{\mu\nu}\right)\right],\label{general_cubic_action}
\end{equation}
where $F\left(R_{\rho\sigma}^{\mu\nu}\right)$ represents the eight
possible cubic curvature terms with no derivatives
\begin{align}
F\left(R_{\rho\sigma}^{\mu\nu}\right)\equiv & c_{1}R_{\rho\sigma}^{\mu\nu}R_{\mu\alpha}^{\rho\beta}R_{\nu\beta}^{\sigma\alpha}+c_{2}R_{\rho\sigma}^{\mu\nu}R_{\mu\nu}^{\alpha\beta}R_{\alpha\beta}^{\rho\sigma}+c_{3}R_{\nu}^{\mu}R_{\alpha\mu}^{\rho\sigma}R_{\rho\sigma}^{\alpha\nu}+c_{4}RR_{\rho\sigma}^{\mu\nu}R_{\mu\nu}^{\rho\sigma}\nonumber \\
 & +c_{5}R_{\nu}^{\mu}R_{\sigma}^{\rho}R_{\mu\rho}^{\nu\sigma}+c_{6}R_{\nu}^{\mu}R_{\mu}^{\rho}R_{\rho}^{\nu}+c_{7}RR_{\nu}^{\mu}R_{\mu}^{\nu}+c_{8}R^{3}.\label{eq:Cubic_curvature_Lagrangian}
\end{align}
The EQCA of (\ref{general_cubic_action}) was calculated in \cite{Sisman-AllUniD}
as 
\begin{equation}
I=\int d^{4}x\sqrt{-g}\,\left[\frac{1}{\tilde{\kappa}}\left(R-2\tilde{\Lambda}_{0}\right)+\tilde{\alpha}R^{2}+\tilde{\beta}R_{ab}^{2}+\tilde{\gamma}\left(R_{abcd}^{2}-4R_{ab}^{2}+R^{2}\right)\right],\label{eq:Equiv_quad_act}
\end{equation}
 with effective parameters 
\begin{align}
\frac{1}{\tilde{\kappa}}\equiv & \frac{1}{\kappa}-\frac{\Lambda^{2}}{3\kappa}\left[c_{1}+4c_{2}+6\left(c_{3}+4c_{4}\right)+9\left(c_{5}+c_{6}+4c_{7}+16c_{8}\right)\right],\nonumber \\
\tilde{\Lambda}_{0}\equiv & \frac{\tilde{\kappa}}{\kappa}\Lambda_{0}+\frac{2\Lambda}{3}\left(1-\frac{\tilde{\kappa}}{\kappa}\right),\nonumber \\
\tilde{\alpha}\equiv & \frac{\alpha}{\kappa}+\frac{\Lambda}{3\kappa}\left[3c_{1}-6c_{2}-8c_{4}+c_{5}+3\left(-c_{3}+2c_{7}+12c_{8}\right)\right],\nonumber \\
\tilde{\beta}\equiv & \frac{\beta}{\kappa}+\frac{\Lambda}{3\kappa}\left[-9c_{1}+24c_{2}+16c_{3}+5c_{5}+3\left(16c_{4}+3c_{6}+4c_{7}\right)\right],\nonumber \\
\tilde{\gamma}\equiv & \frac{\gamma}{\kappa}+\frac{\Lambda}{\kappa}\left[-c_{1}+2c_{2}+\left(c_{3}+4c_{4}\right)\right].\label{geff}
\end{align}
Comparing (\ref{general_cubic_action}) with (\ref{cubic_b=000131}),
one obtains the following parameters for the EQCA from (\ref{geff})
\begin{align}
\frac{1}{\tilde{\kappa}}= & \frac{1}{\kappa},\;\tilde{\Lambda}_{0}=0,\;\tilde{\alpha}=-\frac{\Lambda}{4\kappa}\gamma^{2},\;\tilde{\beta}=\frac{\Lambda}{2\kappa}\gamma^{2},\;\tilde{\gamma}=\frac{3\gamma}{4\kappa}\left(1+\Lambda\gamma\right),
\end{align}
which give the equivalent quadratic curvature action of (\ref{cubic_b=000131})
\begin{equation}
I=\frac{1}{\kappa}\int d^{4}x\sqrt{-\text{det}\, g}\left\{ R+\frac{3}{4}\gamma\chi_{\text{GB}}+\frac{\gamma^{2}\Lambda}{4}\left(3\chi_{\text{GB}}+2R_{\mu\nu}R^{\mu\nu}-R^{2}\right)\right\} .\label{cubic_bi_action}
\end{equation}
Here, we still have to find $\Lambda$ which corresponds to the maximally
symmetric vacuum. There are two ways to do this: One can either derive
the field equations of the cubic theory (\ref{cubic_bi_action}) and
get the vacuum from those equations or one can find the field equations
of the equivalent quadratic theory (\ref{cubic_bi_action}) instead.
Of course the second method is easier and in fact these field equations
were given in \cite{Deser_Tekin-PRL}, and hence, no need to repeat
them here. Inserting $R_{\mu\sigma\nu\rho}=\frac{\Lambda}{3}\left(g_{\mu\nu}g_{\sigma\rho}-g_{\mu\rho}g_{\sigma\nu}\right)$
to the field equations, one finds that $\Lambda=0$. Therefore, flat
space is the unique vacuum. In fact more importantly $\Lambda=0$
also kills the ghost term ($R_{\mu\nu}R^{\mu\nu}$) in the action.
As expected (\ref{cubic_b=000131}) has a unitary massless spin-2
excitation just like its exact ``mother'' (\ref{flatspace_minimal}).
Happily, this state of affairs is intact for any $O\left(R^{i}\right)$
truncation of the exact theory: That is at any order the vacuum is
\emph{uniquely} flat and the theory has a massless unitary graviton.

\section{Unitarity Around (A)dS Backgrounds}

Let us now study the unitarity of the BI gravity around its (A)dS
background. It is important to establish what we mean by the tree-level
unitarity of the BI theory in (A)dS backgrounds: As we noted in the
Introduction, we require that the theory is tree-level unitary at
any finite order in the curvature expansion and at infinite order
in the curvature expansion which is the full theory. Namely, the full
theory or any truncated version of the theory, for example the linear
Einstein theory, quadratic gravity or in general $O\left(R^{i}\right)$
theory should be unitary. Note that this condition on unitarity is
stronger than the unitarity condition in string theory generated effective
gravity models. For example, the full string theory is unitary yet
$O\left(R^{3}\right)$ effective theory is non-unitary for bosonic
string theory \cite{Metsaev_Tseytlin} as shown in \cite{Sisman-AllUniD}.
In (A)dS backgrounds, unlike the flat space case, infinitely many
terms contribute to the propagator and to the free theory; \emph{i.e.}
the vacuum \emph{etc}. Therefore, as explained above, we need the
equivalent quadratic curvature theory of 
\begin{equation}
\kappa\mathcal{L}=\frac{2}{\gamma}\left[\sqrt{\det\left(\delta_{\nu}^{\rho}+\gamma A_{\nu}^{\rho}\right)}-\left(\lambda_{0}+1\right)\right],
\end{equation}
which upon use of (\ref{eq:EQCA_as_Taylor_epansion}) in the Weyl--traceless-Ricci--Ricci
(CSR) basis, reads as
\begin{align}
\kappa\mathcal{L}_{\text{EQCA}}= & \frac{2}{\gamma}\left[\sqrt{\det\left(\delta_{\nu}^{\rho}+\gamma\bar{A}_{\nu}^{\rho}\right)}-\left(\lambda_{0}+1\right)\right]\nonumber \\
 & +\left[\frac{\partial\mathcal{L}}{\partial C_{\alpha\beta}^{\mu\nu}}\right]_{\bar{R}_{\rho\sigma}^{\mu\nu}}C_{\alpha\beta}^{\mu\nu}+\left[\frac{\partial\mathcal{L}}{\partial S_{\nu}^{\mu}}\right]_{\bar{R}_{\rho\sigma}^{\mu\nu}}S_{\nu}^{\mu}+\left[\frac{\partial\mathcal{L}}{\partial R_{\nu}^{\mu}}\right]_{\bar{R}_{\rho\sigma}^{\mu\nu}}\left(R_{\nu}^{\mu}-\bar{R}_{\nu}^{\mu}\right)\nonumber \\
 & +\frac{1}{2}\left[\frac{\partial^{2}\mathcal{L}}{\partial C_{\alpha\beta}^{\mu\nu}C_{\lambda\tau}^{\eta\theta}}\right]_{\bar{R}_{\rho\sigma}^{\mu\nu}}C_{\alpha\beta}^{\mu\nu}C_{\lambda\tau}^{\eta\theta}+\frac{1}{2}\left[\frac{\partial^{2}\mathcal{L}}{\partial S_{\nu}^{\mu}\partial S_{\beta}^{\alpha}}\right]_{\bar{R}_{\rho\sigma}^{\mu\nu}}S_{\nu}^{\mu}S_{\beta}^{\alpha}\nonumber \\
 & +\frac{1}{2}\left[\frac{\partial^{2}\mathcal{L}}{\partial R_{\nu}^{\mu}\partial R_{\beta}^{\alpha}}\right]_{\bar{R}_{\rho\sigma}^{\mu\nu}}\left(R_{\nu}^{\mu}-\bar{R}_{\nu}^{\mu}\right)\left(R_{\beta}^{\alpha}-\bar{R}_{\beta}^{\alpha}\right)\nonumber \\
 & +\left[\frac{\partial^{2}\mathcal{L}}{\partial C_{\alpha\beta}^{\mu\nu}\partial S_{\theta}^{\eta}}\right]_{\bar{R}_{\rho\sigma}^{\mu\nu}}C_{\alpha\beta}^{\mu\nu}S_{\theta}^{\eta}+\left[\frac{\partial^{2}\mathcal{L}}{\partial C_{\alpha\beta}^{\mu\nu}\partial R_{\theta}^{\eta}}\right]_{\bar{R}_{\rho\sigma}^{\mu\nu}}C_{\alpha\beta}^{\mu\nu}\left(R_{\theta}^{\eta}-\bar{R}_{\theta}^{\eta}\right)\nonumber \\
 & +\left[\frac{\partial^{2}\mathcal{L}}{\partial S_{\nu}^{\mu}\partial R_{\beta}^{\alpha}}\right]_{\bar{R}_{\rho\sigma}^{\mu\nu}}S_{\nu}^{\mu}\left(R_{\beta}^{\alpha}-\bar{R}_{\beta}^{\alpha}\right),
\end{align}
where the bracketed and barred quantities denote the maximally symmetric
background values for the corresponding expressions. Note again that
$\bar{C}_{\alpha\beta}^{\mu\nu}=0$ and $\bar{S}_{\nu}^{\mu}=0$.
The terms up to quadratic order are just the ELA given in (\ref{eq:ELA-BI}),
so we just need the quadratic contributions which are again given
in the Appendix-\ref{sec:ELA_and_EQCA_terms}. By using these results,
the equivalent quadratic curvature action of (\ref{eq:Generic_BI_intro})
can be compactly written as
\begin{equation}
\kappa\mathcal{L}_{\text{EQCA}}=\frac{1}{\tilde{\kappa}}\left(R-\frac{2}{\gamma}\tilde{\lambda}_{0}+\ensuremath{\alpha}_{1}C_{\rho\sigma}^{\mu\nu}C_{\mu\nu}^{\rho\sigma}+\ensuremath{\alpha}_{2}R_{\mu}^{\nu}R_{\nu}^{\mu}+\alpha_{3}S_{\mu}^{\nu}S_{\nu}^{\mu}\right),\label{eq:EQCA-BI}
\end{equation}
where the effective Newton's constant and the effective ``bare''
cosmological constant are given as
\begin{equation}
\frac{1}{\tilde{\kappa}}=1+\bar{a}-\lambda\left(2\lambda c+1\right){}^{2},\label{eq:k_eff_ECQA}
\end{equation}
\begin{equation}
\tilde{\lambda}_{0}=\tilde{\kappa}\left[\lambda\left(1+\bar{a}\right)\left(2\lambda c+1\right)-\bar{a}\left(2+\bar{a}\right)+\lambda_{0}\right]+\lambda,\label{eq:L0_eff_ECQA}
\end{equation}
and the quadratic curvature parameters read as
\begin{equation}
\alpha_{1}=\gamma b_{1}\tilde{\kappa}\left(1+\bar{a}\right),\label{eq:alpha_1}
\end{equation}
\begin{equation}
\alpha_{2}=\frac{\gamma}{2\lambda}\left[\tilde{\kappa}\left(1+\bar{a}\right)\left(2\lambda c+1\right)-1\right],\label{eq:alpha_2}
\end{equation}
\begin{equation}
\alpha_{3}=\frac{\gamma}{2\lambda}\left[\tilde{\kappa}\left(\left(1+\bar{a}\right)\left(2\lambda\left(a_{4}+b_{3}\right)-1\right)-\lambda\left(2a_{3}\lambda+\beta+1\right)^{2}\right)+1\right].\label{eq:alpha_3}
\end{equation}
Here, $\bar{a}$ represents the combination
\begin{equation}
\bar{a}=\lambda+\lambda^{2}c.
\end{equation}
Note that in the $\lambda\rightarrow0$ limit, the equivalent quadratic
action of the full theory (\ref{eq:EQCA-BI}) reduces to the second
order of the full theory in small curvature expansion (\ref{flat_space_o(R^2)})
as expected. 

Let us list the conditions that our full theory should satisfy:
\begin{enumerate}
\item It should reduce to the cosmological Einstein or Einstein-Gauss-Bonnet
theory at the lowest order,
\item It should describe unitary massless spin-2 excitations at any finite
order in the curvature expansion and infinite order in the curvature
expansion. We have shown that if the theory is unitary at $O\left(R^{4}\right)$
it is unitary at any order of the form $O\left(R^{4+i}\right)$ including
$i\rightarrow\infty$. Therefore, together with the first condition,
once exact unitarity of the theory is checked all that is required
is to check the unitarity at $O\left(R^{3}\right)$.
\end{enumerate}

\subsection{Reduction to Cosmological Einstein Theory:}

To reduce (\ref{eq:EQCA-BI}) to Einstein's theory one should set
$\alpha_{1}=\alpha_{2}=\alpha_{3}=0$. These conditions together with
the condition that the theory is unitary at $O\left(R^{2}\right)$,
that are
\begin{equation}
b_{1}=0,\qquad a_{3}=-\frac{1}{2}-b_{2},\qquad a_{4}=\frac{\beta\left(\beta+2\right)}{2}+1-b_{3},
\end{equation}
lead to the following relation 
\begin{equation}
0=\tilde{\kappa}\left(1+\bar{a}\right)\left(-\lambda+1\right)-1,
\end{equation}
which is obtained from $\alpha_{2}=0$, and
\begin{equation}
0=\tilde{\kappa}\left\{ \left(1+\bar{a}\right)\left[2\lambda\left(\frac{\beta\left(\beta+2\right)}{2}+1\right)-1\right]-\lambda\left[-2\left(\frac{1}{2}+b_{2}\right)\lambda+\beta+1\right]^{2}\right\} +1,
\end{equation}
which is obtained from $\alpha_{3}=0$. Note that for $c=-\frac{1}{2}$,
$\frac{1}{\tilde{\kappa}}$ and $\overline{a}$ take the forms 
\begin{equation}
\frac{1}{\tilde{\kappa}}=1-\lambda^{3}+\frac{3\lambda^{2}}{2},\qquad\bar{a}=\lambda-\frac{\lambda^{2}}{2}.\label{eq:k_tilde_ads}
\end{equation}
With these results, $\alpha_{2}=0$ condition becomes

\begin{equation}
\lambda\left(\lambda-2\right)=0,
\end{equation}
which is consistent only if $\lambda_{0}=2$ or $\lambda_{0}=0$,
namely $\lambda=\lambda_{0}$. We have studied the $\lambda=0$ case
before. For the other case, that is $\lambda=2$, the theory is not
unitary since $\tilde{\kappa}=-1$ as follows from (\ref{eq:k_tilde_ads}).
This says that we cannot reduce our theory to the cosmological Einstein
theory.

\subsection{Reduction to Einstein-Gauss-Bonnet Theory:}

The next possible option is to try to reduce it to the EGB theory.
Unitarity of the theory at $O\left(R^{2}\right)$ yields
\begin{equation}
a_{3}=\frac{2}{3}b_{1}-b_{2}-\frac{1}{2},\qquad a_{4}=\frac{\beta\left(\beta+2\right)}{2}-\frac{8}{3}b_{1}-b_{3}+1,\label{eq:a3_a4_of_EGB_reduction}
\end{equation}
and using in (\ref{eq:k_eff_ECQA}) and (\ref{eq:L0_eff_ECQA}), one
gets the effective Newton's constant and the effective ``bare''
cosmological constant as

\begin{equation}
\frac{1}{\tilde{\kappa}}=1+\bar{a}-\lambda\left(2\lambda\left(\frac{2}{3}b_{1}-\frac{1}{2}\right)+1\right)^{2},\label{eq:k_tilde_of_EGB_reduction}
\end{equation}
\begin{equation}
\tilde{\lambda}_{0}=\tilde{\kappa}\left[\lambda\left(1+\bar{a}\right)\left(2\lambda\left(\frac{2}{3}b_{1}-\frac{1}{2}\right)+1\right)-\bar{a}\left(2+\bar{a}\right)+\lambda_{0}\right]+\lambda.
\end{equation}
In addition, the quadratic curvature parameters of EQCA also become
\begin{equation}
\alpha_{1}=\gamma b_{1}\tilde{\kappa}\left(1+\bar{a}\right),
\end{equation}
\begin{equation}
\alpha_{2}=\frac{\gamma}{2\lambda}\left[\tilde{\kappa}\left(1+\bar{a}\right)\left(2\lambda\left(\frac{2}{3}b_{1}-\frac{1}{2}\right)+1\right)-1\right],
\end{equation}
\begin{equation}
\alpha_{3}=\frac{\gamma}{2\lambda}\left[\tilde{\kappa}\left(\left(1+\bar{a}\right)\left(2\lambda\left(\frac{\beta\left(\beta+2\right)}{2}-\frac{8}{3}b_{1}+1\right)-1\right)-\lambda\left(2a_{3}\lambda+\beta+1\right)^{2}\right)+1\right].
\end{equation}
Here, $\bar{a}$ represents the combination
\begin{equation}
\bar{a}=\lambda+\lambda^{2}\left(\frac{2}{3}b_{1}-\frac{1}{2}\right).\label{bar_a_}
\end{equation}
To reduce our theory to the EGB theory, we must impose two conditions
that are 
\[
\frac{\alpha_{2}}{\alpha_{1}}=\frac{2}{3},\qquad\frac{\alpha_{3}}{\alpha_{1}}=-\frac{8}{3},
\]
which respectively lead to the following two equations
\begin{align}
\frac{3\gamma}{2\lambda}\left[\tilde{\kappa}\left(1+\bar{a}\right)\left(2\lambda\left(\frac{2}{3}b_{1}-\frac{1}{2}\right)+1\right)-1\right]= & 2\gamma b_{1}\tilde{\kappa}\left(1+\bar{a}\right),\label{EGB_uni_cond_1_ads}
\end{align}
\begin{multline}
\frac{3\gamma}{2\lambda}\left[\tilde{\kappa}\left(\left(1+\bar{a}\right)\left(2\lambda\left(\frac{\beta\left(\beta+2\right)}{2}-\frac{8}{3}b_{1}+1\right)-1\right)-\lambda\left(2a_{3}\lambda+\beta+1\right)^{2}\right)+1\right]\\
=-8\gamma b_{1}\tilde{\kappa}\left(1+\bar{a}\right).\label{EGB_uni_cond_2_ads}
\end{multline}
Simplification of (\ref{EGB_uni_cond_1_ads}) yields 
\begin{align}
\left(b_{1}-\frac{9}{8}\right)\left(b_{1}-\frac{3}{4}\right)\lambda= & -\frac{3}{2}\left(b_{1}-\frac{9}{8}\right).\label{ads_cond_1}
\end{align}
Note that it is immediately clear that $b_{1}\ne\frac{3}{4}$. Here,
the discussion bifurcates: Either $b_{1}\ne\frac{9}{8}$ or $b_{1}=\frac{9}{8}$.
We have to study both cases:

\subsection*{Case 1: $b_{1}\protect\ne\frac{9}{8}$ }

Then, from (\ref{ads_cond_1}), one gets 
\begin{align}
\lambda= & \frac{-3}{2\left(b_{1}-\frac{3}{4}\right)},\label{ads_lambda}
\end{align}
and inserting this in (\ref{bar_a_}) leads to $\bar{a}=0$ which
also gives $\frac{1}{\tilde{\kappa}}=1-\lambda$; therefore, one has
the constraint $\lambda<1$ for the unitarity of the theory. From
(\ref{ads_lambda}) $\lambda<1$ yields $\left|b_{1}\right|>\frac{3}{4}$.
Note that, this condition on $b_{1}$also guarantees that $\lambda\ne1$.
Now, let us look at the second constraint (\ref{EGB_uni_cond_2_ads})
which simplifies to 

\begin{align}
-4a_{3}\lambda\left(a_{3}\lambda+\beta+1\right)= & 0.
\end{align}
Since we are studying the $\lambda\ne0$ case, this equation is satisfied
when either $a_{3}=0$ or $a_{3}\lambda+\beta+1=0$. We must consider
these subclasses separately.

\subsubsection*{Case 1a: $a_{3}=0$}

Using (\ref{EGB_reduction_conditions}) $b_{2}$ can be determined
as
\begin{equation}
b_{2}=\frac{2}{3}\left(b_{1}-\frac{3}{4}\right).
\end{equation}
Making use of (\ref{ads_lambda}) one obtains $b_{2}=-\frac{1}{\lambda}$.
Since $b_{2}=c$ in this case the vacuum equation (\ref{vacuum_eq})
leads to $\lambda=\lambda_{0}$. Then we have the following $A_{\mu\nu}$
tensor
\begin{align}
A_{\mu\nu}= & R_{\mu\nu}+\beta S_{\mu\nu}\nonumber \\
 & +\gamma\left(a_{2}C_{\mu\rho\nu\sigma}R^{\rho\sigma}+\left(\frac{\beta\left(\beta+2\right)}{2}+\frac{4}{\lambda_{0}}-b_{3}-1\right)S_{\mu\rho}S_{\nu}^{\rho}\right)\nonumber \\
 & +\frac{\gamma}{4}g_{\mu\nu}\left(\left(\frac{3}{4}-\frac{3}{2\lambda_{0}}\right)C_{\rho\sigma\lambda\gamma}C^{\rho\sigma\lambda\gamma}-\frac{1}{\lambda_{0}}R_{\rho\sigma}R^{\rho\sigma}+b_{3}S_{\rho\sigma}S^{\rho\sigma}\right).\label{EGB_Amn_2}
\end{align}

\subsubsection*{Case 1b: $a_{3}\lambda+\beta+1=0$}

Together with (\ref{ads_lambda}) one has 
\begin{equation}
a_{3}=\frac{2}{3}\left(\beta+1\right)\left(b_{1}-\frac{3}{4}\right).\label{EGB_a3}
\end{equation}
Then from (\ref{EGB_reduction_conditions}) $b_{2}$ can be determined
as 
\begin{align}
b_{2}= & -\frac{2}{3}\beta\left(b_{1}-\frac{3}{4}\right),\label{EGB_b2}
\end{align}
which leads to 
\begin{align}
c & =\frac{2}{3}\left(b_{1}-\frac{3}{4}\right).\label{EGB_c}
\end{align}
From (\ref{vacuum_eq}) these lead to $\lambda=\lambda_{0}$, $b_{1}=\frac{3}{4}-\frac{3}{2\lambda_{0}}$,
$b_{2}=\frac{\beta}{\lambda_{0}}$ and $a_{3}=-\frac{\beta+1}{\lambda_{0}}$
yielding 
\begin{align}
A_{\mu\nu}= & R_{\mu\nu}+\beta S_{\mu\nu}\nonumber \\
 & +\gamma\left(a_{2}C_{\mu\rho\nu\sigma}R^{\rho\sigma}-\frac{\beta+1}{\lambda_{0}}R_{\mu\rho}R_{\nu}^{\rho}+\left(\frac{\beta\left(\beta+2\right)}{2}+\frac{4}{\lambda_{0}}-b_{3}-1\right)S_{\mu\rho}S_{\nu}^{\rho}\right)\nonumber \\
 & +\frac{\gamma}{4}g_{\mu\nu}\left(\left(\frac{3}{4}-\frac{3}{2\lambda_{0}}\right)C_{\rho\sigma\lambda\gamma}C^{\rho\sigma\lambda\gamma}+\frac{\beta}{\lambda_{0}}R_{\rho\sigma}R^{\rho\sigma}+b_{3}S_{\rho\sigma}S^{\rho\sigma}\right).\label{EGB_Amn_1}
\end{align}
Note that, since $\lambda_{0}$ appears in the inverse power there
is no $\lambda_{0}\rightarrow0$ limit for (\ref{EGB_Amn_2}) and
(\ref{EGB_Amn_1}). Therefore, we will not study these theories anymore
even though they describe unitary massless spin-2 excitations at all
orders in the curvature expansion about their (A)dS vacuum. Let us
study the second case.

\subsection*{Case 2: $b_{1}=\frac{9}{8}$}

In this case, $\lambda$ is not determined from (\ref{ads_cond_1}).
This choice reduces (\ref{eq:a3_a4_of_EGB_reduction}) and (\ref{eq:k_tilde_of_EGB_reduction})
to 
\begin{equation}
a_{3}=\frac{1}{4}-b_{2},\qquad a_{4}=\frac{\beta\left(\beta+2\right)}{2}-2-b_{3},
\end{equation}
\begin{align}
\frac{1}{\tilde{\kappa}} & =\left(1-\lambda\right)\left(1+\frac{\lambda}{2}\right)^{2}.\label{9/8_tilde_k}
\end{align}
Again, positivity of the Newton's constant leads to $\lambda<1$,
and we demand that $\lambda\ne-2$, so that the Newton's constant
does not vanish. The vacuum equation (\ref{vacuum_eq}) boils down
to 
\begin{align}
\frac{\lambda^{4}}{16}+\frac{\lambda^{3}}{4}-\lambda+\lambda_{0} & =0.\label{9/8_case_vacuum}
\end{align}
The solutions of this equations were discussed in Section-IV hence
we do not repeat here, but just note that there is a unique viable
solution with $\lambda<1$ as long as $\lambda_{0}<\frac{11}{16}$.
Note also that for $\lambda\ne-2$, one must have $\lambda_{0}\ne-1$.
 The second condition (\ref{EGB_uni_cond_2_ads}) gives 
\begin{align}
\frac{1}{2}\left(\lambda+2\right)\left(\beta+1\right)= & \pm\left(2a_{3}\lambda+\beta+1\right).\label{9/8_case}
\end{align}
We must study both signs separately.

Let us consider the minus sign case which yields 
\begin{equation}
a_{3}=-\frac{\left(\lambda+4\right)\left(\beta+1\right)}{4\lambda}.
\end{equation}
Since we would like to have a smooth $\lambda\rightarrow0$ limit,
we must have $\beta=-1$ and $a_{3}=0$. Then, the theory is 
\begin{align}
A_{\mu\nu}= & R_{\mu\nu}-S_{\mu\nu}\nonumber \\
 & +\gamma\left(a_{2}C_{\mu\rho\nu\sigma}R^{\rho\sigma}-\left(b_{3}+\frac{5}{2}\right)S_{\mu\rho}S_{\nu}^{\rho}\right)\nonumber \\
 & +\frac{\gamma}{4}g_{\mu\nu}\left(\frac{9}{8}C_{\rho\sigma\lambda\gamma}C^{\rho\sigma\lambda\gamma}+\frac{1}{4}R_{\rho\sigma}R^{\rho\sigma}+b_{3}S_{\rho\sigma}S^{\rho\sigma}\right),\label{EGB_Amn_4}
\end{align}
which will also appear as a subcase below.

Let us consider the final case that is choosing the plus sign in (\ref{9/8_case})
which leads to $a_{3}=\frac{\beta+1}{4}$. The $A_{\mu\nu}$ tensor
reads\begin{empheq}[box=\widefbox]{align}
A_{\mu\nu}= & R_{\mu\nu} + \beta S_{\mu\nu} \nonumber \\
& + \gamma \left( a_{2} C_{\mu\rho\nu\sigma} R^{\rho\sigma} + \frac{\beta + 1}{4} R_{\mu\rho} R_{\nu}^{\rho} + \left( \frac{\beta\left(\beta+2\right)}{2} - 2 - b_{3} \right) S_{\mu\rho} S_{\nu}^{\rho} \right) \nonumber \\
& + \frac{\gamma}{4} g_{\mu\nu} \left( \frac{9}{8} C_{\rho\sigma\lambda\gamma} C^{\rho\sigma\lambda\gamma} - \frac{\beta}{4} R_{\rho\sigma} R^{\rho\sigma} + b_{3} S_{\rho\sigma} S^{\rho\sigma} \right), \label{EGB_Amn_3} 
\end{empheq}with $a_{2}$, $b_{3}$ and $\beta$ are arbitrary real parameters. 

Let us summarize the properties of this theory:
\begin{enumerate}
\item With a given $\lambda_{0}<\frac{11}{16}$, it has a unique viable
maximally symmetric vacuum with a cosmological parameter $\lambda<1$,
and an effective Newton's constant $\frac{1}{\tilde{\kappa}}=\left(1-\lambda\right)\left(1+\frac{\lambda}{2}\right)^{2}.$
\item It describes a unitary massless spin-2 excitation around this vacuum
for any value of $\lambda_{0}<\frac{11}{16}$ including $\lambda_{0}=0$,
except $\lambda_{0}=-1$, which yields $\lambda=-2$ and so ruled
out by the requirement of a non-zero effective Newton's constant.
This statement means that the theory has the same propagator structure
as Einstein's gravity in (A)dS and flat backgrounds.
\item It provides an infinite order unitary extension of Einstein's gravity.
\end{enumerate}
All these features are quite attractive but we still have to show
that the theory is also healthy at the truncated orders $O\left(R^{2}\right)$
and $O\left(R^{3}\right)$. At $O\left(R^{2}\right)$ since the theory
is equivalent to the Einstein-Gauss-Bonnet theory it is unitary as
long as $\kappa$ is positive. Let us now check the $O\left(R^{3}\right)$
theory. Expanding the Lagrangian density built with (\ref{EGB_Amn_3})
up to $O\left(R^{3}\right)$ we arrive at 
\begin{align}
\kappa\mathcal{L}_{O\left(R^{3}\right)}= & R-2\Lambda_{0}+\gamma\left(-\frac{2b_{1}}{3}-\frac{a_{4}+b_{3}}{4}+\frac{\left(\beta+1\right)^{2}}{8}+\frac{1}{8}\right)R^{2}\nonumber \\
+ & \gamma\left(2b_{1}+b_{2}+a_{3}+b_{3}+a_{4}-\frac{\left(\beta+1\right)^{2}}{2}\right)R_{\mu\nu}R^{\mu\nu}+\gamma b_{1}\chi_{\text{GB}}\nonumber \\
+ & \gamma^{2}\frac{b_{1}}{4}RR_{\mu\rho\nu\sigma}R^{\mu\rho\nu\sigma}-\gamma^{2}\left(\beta+1\right)a_{2}R_{\mu\nu}R^{\mu\sigma\nu\rho}R_{\sigma\rho}\nonumber \\
+ & \gamma^{2}\left(\frac{7\left(\beta+1\right)a_{2}}{6}+\frac{b_{2}+b_{3}-2b_{1}}{4}+\frac{\left(3\beta+4\right)a_{4}}{4}+\frac{\left(\beta+2\right)a_{3}}{4}-\frac{\left(\beta+1\right)^{3}}{4}\right)RR_{\mu\nu}R^{\mu\nu}\nonumber \\
+ & \gamma^{2}\left(\frac{b_{1}}{12}-\frac{\left(2\beta+3\right)a_{4}}{16}-\frac{b_{3}}{16}-\frac{\left(\beta+1\right)a_{2}}{6}+\frac{\left(\beta+1\right)^{3}}{24}\right)R^{3}\nonumber \\
+ & \gamma^{2}\left(\beta+1\right)\left(-a_{2}-a_{3}-a_{4}+\frac{\left(\beta+1\right)^{2}}{3}\right)R_{\mu\nu}R_{\rho}^{\nu}R^{\rho\mu}.\label{gen_cubic_lagrangian}
\end{align}
The question is if one takes this theory as the full theory what kind
of excitations will it have? We can answer this question with the
methods we have employed several times in this work. Namely we can
construct an equivalent quadratic action that has the same vacuum
and excitations as this theory. Using the above cubic curvature parameters
in (\ref{geff}) yields the EQCA parameters for (\ref{gen_cubic_lagrangian})
as
\begin{align}
\frac{1}{\tilde{\kappa}}\equiv & \frac{1}{\kappa}\left[1-3\lambda^{2}\left(a_{3}+b_{2}\right)\right],\\
\tilde{\lambda}_{0}\equiv & \frac{\tilde{\kappa}}{\kappa}\lambda_{0}+\frac{2\lambda}{3}\left(1-\frac{\tilde{\kappa}}{\kappa}\right),\\
\kappa\tilde{\alpha}\equiv & \gamma\left(-\frac{2b_{1}}{3}-\frac{a_{4}+b_{3}}{4}+\frac{\left(\beta+1\right)^{2}}{8}+\frac{1}{8}\right)+\frac{\lambda\gamma}{3}\left(-2b_{1}+\frac{3b_{2}}{2}-\frac{3b_{3}}{4}+\frac{3\left(\beta+2\right)a_{3}}{2}-\frac{3a_{4}}{4}\right),\\
\kappa\tilde{\beta}\equiv & \gamma\left(2b_{1}+b_{2}+a_{3}+b_{3}+a_{4}-\frac{\left(\beta+1\right)^{2}}{2}\right)+\lambda\gamma\left(2b_{1}+b_{2}-\left(2\beta+1\right)a_{3}+b_{3}+a_{4}\right),\\
\kappa\tilde{\gamma}\equiv & \gamma\left(1+\lambda\right)b_{1}.
\end{align}
For this theory to describe unitary massless spin-2 excitations we
must set $\tilde{\alpha}=0=\tilde{\beta}$. These conditions are automatically
satisfied because of the conditions (\ref{EGB_uni_cond_1_ads}) and
(\ref{EGB_uni_cond_2_ads}) of the full unitary theory. We only need
to show that the effective Newton's constant remains positive: Thus
we have
\begin{equation}
\frac{1}{\kappa}\left(1-\frac{3\lambda^{2}}{4}\right)>0,
\end{equation}
which is satisfied only if $-\frac{2}{\sqrt{3}}<\lambda<\frac{2}{\sqrt{3}}$.
The upper bound is weaker than $\lambda<1$ but a lower bound is introduced.
Thus, unitarity of our theory at $O\left(R^{3}\right)$ is achieved
if $-\frac{2}{\sqrt{3}}<\lambda<1$. Of course now the vacuum equation
should allow such a solution. Here the vacuum equation at this order
is 
\begin{equation}
\lambda^{3}-4\lambda+4\lambda_{0}=0,
\end{equation}
and if $-\frac{4}{3\sqrt{3}}<\lambda_{0}<\frac{4}{3\sqrt{3}}$ then
there is such a real $\lambda$. Observe that the upper bound is larger
than $\frac{11}{16}$. Hence, the condition on $\lambda_{0}$ is $-\frac{4}{3\sqrt{3}}<\lambda_{0}<\frac{11}{16}$. 

Therefore, with these constraints coming from the unitarity of the
theory at $O\left(R^{3}\right)$ we can now summarize the properties
of the theory (\ref{EGB_Amn_3}) as: It describes a unitary massless
spin-2 excitation about its unique viable vacuum (with $-\frac{2}{\sqrt{3}}<\lambda<1$)
at every order in the curvature expansion including the infinite order
expansion as long as $\kappa>0$, $-\frac{4}{3\sqrt{3}}<\lambda_{0}<\frac{11}{16}$
for arbitrary real $\beta$, $a_{2}$ and $b_{3}$. What is fascinating
is that no new condition arises at any $O\left(R^{4+i}\right)$ expansion.
Namely, at every such order, contributions to the effective parameters
vanish among each other, therefore, for example effective Newton's
constant or the vacuum equation do not receive any corrections from
the terms of the $O\left(R^{4+i}\right)$ theory. This is the first
known theory in four dimensions which is unitary at every order in
the curvature expansion in its (A)dS vacuum. 

Having three arbitrary parameters at our disposal we can define various
minimal theories out of which one is particularly interesting: For
$\beta=-1$, $a_{2}=0$ and $b_{3}=-\frac{5}{2}$, one has the BI
action 
\begin{equation}
I=\frac{2}{\kappa\gamma}\int d^{4}x\left\{ \sqrt{\text{-det}\left[g_{\mu\nu}+\frac{\gamma}{4}g_{\mu\nu}R+\frac{9\gamma^{2}}{32}g_{\mu\nu}\left(\chi_{\text{GB}}-\frac{1}{9}R^{2}\right)\right]}-\left(\lambda_{0}+1\right)\sqrt{-\text{det}\, g}\right\} ,
\end{equation}
which actually can be recast as 
\begin{equation}
I=\frac{2}{\kappa\gamma}\int d^{4}x\sqrt{-\text{det}\, g}\left\{ \left[1+\frac{\gamma}{4}R+\frac{9\gamma^{2}}{32}\left(\chi_{\text{GB}}-\frac{1}{9}R^{2}\right)\right]^{2}-\left(\lambda_{0}+1\right)\right\} ,\label{final_action}
\end{equation}
or more explicitly 
\begin{equation}
I=\frac{1}{\kappa}\int d^{4}x\sqrt{-\text{det}\, g}\left\{ R-2\Lambda_{0}-\frac{\gamma^{2}}{32}R^{3}+\frac{9\gamma^{2}}{32}R\chi_{\text{GB}}+\frac{\gamma^{3}}{512}R^{4}-\frac{9\gamma^{3}}{256}R^{2}\chi_{\text{GB}}+\frac{81\gamma^{3}}{512}\chi_{\text{GB}}^{2}\right\} ,
\end{equation}
where we dropped the boundary term. The important point here is that
as an $O\left(R^{4}\right)$ theory, this describes massless unitary
excitations about its (A)dS vacuum; but, it also describes massless
unitary excitations at order $O\left(R^{i}\right)$ for $i\leq4$,
when expanded in small curvature.

\section{Conclusion and Further Discussions}

Using physical requirements such as the existence of a unique viable
maximally symmetric vacuum with a zero or a non-zero curvature, unitary
massless spin-2 excitations about this vacuum at tree-level, and the
reduction to the cosmological Einstein theory for weak field gravity,
we have constructed Born-Infeld gravity actions with the metric being
the only independent variable following the route of \cite{Deser_Gibbons}.
To the best of our knowledge, the theory we have constructed is the
only known theory in four dimensions that is unitary at \emph{every
order} in the curvature expansion about its (A)dS vacuum. 

One interesting observation is that four dimensional Gauss-Bonnet
term, being a total derivative, which has no classical effect, plays
an important role in the construction of the actions: Namely, at the
lowest order BI gravity reduces to the Einstein-Gauss-Bonnet theory
and not to its classically equivalent partner the Einstein's theory.
In addition to the above mentioned physical requirements, we have
also employed the notion of minimality which is essentially constructing
determinantal actions which are as simple as possible, and that do
not involve many powers of curvature and derivatives of curvature.
This leads to a quadratic theory inside the determinant. In the most
general form, the set of such theories has three dimensionless and
one dimensionful parameter which is the BI parameter that comes from
the coefficient of the Gauss-Bonnet term. To further restrict the
viable BI theories, one must turn to their phenomenological applications.
By construction, the theory matches Einstein's gravity for small curvature;
and hence, deviations from the results of Einstein's theory should
be expected at the strong gravity regime.

In this work, we have concentrated in pure gravity and not worried
about matter couplings which can be either done with the usual way
of assuming a $\int d^{4}x\sqrt{-g}g^{\mu\nu}T_{\mu\nu}$ type interaction
in the action or in the non-minimal way by inserting matter fields
into the determinant. As an example to the latter case, one can couple
Maxwell theory by simply taking $A_{\mu\nu}\rightarrow A_{\mu\nu}+\alpha F_{\mu\nu}$,
with $F_{\mu\nu}$ being the field strength tensor. Conformally invariant
versions of the actions can also be found following \cite{Dengiz,Dengiz_Tanhayi}.

We shall study cosmological and black hole type solutions in a separate
work, but here with the tools in our hands, we can find some exact
solutions of the BI gravity (\ref{EGB_Amn_3}). These solutions are
the AdS-wave solutions of the cosmological Einstein's theory \cite{GursesPRL,Gurses-Spherical,Gurses-AdSandpp}.
These solutions not only solve the exact cosmological Einstein's theory,
but also its linearized version. These solutions remain intact in
the BI gravity (\ref{EGB_Amn_3}), and the only thing that one needs
to change is the effective cosmological constant which can be found
from the vacuum equation (\ref{9/8_case_vacuum}). This comes from
the fact that the equivalent quadratic curvature action of a theory
determines the linearized field equations which in turn determine
the properties of its AdS-wave solutions \cite{GursesPRL,Gurses-AdSandpp},
and here we have shown that for BI-gravity this action is the Einstein--Gauss-Bonnet
action whose linearized field equations are the same as the Einstein's
theory. The same fact gives a way to construct the conserved charges
of the BI theory which we now do.

\subsection*{Conserved Charges in the BI Gravity}

The conserved charges of a given $f\left(R_{\alpha\beta}^{\mu\nu}\right)$
theory can be written in terms of the conserved charges of cosmological
Einstein's gravity as was shown in \cite{Senturk}. This follows from
the linearized field equations of the generic $f\left(R_{\alpha\beta}^{\mu\nu}\right)$
theory given in Appendix-\ref{sec:Linearization-of-the} and the charge
construction in \cite{Deser_Tekin-PRL,Deser_Tekin}. Without going
into further details, let us recall the expression in \cite{Senturk}
\begin{align}
Q_{f}^{0}(\bar{\xi})= & \left(\frac{1}{\tilde{\kappa}}+\frac{4\Lambda n}{n-2}\alpha+\frac{4\Lambda}{n-2}\beta+\frac{4\Lambda\left(n-3\right)\left(n-4\right)}{\left(n-1\right)\left(n-2\right)}\gamma\right)Q_{\text{Einstein}}^{0}(\bar{\xi}),\label{eq:Q_F}
\end{align}
where $\bar{\xi}$ is the background Killing vector which for energy
reads $\bar{\xi}^{\mu}=\left(-1,0,0,0\right)$. $Q_{\text{Einstein}}^{0}$
is the Abbott-Deser charge for asymptotically (A)dS spacetimes in
cosmological Einstein's gravity \cite{Abbott-Deser}. For the viable
BI-gravity theory given in (\ref{EGB_Amn_3}), $\alpha=0$, $\beta=0$,
and $n=4$; hence, the conserved charges of asymptotically (A)dS spacetimes
read 
\begin{equation}
Q_{{\rm BI}}^{0}(\bar{\xi})=\left(1-\lambda\right)\left(1+\frac{\lambda}{2}\right)^{2}Q_{\text{Einstein}}^{0}(\bar{\xi}).
\end{equation}
For example, for asymptotically rotating (A)dS-Schwarzschild black
hole%
\footnote{By construction, this solution asymptotically exists in BI gravity
since at large distances cosmological Einstein's theory is dominant.%
}, the energy and the angular momentum read 
\[
E=\left(1-\lambda\right)\left(1+\frac{\lambda}{2}\right)^{2}m,\qquad J=\left(1-\lambda\right)\left(1+\frac{\lambda}{2}\right)^{2}ma,
\]
where $m$ is the mass parameter and $a$ is the rotation parameter.
It is also clear that the black hole has a positive mass when the
graviton has a positive kinetic energy, that is $\lambda<1$.

\section{Acknowledgment}

I.~G. and B.~T. are supported by the T\"{U}B\.{I}TAK grant 113F155.
T.~C.~S. is supported by the Fondecyt grant 3140127. The Centro
de Estudios Cient\'{i}ficos (CECS) is funded by the Chilean Government
through the Centers of Excellence Base Financing Program of Conicyt.

\appendix

\section{Naive BI Gravity\label{sec:Naive-BI-Gravity}}

Let us consider the following BI-gravity action:
\begin{equation}
I=-\frac{2}{\kappa\gamma}\int d^{n}x\,\left[\sqrt{-\det\left(g_{\mu\nu}+\gamma G_{\mu\nu}+\gamma\beta g_{\mu\nu}R\right)}-\sqrt{-\det g}\right],
\end{equation}
Expanding to the $O\left(R^{2}\right)$ yields
\begin{align}
\mathcal{L}_{O\left(R^{2}\right)} & =\frac{n-2}{2\kappa}\left[R+\frac{\gamma}{n-2}\left(R_{\mu\nu}R^{\mu\nu}-\frac{n^{2}-6n+12}{8}R^{2}\right)\right]-\beta\frac{n}{\kappa}\left[R+\frac{n-2}{4}\gamma\beta R^{2}\right].
\end{align}
For $n=3$ and $\beta=0$ one gets the BINMG action which describes
a unitary massive spin-2 graviton but for any other dimension there
is a massive spin-2 ghost due to $R_{\mu\nu}R^{\mu\nu}$ term \cite{Stelle}.

\section{Conversions Between CSR Basis and RRR Basis\label{sec:Conversions-Between-Basis}}

In this Appendix, we discuss the conversions between the Weyl--traceless-Ricci--Ricci
(CSR) basis and Riemann--Ricci--curvature-scalar (RRR) basis.

The $A_{\mu\nu}$ tensor written in the CSR basis, that is
\begin{align}
A_{\mu\nu}= & R_{\mu\nu}+\beta S_{\mu\nu}\nonumber \\
 & +\gamma\left(a_{1}C_{\mu\rho\sigma\lambda}C_{\nu}^{\phantom{\nu}\rho\sigma\lambda}+a_{2}C_{\mu\rho\nu\sigma}R^{\rho\sigma}+a_{3}R_{\mu\rho}R_{\nu}^{\rho}+a_{4}S_{\mu\rho}S_{\nu}^{\rho}\right)\nonumber \\
 & +\frac{\gamma}{4}g_{\mu\nu}\left(b_{1}C_{\rho\sigma\lambda\gamma}C^{\rho\sigma\lambda\gamma}+b_{2}R_{\rho\sigma}R^{\rho\sigma}+b_{3}S_{\rho\sigma}S^{\rho\sigma}\right),
\end{align}
can be converted to the RRR basis, that is

\begin{align}
A_{\mu\nu}= & \left(1+\tilde{\beta}\right)R_{\mu\nu}-\frac{\tilde{\beta}}{4}g_{\mu\nu}R+c_{1}g_{\mu\nu}R^{2}+c_{2}RR_{\mu\nu}+c_{3}g_{\mu\nu}R_{\rho\sigma}R^{\rho\sigma}\nonumber \\
 & +c_{4}R_{\phantom{\sigma}\mu}^{\sigma}R_{\nu\sigma}+c_{5}R_{\mu\sigma\nu\rho}R^{\sigma\rho}+c_{6}g_{\mu\nu}R_{\rho\sigma\lambda\gamma}R^{\rho\sigma\lambda\gamma}+c_{7}R_{\mu}^{\phantom{\mu}\sigma\rho\tau}R_{\nu\sigma\rho\tau},\label{generic_A}
\end{align}
by using $S_{\mu\nu}=R_{\mu\nu}-\frac{1}{4}g_{\mu\nu}R$ and the definition
of the Weyl tensor in four dimensions 
\begin{equation}
C_{\mu\alpha\nu\beta}=R_{\mu\alpha\nu\beta}-g_{\mu[\nu}R_{\beta]\alpha}+g_{\alpha[\nu}R_{\beta]\mu}+\frac{R}{3}g_{\mu[\nu}g_{\beta]\alpha},
\end{equation}
in place. Then, the coefficients in (\ref{generic_A}) becomes 
\begin{align}
\tilde{\beta}= & \beta,\quad c_{1}=\frac{\gamma}{48}\left(-8a_{1}+8a_{2}+3a_{4}+4b_{1}-3b_{3}\right),\nonumber \\
c_{2}= & \gamma\left(a_{1}-\frac{2}{3}a_{2}-\frac{1}{2}a_{4}\right),\quad c_{3}=\frac{\gamma}{4}\left(2a_{1}-2a_{2}-2b_{1}+b_{2}+b_{3}\right),\nonumber \\
c_{4}= & \gamma\left(-2a_{1}+a_{2}+a_{3}+a_{4}\right),\quad c_{5}=\gamma\left(-2a_{1}+a_{2}\right),\nonumber \\
c_{6}= & \frac{\gamma}{4}b_{1},\quad c_{7}=\gamma a_{1}.\label{conversion_relations}
\end{align}
Sometimes the inverse transformation is also needed; therefore, we
shall give it here
\begin{align}
\beta= & \tilde{\beta},\quad a_{1}=\frac{c_{7}}{\gamma},\quad a_{2}=\frac{1}{\gamma}\left(c_{5}+2c_{7}\right),\nonumber \\
a_{3}= & \frac{1}{\gamma}\left(2c_{2}+c_{4}+\frac{c_{5}}{3}+\frac{2c_{7}}{3}\right),\quad a_{4}=\frac{1}{\gamma}\left(-2c_{2}-\frac{4c_{5}}{3}-\frac{2c_{7}}{3}\right),\nonumber \\
b_{1}= & \frac{4c_{6}}{\gamma},\quad b_{2}=\frac{1}{\gamma}\left(16c_{1}+2c_{2}+4c_{3}+\frac{2c_{5}}{3}+\frac{8}{3}c_{6}\right)\nonumber \\
b_{3}= & \frac{1}{\gamma}\left(-16c_{1}-2c_{2}+\frac{4c_{5}}{3}+\frac{16c_{6}}{3}+2c_{7}\right).\label{Inverse_conversion_relations}
\end{align}

In the RRR basis, the EQCA takes the form

\begin{align}
\mathcal{L}_{\text{EQCA}}= & -2\Lambda_{0}+\frac{2}{3}\left(2+\frac{l_{1}\Lambda}{\gamma}\right)l_{1}\Lambda^{3}+\left\{ 1-l_{1}\Lambda^{2}-\frac{4}{9\gamma}l_{1}^{2}\Lambda^{3}\right\} R\nonumber \\
 & +\frac{1}{\gamma}\left\{ \left(4c_{1}+c_{2}\right)\left(1+\gamma\Lambda+\frac{l_{1}}{3}\Lambda^{2}\right)+\frac{1}{8}\left(\gamma\left(\tilde{\beta}+1\right)+2l_{2}\Lambda\right)^{2}+\frac{1}{2}\left(\frac{\gamma}{2}+\frac{l_{1}}{3}\Lambda\right)^{2}\right\} R^{2}\nonumber \\
 & +\frac{1}{\gamma}\left\{ \left(4c_{3}+c_{4}+c_{5}\right)\left(1+\gamma\Lambda+\frac{l_{1}}{3}\Lambda^{2}\right)-\frac{1}{2}\left(\gamma\left(\tilde{\beta}+1\right)+2l_{2}\Lambda\right)^{2}\right\} R_{\mu\nu}^{2}\nonumber \\
 & +\frac{1}{\gamma}\left(4c_{6}+c_{7}\right)\left(1+\gamma\Lambda+\frac{l_{1}}{3}\Lambda^{2}\right)R_{\mu\sigma\nu\rho}^{2},\label{eq_quad_lag}
\end{align}
where the coefficients read
\begin{equation}
l_{1}=48c_{1}+12c_{2}+12c_{3}+3c_{4}+3c_{5}+8c_{6}+2c_{7},
\end{equation}
\begin{equation}
l_{2}=\frac{1}{3}\left(6c_{2}+3c_{4}+c_{5}+2c_{7}\right).
\end{equation}

\section{An Example on EQCA Construction}

The unitarity discussions using the EQCA construction involve various
Taylor series expansions of functions depending on tensor quantities
which sometimes complicate the inherent physical meaning. To understand
the basic idea of the EQCA and the relation between various expansions,
it may be worth to consider analogue expansions for a function with
a single scalar variable. First, remember that the EQCA of a gravity
theory is given with the second order Taylor series expansion of the
Lagrangian in the curvature around the maximally symmetric background,
$\bar{R}_{\rho\sigma}^{\mu\nu}$, which is either already determined
by using the ELA of the theory or will be determined by using the
EQCA of the theory. Then, for a function $f\left(x\right)$, the analogue
of EQCA is the following second order Taylor series expansion around
$x=\overline{x}$;
\begin{equation}
f_{\text{EQCA}}\left(x\right)=f\left(\bar{x}\right)+f^{\prime}\left(\bar{x}\right)\left(x-\bar{x}\right)+\frac{1}{2}f^{\prime\prime}\left(\bar{x}\right)\left(x-\bar{x}\right)^{2},
\end{equation}
which can be recast in the form
\begin{align}
f_{\text{EQCA}}\left(x\right)= & f\left(\bar{x}\right)-f^{\prime}\left(\bar{x}\right)\bar{x}+\frac{1}{2}f^{\prime\prime}\left(\bar{x}\right)\bar{x}^{2}\nonumber \\
 & +\left[f^{\prime}\left(\bar{x}\right)-f^{\prime\prime}\left(\bar{x}\right)\bar{x}\right]x+\frac{1}{2}f^{\prime\prime}\left(\bar{x}\right)x^{2}.\label{eq:f_EQCA}
\end{align}
Here, note that in the gravitational setting, $O\left(x\right)$ term
represents the Einstein-Hilbert piece and its coefficient is the effective
Newton's constant of the theory whose positivity puts a constraint
on the theory. The $O\left(1\right)$ term determines the effective
``bare'' cosmological constant while the $O\left(x^{2}\right)$
term are the quadratic curvature terms.

In addition to this EQCA expansion, we also discussed the small curvature
expansion of a gravitational theory which corresponds to the Taylor
series expansion of $f\left(x\right)$ around $x=0$ as
\begin{equation}
f\left(x\right)=\sum_{n=0}^{\infty}\frac{f^{\left(n\right)}\left(0\right)}{n!}x^{n}.\label{eq:Maclaurin_of_f}
\end{equation}
Note that unless $f^{\left(n\right)}\left(0\right)$ is zero, each
order in (\ref{eq:Maclaurin_of_f}) will contribute to EQCA as
\begin{equation}
\left(x^{n}\right)_{\text{EQCA}}=\left(1-\frac{3n}{2}+\frac{n^{2}}{2}\right)\bar{x}^{n}+n^{2}\bar{x}^{n-1}x+\frac{n\left(n-1\right)}{2}\bar{x}^{n-2}x^{2}.
\end{equation}
This result implies that to see the contributions to the EQCA of a
gravity theory coming from the $O\left(R^{i}\right)$ terms in the
small curvature expansion of the theory, one needs to look at the
$\Lambda^{i-2}$ terms at the quadratic curvature level, the $\Lambda^{i-1}$
terms in the effective Newton's constant part $\frac{1}{\tilde{\kappa}}$,
and $\Lambda^{i}$ terms in the effective bare cosmological part $\frac{\tilde{\Lambda}_{0}}{\tilde{\kappa}}$.

Another implication of this result is that once the EQCA analogue
of $f\left(x\right)$ is found, there is no need to calculate the
EQCA analogue of any finite order truncation of (\ref{eq:Maclaurin_of_f})
separately. One just needs to have $O\left(\bar{x}^{i}\right)$, $O\left(\bar{x}^{i-1}\right)$,
and $O\left(\bar{x}^{i-2}\right)$ expansions of $O\left(1\right)$,
$O\left(x\right)$, and $O\left(x^{2}\right)$ terms in (\ref{eq:f_EQCA}),
respectively, around $\bar{x}=0$. For example, let us write the $O\left(x^{3}\right)$
truncation of (\ref{eq:Maclaurin_of_f}); 
\[
f_{x^{3}}\left(x\right)=f\left(0\right)+f^{\prime}\left(0\right)x+\frac{1}{2}f^{\prime\prime}\left(0\right)x^{2}+\frac{1}{6}f^{\prime\prime\prime}\left(0\right)x^{3},
\]
whose EQCA analogue expansion is
\begin{align}
f_{x^{3}-\text{EQCA}}\left(x\right)= & f\left(0\right)+\frac{1}{6}f^{\prime\prime\prime}\left(0\right)\bar{x}^{3}\nonumber \\
 & +\left[f^{\prime}\left(0\right)-\frac{1}{2}f^{\prime\prime\prime}\left(0\right)\bar{x}^{2}\right]x\nonumber \\
 & +\frac{1}{2}\left[f^{\prime\prime}\left(0\right)+f^{\prime\prime\prime}\left(0\right)\bar{x}\right]x^{2},\label{eq:EQCA_of_Ox3}
\end{align}
Up to $O\left(\bar{x}^{3}\right)$ expansion of the $O\left(1\right)$
term in (\ref{eq:f_EQCA}), that is $f\left(\bar{x}\right)-f^{\prime}\left(\bar{x}\right)\bar{x}+\frac{1}{2}f^{\prime\prime}\left(\bar{x}\right)\bar{x}^{2}$,
around $\overline{x}=0$ gives the first line of (\ref{eq:EQCA_of_Ox3}).
Then, up to $O\left(\bar{x}^{2}\right)$ expansion of the $O\left(x\right)$
term in (\ref{eq:f_EQCA}), that is $f^{\prime}\left(\bar{x}\right)-f^{\prime\prime}\left(\bar{x}\right)\bar{x}$,
around $\overline{x}=0$ gives the coefficient of $x$ in (\ref{eq:EQCA_of_Ox3}).
Finally, up to $O\left(\bar{x}\right)$ expansion of the $O\left(x^{2}\right)$
term in (\ref{eq:f_EQCA}), that is $\frac{1}{2}f^{\prime\prime}\left(\bar{x}\right)$,
around $\overline{x}=0$ gives the coefficient of $x^{2}$ in (\ref{eq:EQCA_of_Ox3}).

The same approach can be used in the gravitational setting. For example,
the EQCA of the BI-theory defined by (\ref{EGB_Amn_3}) can be obtained
by using (\ref{eq:EQCA-BI}) as 
\begin{equation}
\kappa\mathcal{L}_{\text{EQCA}}=-2\frac{\lambda_{0}}{\gamma}+\frac{\lambda^{3}}{\gamma}\left(1+\frac{3\lambda}{8}\right)+R\left(1-\lambda\right)\left(1+\frac{\lambda}{2}\right)^{2}+\frac{9}{8}\gamma\left(1+\frac{\lambda}{2}\right)^{2}\chi_{\text{GB}}.\label{eq:EQCA_BI-theory_choice}
\end{equation}
Then, the EQCA of the $O\left(R^{3}\right)$ expansion of the BI-theory
defined by (\ref{EGB_Amn_3}) can be obtained by taking up to $\frac{\lambda^{3}}{\gamma}$
order in the cosmological constant term, up to $\lambda^{2}$ order
in the effective Newton's constant term, and up to $\lambda$ order
in the quadratic curvature parameters in (\ref{eq:EQCA_BI-theory_choice})
as 
\begin{equation}
\kappa\mathcal{L}_{\text{EQCA}-O\left(R^{3}\right)}=-2\frac{\lambda_{0}}{\gamma}+\frac{\lambda^{3}}{\gamma}+R\left(1-\frac{3}{4}\lambda^{2}\right)+\frac{9}{8}\gamma\left(1+\lambda\right)\chi_{\text{GB}}.\label{eq:EQCA_of_OR3_of_BI-theory_choice}
\end{equation}
which can also be obtained by using the EQCA result for $\kappa\mathcal{L}_{O\left(R^{3}\right)}$
given in (\ref{gen_cubic_lagrangian}). In addition, the vacuum equation
for $O\left(R^{3}\right)$ truncation can be calculated from (\ref{eq:EQCA_of_OR3_of_BI-theory_choice})
as 
\begin{equation}
\lambda=\frac{\lambda_{0}-\frac{\lambda^{3}}{2}}{\left(1-\frac{3}{4}\lambda^{2}\right)}\Rightarrow\lambda-\lambda_{0}-\frac{1}{4}\lambda^{3}=0,
\end{equation}
which can again be obtained from the vacuum equation of the whole
theory which is
\begin{equation}
\lambda-\lambda_{0}-\frac{\lambda^{3}}{16}\left(4+\lambda\right)=0,
\end{equation}
by eliminating the highest power coming from the $O\left(R^{4}\right)$
truncation.

\section{Linearization of the Field Equations of $f\left(R_{\alpha\beta}^{\mu\nu}\right)$\label{sec:Linearization-of-the}}

In this Appendix, we carry out the linearization of the field equations
of a $f\left(R_{\alpha\beta}^{\mu\nu}\right)$ theory, that is a gravity
theory whose Lagrangian is constructed from the contractions of the
Riemann tensor but not its derivatives, and show that this linearized
field equations are the same as those of a quadratic curvature gravity
theory with redefined parameters. First, note that the field equations
of a $f\left(R_{\alpha\beta}^{\mu\nu}\right)$ theory is 
\begin{align}
\frac{1}{2}\left(g_{\nu\rho}\nabla^{\lambda}\nabla_{\sigma}-g_{\nu\sigma}\nabla^{\lambda}\nabla_{\rho}\right)\frac{\partial f}{\partial R_{\rho\sigma}^{\mu\lambda}}-\frac{1}{2}\left(g_{\mu\rho}\nabla^{\lambda}\nabla_{\sigma}-g_{\mu\sigma}\nabla^{\lambda}\nabla_{\rho}\right)\frac{\partial f}{\partial R_{\rho\sigma}^{\lambda\nu}}\nonumber \\
-\frac{1}{2}\left(\frac{\partial f}{\partial R_{\rho\sigma}^{\mu\lambda}}R_{\rho\sigma\phantom{\lambda}\nu}^{\phantom{\rho\sigma}\lambda}-\frac{\partial f}{\partial R_{\rho\sigma}^{\lambda\nu}}R_{\rho\sigma\phantom{\lambda}\mu}^{\phantom{\rho\sigma}\lambda}\right)-\frac{1}{2}g_{\mu\nu}f\left(R_{\alpha\beta}^{\mu\nu}\right) & =0.\label{field_eq}
\end{align}
In Sec.~\ref{sec:Equivalent-Linear-Action}, we showed that the (A)dS
spacetime solutions of this theory satisfy 
\begin{equation}
-2\zeta\bar{R}_{\mu\nu}+g_{\mu\nu}f\left(\bar{R}_{\rho\sigma}^{\alpha\beta}\right)=0,\label{eq:AdS_back}
\end{equation}
where $\zeta$ is defined in (\ref{zeta_defn}) as $\left[\frac{\partial f}{\partial R_{\rho\sigma}^{\mu\nu}}\right]_{\bar{R}_{\rho\sigma}^{\mu\lambda}}R_{\rho\sigma}^{\mu\nu}\equiv\zeta R$.
Let us linearize (\ref{field_eq}) in the metric perturbation $h_{\mu\nu}\equiv g_{\mu\nu}-\bar{g}_{\mu\nu}$
where $\bar{g}_{\mu\nu}$ is the (A)dS background solving (\ref{eq:AdS_back}).
Starting with the last term in (\ref{field_eq}), which becomes
\begin{equation}
\left[g_{\mu\nu}f\left(R_{\alpha\beta}^{\mu\nu}\right)\right]_{L}=h_{\mu\nu}f\left(\bar{R}_{\alpha\beta}^{\mu\nu}\right)+\bar{g}_{\mu\nu}\left[\frac{\partial f}{\partial R_{\rho\sigma}^{\alpha\beta}}\right]_{\bar{R}_{\rho\sigma}^{\alpha\beta}}\left(R_{\rho\sigma}^{\alpha\beta}\right)_{L},\label{gf}
\end{equation}
and using the equation defining $\zeta$, one finds 
\begin{equation}
\left[\frac{\partial f}{\partial R_{\rho\sigma}^{\mu\lambda}}\right]_{\bar{R}_{\rho\sigma}^{\mu\lambda}}=\zeta\delta_{\mu}^{[\rho}\delta_{\nu}^{\sigma]},\label{eq:Background_value_of_df/dR-App}
\end{equation}
as discussed in Sec.~\ref{sec:Equivalent-Linear-Action}, one gets
\begin{equation}
\left[g_{\mu\nu}f\left(R_{\alpha\beta}^{\mu\nu}\right)\right]_{L}=h_{\mu\nu}f\left(\bar{R}_{\rho\sigma}^{\alpha\beta}\right)+\bar{g}_{\mu\nu}\zeta R_{L}.
\end{equation}
Moving to the first term in the second line of (\ref{field_eq}):
One has the linearization
\begin{equation}
\left(\frac{\partial f}{\partial R_{\rho\sigma}^{\mu\lambda}}R_{\rho\sigma\phantom{\lambda}\nu}^{\phantom{\rho\sigma}\lambda}\right)_{L}=\left[\frac{\partial^{2}f}{\partial R_{\alpha\tau}^{\eta\theta}\partial R_{\rho\sigma}^{\mu\lambda}}\right]_{\bar{R}_{\rho\sigma}^{\mu\lambda}}\left(R_{\alpha\tau}^{\eta\theta}\right)_{L}\bar{R}_{\rho\sigma\phantom{\lambda}\nu}^{\phantom{\rho\sigma}\lambda}+\left[\frac{\partial f}{\partial R_{\rho\sigma}^{\mu\lambda}}\right]_{\bar{R}_{\rho\sigma}^{\mu\lambda}}\left(R_{\rho\sigma\phantom{\lambda}\nu}^{\phantom{\rho\sigma}\lambda}\right)_{L}.\label{eq:Middle_structure}
\end{equation}
Here, remember that $\left[\frac{\partial^{2}f}{\partial R_{\alpha\tau}^{\eta\theta}\partial R_{\rho\sigma}^{\mu\lambda}}\right]_{\bar{R}_{\rho\sigma}^{\mu\lambda}}$
has the following form as we discussed in Sec.~\ref{sec:Equivalent-Linear-Action}
\begin{equation}
\left[\frac{\partial^{2}f}{\partial R_{\alpha\tau}^{\eta\theta}\partial R_{\rho\sigma}^{\mu\lambda}}\right]_{\bar{R}_{\rho\sigma}^{\mu\nu}}=2\alpha\delta_{\eta}^{[\alpha}\delta_{\theta}^{\tau]}\delta_{\mu}^{[\rho}\delta_{\lambda}^{\sigma]}+\beta\left(\delta_{[\eta}^{\alpha}\delta_{\theta]}^{[\rho}\delta_{[\mu}^{\left|\tau\right|}\delta_{\lambda]}^{\sigma]}-\delta_{[\eta}^{\tau}\delta_{\theta]}^{[\rho}\delta_{[\mu}^{\left|\alpha\right|}\delta_{\lambda]}^{\sigma]}\right)+12\gamma\delta_{\eta}^{[\alpha}\delta_{\theta}^{\tau}\delta_{\mu}^{\rho}\delta_{\lambda}^{\sigma]}.\label{eq:Background_value_of_d2f/dR2}
\end{equation}
Using this result together with (\ref{eq:Background_value_of_df/dR-App})
and $\bar{R}_{\rho\sigma}^{\mu\lambda}=\frac{2\Lambda}{\left(n-1\right)\left(n-2\right)}\left(\delta_{\rho}^{\mu}\delta_{\sigma}^{\lambda}-\delta_{\sigma}^{\mu}\delta_{\rho}^{\lambda}\right)$,
one has
\begin{align*}
\left(\frac{\partial f}{\partial R_{\rho\sigma}^{\mu\lambda}}R_{\rho\sigma\phantom{\lambda}\nu}^{\phantom{\rho\sigma}\lambda}\right)_{L}= & -\left(\alpha\frac{4\Lambda}{\left(n-2\right)}+\beta\frac{n\Lambda}{\left(n-1\right)\left(n-2\right)}\right)R_{L}\bar{g}_{\mu\nu}\\
 & +\left(\gamma\frac{8\Lambda\left(n-3\right)}{\left(n-1\right)\left(n-2\right)}-\beta\frac{2\Lambda}{n-1}\right)\left(R_{\mu\nu}^{L}-\frac{1}{2}\bar{g}_{\mu\nu}R_{L}-\frac{2\Lambda}{n-2}h_{\mu\nu}\right)\\
 & -\zeta R_{\mu\nu}^{L}.
\end{align*}
Now, let us linearize the first term in (\ref{field_eq}), $g_{\nu\rho}\nabla^{\lambda}\nabla_{\sigma}\frac{\partial f}{\partial R_{\rho\sigma}^{\mu\lambda}}$,
and to do this, first note that the linearization of the metric compatibility,
$\nabla_{\mu}g_{\nu\rho}=0$, yields
\begin{equation}
\bar{\nabla}_{\mu}h_{\nu\rho}=\left(\Gamma_{\mu\nu}^{\sigma}\right)_{L}\bar{g}_{\sigma\rho}+\left(\Gamma_{\mu\rho}^{\sigma}\right)_{L}\bar{g}_{\nu\sigma}.\label{eq:lin_metric_comp_rel}
\end{equation}
Then, for a two tensor $A_{\mu\nu}$ with the background value $\bar{A}_{\mu\nu}=\bar{a}\bar{g}_{\mu\nu}$,
the linearization of $\nabla_{\mu}A_{\nu\rho}$ yields
\begin{equation}
\left(\nabla_{\mu}A_{\nu\rho}\right)_{L}=\bar{\nabla}_{\mu}A_{\nu\rho}^{L}-\bar{a}\left[\left(\Gamma_{\mu\nu}^{\sigma}\right)_{L}\bar{g}_{\sigma\rho}+\left(\Gamma_{\mu\rho}^{\sigma}\right)_{L}\bar{g}_{\nu\sigma}\right],
\end{equation}
and with (\ref{eq:lin_metric_comp_rel}), one has
\begin{equation}
\left(\nabla_{\mu}A_{\nu\rho}\right)_{L}=\bar{\nabla}_{\mu}A_{\nu\rho}^{L}-\bar{a}\bar{\nabla}_{\mu}h_{\nu\rho}.
\end{equation}
Next, let us consider the linearization of $\left(\nabla_{\sigma}\nabla_{\mu}A_{\nu\rho}\right)_{L}$;
\begin{align}
\left(\nabla_{\sigma}\nabla_{\mu}A_{\nu\rho}\right)_{L}= & \bar{\nabla}_{\sigma}\bar{\nabla}_{\mu}A_{\nu\rho}^{L}-\left(\Gamma_{\sigma\mu}^{\lambda}\right)_{L}\bar{\nabla}_{\lambda}\bar{A}_{\nu\rho}-\left(\Gamma_{\sigma\nu}^{\lambda}\right)_{L}\bar{\nabla}_{\mu}\bar{A}_{\lambda\rho}-\left(\Gamma_{\sigma\rho}^{\lambda}\right)_{L}\bar{\nabla}_{\mu}\bar{A}_{\nu\lambda}\nonumber \\
 & -\bar{\nabla}_{\sigma}\left(\left(\Gamma_{\mu\nu}^{\lambda}\right)_{L}\bar{A}_{\lambda\rho}\right)-\bar{\nabla}_{\sigma}\left(\left(\Gamma_{\mu\rho}^{\lambda}\right)_{L}\bar{A}_{\nu\lambda}\right),
\end{align}
which boils down to
\begin{equation}
\left(\nabla_{\sigma}\nabla_{\mu}A_{\nu\rho}\right)_{L}=\bar{\nabla}_{\sigma}\left(\bar{\nabla}_{\mu}A_{\nu\rho}^{L}-\bar{a}\bar{\nabla}_{\mu}h_{\nu\rho}\right).
\end{equation}
Finally, we have 
\begin{align}
\left(g_{\nu\rho}\nabla^{\lambda}\nabla_{\sigma}\frac{\partial f}{\partial R_{\rho\sigma}^{\mu\lambda}}\right)_{L} & =\bar{g}_{\nu\rho}\bar{g}^{\lambda\beta}\bar{\nabla}_{\beta}\bar{\nabla}_{\sigma}\left(\frac{\partial f}{\partial R_{\rho\sigma}^{\mu\lambda}}\right)_{L}+\bar{g}_{\nu\rho}\bar{g}^{\lambda\beta}\bar{\nabla}_{\beta}\left[\left(\Gamma_{\sigma}^{\alpha}\right)_{L}\left(\frac{\partial f}{\partial R_{\rho\sigma}^{\mu\lambda}}\right)_{\bar{R}_{\rho\sigma}^{\mu\lambda}}\right],
\end{align}
where the second term represents
\begin{align}
\left(\Gamma_{\sigma}^{\alpha}\right)_{L}\left(\frac{\partial f}{\partial R_{\rho\sigma}^{\mu\lambda}}\right)_{\bar{R}_{\rho\sigma}^{\mu\lambda}}\equiv & \left(\Gamma_{\sigma\alpha}^{\rho}\right)_{L}\left(\frac{\partial f}{\partial R_{\alpha\sigma}^{\mu\lambda}}\right)_{\bar{R}_{\rho\sigma}^{\mu\lambda}}+\left(\Gamma_{\sigma\alpha}^{\sigma}\right)_{L}\left(\frac{\partial f}{\partial R_{\rho\alpha}^{\mu\lambda}}\right)_{\bar{R}_{\rho\sigma}^{\mu\lambda}}\nonumber \\
 & -\left(\Gamma_{\sigma\mu}^{\alpha}\right)_{L}\left(\frac{\partial f}{\partial R_{\rho\sigma}^{\alpha\lambda}}\right)_{\bar{R}_{\rho\sigma}^{\mu\lambda}}-\left(\Gamma_{\sigma\lambda}^{\alpha}\right)_{L}\left(\frac{\partial f}{\partial R_{\rho\sigma}^{\mu\alpha}}\right)_{\bar{R}_{\rho\sigma}^{\mu\lambda}},
\end{align}
and using (\ref{eq:Background_value_of_df/dR-App}), one has
\begin{equation}
\left(\Gamma_{\sigma}^{\alpha}\right)_{L}\left(\frac{\partial f}{\partial R_{\rho\sigma}^{\mu\lambda}}\right)_{\bar{R}_{\rho\sigma}^{\mu\lambda}}=0.
\end{equation}
On the other hand, $\left(\frac{\partial f}{\partial R_{\rho\sigma}^{\mu\lambda}}\right)_{L}$
takes the following form by using (\ref{eq:Background_value_of_d2f/dR2});
\begin{align}
\left(\frac{\partial f}{\partial R_{\rho\sigma}^{\mu\lambda}}\right)_{L}= & \left[\frac{\partial^{2}f}{\partial R_{\alpha\tau}^{\eta\theta}\partial R_{\rho\sigma}^{\mu\lambda}}\right]_{\bar{R}_{\rho\sigma}^{\mu\lambda}}\left(R_{\alpha\tau}^{\eta\theta}\right)_{L}\nonumber \\
= & 2\alpha\delta_{\mu}^{[\rho}\delta_{\lambda}^{\sigma]}R_{L}+2\beta\delta_{\theta}^{[\rho}\delta_{[\mu}^{\left|\tau\right|}\delta_{\lambda]}^{\sigma]}\left(R_{\tau}^{\theta}\right)_{L}\nonumber \\
 & +2\gamma\delta_{\mu}^{[\rho}\delta_{\lambda}^{\sigma]}R_{L}-8\gamma\delta_{\theta}^{[\rho}\delta_{[\mu}^{\left|\tau\right|}\delta_{\lambda]}^{\sigma]}\left(R_{\tau}^{\theta}\right)_{L}+2\gamma\delta_{\eta}^{[\rho}\delta_{\theta}^{\sigma]}\left(R_{\mu\lambda}^{\eta\theta}\right)_{L},
\end{align}
Using these results, one arrives at
\begin{align}
\left(g_{\nu\rho}\nabla^{\lambda}\nabla_{\sigma}\frac{\partial f}{\partial R_{\rho\sigma}^{\mu\lambda}}\right)_{L}= & \alpha\left(\bar{g}_{\mu\nu}\bar{\nabla}^{\lambda}\bar{\nabla}_{\lambda}R_{L}-\bar{\nabla}_{\nu}\bar{\nabla}_{\mu}R_{L}\right)\nonumber \\
 & +\frac{\beta}{2}\Biggl[\bar{g}_{\nu\rho}\bar{\nabla}^{\lambda}\bar{\nabla}_{\lambda}\left(R_{\mu}^{\rho}\right)_{L}-\bar{g}_{\nu\rho}\bar{\nabla}^{\lambda}\bar{\nabla}_{\mu}\left(R_{\lambda}^{\rho}\right)_{L}\nonumber \\
 & \phantom{+\frac{\beta}{2}\Biggl[}-\bar{\nabla}_{\nu}\bar{\nabla}_{\sigma}\left(R_{\mu}^{\sigma}\right)_{L}+\bar{g}_{\mu\nu}\bar{\nabla}^{\lambda}\bar{\nabla}_{\sigma}\left(R_{\lambda}^{\sigma}\right)_{L}\Biggr]\nonumber \\
 & +\gamma\Biggl[\bar{g}_{\mu\nu}\bar{\nabla}^{\lambda}\bar{\nabla}_{\lambda}R_{L}-2\bar{g}_{\nu\rho}\bar{\nabla}^{\sigma}\bar{\nabla}_{\sigma}\left(R_{\mu}^{\rho}\right)_{L}\nonumber \\
 & \phantom{+\gamma\Biggl[}+2\bar{g}_{\nu\rho}\bar{\nabla}^{\lambda}\bar{\nabla}_{\mu}\left(R_{\lambda}^{\rho}\right)_{L}+2\bar{g}_{\nu\rho}\bar{\nabla}^{\lambda}\bar{\nabla}_{\sigma}\left(R_{\mu\lambda}^{\rho\sigma}\right)_{L}\Biggr]\nonumber \\
 & -\gamma\left[\bar{\nabla}_{\nu}\bar{\nabla}_{\mu}R_{L}-2\bar{\nabla}_{\nu}\bar{\nabla}_{\sigma}\left(R_{\mu}^{\sigma}\right)_{L}+2\bar{g}_{\mu\nu}\bar{\nabla}^{\lambda}\bar{\nabla}_{\sigma}\left(R_{\lambda}^{\sigma}\right)_{L}\right].
\end{align}
Let us recap the definitions of the linearized Ricci tensor $\left(R_{\mu}^{\rho}\right)_{L}$,
linearized Ricci scalar; 
\begin{equation}
\left(R_{\mu}^{\rho}\right)_{L}=\left(g^{\rho\alpha}R_{\mu\alpha}\right)_{L}=\bar{g}^{\rho\alpha}R_{\mu\alpha}^{L}-\frac{2\Lambda}{n-2}h_{\mu}^{\rho},\qquad R_{L}=\left(R_{\rho}^{\rho}\right)_{L},
\end{equation}
and the linearized Einstein tensor;
\begin{equation}
\mathcal{G}_{\mu\nu}^{L}\equiv R_{\mu\nu}^{L}-\frac{1}{2}\bar{g}_{\mu\nu}R_{L}-\frac{2\Lambda}{n-2}h_{\mu\nu},
\end{equation}
which satisfies the linearized Bianchi identity $\bar{\nabla}^{\mu}\mathcal{G}_{\mu\nu}^{L}=0.$
With these two background tensors and $R_{L}$, one has
\begin{align}
\left(g_{\nu\rho}\nabla^{\lambda}\nabla_{\sigma}\frac{\partial f}{\partial R_{\rho\sigma}^{\mu\lambda}}\right)_{L}= & \alpha\left(\bar{g}_{\mu\nu}\bar{\square}R_{L}-\bar{\nabla}_{\mu}\bar{\nabla}_{\nu}R_{L}\right)\nonumber \\
 & +\frac{\beta}{2}\left[\bar{\square}\mathcal{G}_{\mu\nu}^{L}-\bar{g}_{\nu\rho}\bar{\nabla}^{\lambda}\bar{\nabla}_{\mu}\left(R_{\lambda}^{\rho}\right)_{L}-\frac{1}{2}\bar{\nabla}_{\nu}\bar{\nabla}_{\mu}R_{L}+\bar{g}_{\mu\nu}\bar{\square}R_{L}\right]\nonumber \\
 & +\gamma\left[-2\bar{\square}R_{\mu\nu}^{L}+\frac{4\Lambda}{n-2}\bar{\square}h_{\mu\nu}+2\bar{g}_{\nu\rho}\bar{\nabla}^{\lambda}\bar{\nabla}_{\mu}\left(R_{\lambda}^{\rho}\right)_{L}+2\bar{g}_{\nu\rho}\bar{\nabla}^{\lambda}\bar{\nabla}_{\sigma}\left(R_{\mu\lambda}^{\rho\sigma}\right)_{L}\right],\label{eq:Intermediate_eqn_in_lin}
\end{align}
where $\bar{g}_{\nu\rho}\bar{\nabla}^{\lambda}\bar{\nabla}_{\mu}\left(R_{\lambda}^{\rho}\right)_{L}$
can be calculated as
\begin{equation}
\bar{g}_{\nu\rho}\bar{\nabla}^{\lambda}\bar{\nabla}_{\mu}\left(R_{\lambda}^{\rho}\right)_{L}=\frac{1}{2}\bar{\nabla}_{\mu}\bar{\nabla}_{\nu}R_{L}+\frac{2n\Lambda}{\left(n-1\right)\left(n-2\right)}\mathcal{G}_{\mu\nu}^{L}+\frac{\Lambda}{n-1}\bar{g}_{\mu\nu}R_{L}.
\end{equation}
Finally, the last term in (\ref{eq:Intermediate_eqn_in_lin}) requires
linearized form of 
\begin{equation}
\nabla^{\mu}\nabla_{\nu}R_{\mu\alpha}^{\nu\beta}=\square R_{\alpha}^{\beta}-\nabla^{\mu}\nabla_{\alpha}R_{\mu}^{\beta},
\end{equation}
which can be obtained from the once-contracted Bianchi identity
\begin{equation}
\nabla^{\nu}R_{\mu\alpha\nu\beta}=\nabla_{\mu}R_{\alpha\beta}-\nabla_{\alpha}R_{\mu\beta},
\end{equation}
and the linearization yields
\begin{equation}
\bar{\nabla}^{\mu}\bar{\nabla}_{\nu}\left(R_{\mu\alpha}^{\nu\beta}\right)_{L}=\bar{\square}\left(R_{\alpha}^{\beta}\right)_{L}-\bar{\nabla}^{\mu}\bar{\nabla}_{\alpha}\left(R_{\mu}^{\beta}\right)_{L}.
\end{equation}
Then, putting the pieces together, one arrives at the desired expression
\begin{align}
\left(g_{\nu\rho}\nabla^{\lambda}\nabla_{\sigma}\frac{\partial f}{\partial R_{\rho\sigma}^{\mu\lambda}}\right)_{L}= & \frac{\left(2\alpha+\beta\right)}{2}\left(\bar{g}_{\mu\nu}\bar{\square}R_{L}-\bar{\nabla}_{\mu}\bar{\nabla}_{\nu}R_{L}\right)+\frac{\beta}{2}\bar{\square}\mathcal{G}_{\mu\nu}^{L}\nonumber \\
 & -\frac{\beta}{2}\left(\frac{2n\Lambda}{\left(n-1\right)\left(n-2\right)}\mathcal{G}_{\mu\nu}^{L}+\frac{\Lambda}{n-1}\bar{g}_{\mu\nu}R_{L}\right).
\end{align}

Now, let us start collecting terms in the linearization of the field
equations (\ref{field_eq}). Note that the linearization of the other
three terms in the first line of (\ref{field_eq}) yields the same
contribution as $\left(g_{\nu\rho}\nabla^{\lambda}\nabla_{\sigma}\frac{\partial f}{\partial R_{\rho\sigma}^{\mu\lambda}}\right)_{L}$.
In addition, the linearization of the first two terms in the second
line of (\ref{field_eq}) gives the same contribution. As a result,
the linearized field equations become
\begin{align}
\left[\zeta-\beta\frac{2\Lambda}{\left(n-1\right)\left(n-2\right)}-\gamma\frac{4\Lambda\left(n-3\right)}{\left(n-1\right)\left(n-2\right)}\right]\mathcal{G}_{\mu\nu}^{L}\nonumber \\
+\left(2\alpha+\beta\right)\left(\bar{g}_{\mu\nu}\bar{\square}R_{L}-\bar{\nabla}_{\mu}\bar{\nabla}_{\nu}R_{L}\right)+\beta\bar{\square}\mathcal{G}_{\mu\nu}^{L}\\
+\left(\alpha\frac{4\Lambda}{\left(n-2\right)}+\beta\frac{2\Lambda}{\left(n-1\right)\left(n-2\right)}\right)R_{L}\bar{g}_{\mu\nu}\nonumber \\
-h_{\mu\nu}\left[\frac{1}{2}f\left(\bar{R}_{\rho\sigma}^{\alpha\beta}\right)-\frac{2\Lambda}{n-2}\right] & =0.\label{eq:Lin_field_eqn_of_fRiem}
\end{align}
The last line vanishes because of the background equation (\ref{eq:AdS_back}).
The final equation can be recast in the form of the linearized field
equations coming from the quadratic gravity theory
\begin{equation}
\mathcal{L}=\frac{1}{\tilde{\kappa}}\left(R-2\tilde{\Lambda}_{0}\right)+\alpha R^{2}+\beta R_{\sigma}^{\lambda}R_{\lambda}^{\sigma}+\gamma\chi_{\text{GB}},\label{eq:Equiv_quad}
\end{equation}
 given in \cite{Deser_Tekin} as
\begin{align}
\left[\frac{1}{\tilde{\kappa}}+\frac{4\Lambda n\alpha}{n-2}+\frac{4\Lambda\beta}{n-1}+\frac{4\Lambda\gamma(n-4)(n-3)}{(n-2)(n-1)}\right]\mathcal{G}_{\mu\nu}^{L}\nonumber \\
+(2\alpha+\beta)(\bar{g}_{\mu\nu}\bar{\Box}-\bar{\nabla}_{\mu}\bar{\nabla}_{\nu}+\frac{2\Lambda}{n-2}g_{\mu\nu})R_{L}+\beta(\bar{\Box}\mathcal{G}_{\mu\nu}^{L}-\frac{2\Lambda}{n-1}\bar{g}_{\mu\nu}R_{L}) & =0.\label{eq:Lin_field_eqn_of_quad_grav}
\end{align}
To match (\ref{eq:Lin_field_eqn_of_fRiem}) and (\ref{eq:Lin_field_eqn_of_quad_grav}),
one must have 
\begin{equation}
\frac{1}{\tilde{\kappa}}=2\zeta-\frac{4\Lambda}{n-2}\left[\left(n\alpha+\beta\right)+\gamma\frac{\left(n-2\right)\left(n-3\right)}{\left(n-1\right)}\right].
\end{equation}
In addition, we have to require that both theories have the same vacua
which determines $\tilde{\Lambda}_{0}$ as 
\begin{equation}
\frac{\tilde{\Lambda}_{0}}{\tilde{\kappa}}=-f\left(\bar{R}_{\rho\sigma}^{\alpha\beta}\right)+\frac{2n\Lambda}{n-2}\zeta-\frac{2\Lambda^{2}n}{\left(n-2\right)^{2}}\left[\left(n\alpha+\beta\right)+\gamma\frac{\left(n-2\right)\left(n-3\right)}{\left(n-1\right)}\right],
\end{equation}
which follows from the vacuum field equation of (\ref{eq:Equiv_quad})
\begin{equation}
\frac{\Lambda-\tilde{\Lambda}_{0}}{2\tilde{\kappa}}+\Lambda^{2}\left[\left(n\alpha+\beta\right)\frac{\left(n-4\right)}{\left(n-2\right)^{2}}+\gamma\frac{\left(n-3\right)\left(n-4\right)}{\left(n-1\right)\left(n-2\right)}\right]=0,
\end{equation}
and (\ref{eq:AdS_back}).

\section{Terms in ELA and EQCA\label{sec:ELA_and_EQCA_terms}}

In order to calculate ELA and EQCA for the BI gravity theory defined
by the Lagrangian density 
\begin{equation}
\mathcal{L}\left(C_{\alpha\beta}^{\mu\nu},R_{\nu}^{\mu},R_{\nu}^{\mu}\right)=\frac{2}{\gamma}\left[\sqrt{\det\left(\delta_{\sigma}^{\rho}+\gamma A_{\sigma}^{\rho}\right)}-\left(\lambda_{0}+1\right)\right],\label{eq:BI_Lagrangian}
\end{equation}
one needs to calculate the background values of $\mathcal{L}$, and
its first and second order derivatives. To find the background value
of $\mathcal{L}$, one needs the background value of $A_{\sigma}^{\rho}$
which can be found as 
\begin{equation}
\gamma\bar{A}_{\sigma}^{\rho}\equiv\delta_{\sigma}^{\rho}\bar{a}=\delta_{\sigma}^{\rho}\lambda\left[1+\lambda\left(a_{3}+b_{2}\right)\right].
\end{equation}
In calculating the first order derivatives of (\ref{eq:BI_Lagrangian}),
we use 
\begin{equation}
\partial\left(\sqrt{\det\left(\delta_{\nu}^{\mu}+\gamma A_{\nu}^{\mu}\right)}\right)=\frac{\gamma}{2}\sqrt{\det\left(\delta_{\nu}^{\mu}+\gamma A_{\nu}^{\mu}\right)}B_{\rho}^{\sigma}\partial A_{\sigma}^{\rho},
\end{equation}
where $B_{\rho}^{\sigma}$ is defined as $B_{\rho}^{\alpha}\left(\delta_{\beta}^{\rho}+\gamma A_{\beta}^{\rho}\right)=\delta_{\beta}^{\alpha}$
with the background value $\bar{B}_{\rho}^{\sigma}=\left(1+\bar{a}\right)^{-1}\delta_{\rho}^{\sigma}$.
Thus, we just need the derivatives of $A_{\sigma}^{\rho}$ which can
be found as
\begin{align}
\frac{\partial A_{\sigma}^{\rho}}{\partial C_{\alpha\beta}^{\mu\nu}}= & \gamma a_{1}\left(C_{\sigma\nu}^{\alpha\beta}\delta_{\mu}^{\rho}+C_{\mu\nu}^{\rho\beta}\delta_{\sigma}^{\alpha}\right)+\frac{\gamma b_{1}}{2}C_{\mu\nu}^{\alpha\beta}\delta_{\sigma}^{\rho}+\gamma a_{2}R_{\nu}^{\beta}\delta_{\sigma}^{\alpha}\delta_{\mu}^{\rho},\\
\frac{\partial A_{\sigma}^{\rho}}{\partial S_{\nu}^{\mu}}= & \beta\delta_{\sigma}^{\nu}\delta_{\mu}^{\rho}+\gamma a_{4}\left(S_{\sigma}^{\nu}\delta_{\mu}^{\rho}+S_{\mu}^{\rho}\delta_{\sigma}^{\nu}\right)+\frac{\gamma b_{3}}{2}S_{\mu}^{\nu}\delta_{\sigma}^{\rho},\\
\frac{\partial A_{\sigma}^{\rho}}{\partial R_{\nu}^{\mu}}= & \delta_{\sigma}^{\nu}\delta_{\mu}^{\rho}+\gamma a_{2}C_{\sigma\mu}^{\rho\nu}+\gamma a_{3}\left(R_{\sigma}^{\nu}\delta_{\mu}^{\rho}+R_{\mu}^{\rho}\delta_{\sigma}^{\nu}\right)+\frac{\gamma b_{2}}{2}R_{\mu}^{\nu}\delta_{\sigma}^{\rho},
\end{align}
and their background values are 
\begin{align}
\left[\frac{\partial A_{\sigma}^{\rho}}{\partial C_{\alpha\beta}^{\mu\nu}}\right]_{\bar{R}_{\rho\sigma}^{\mu\nu}} & =\lambda a_{2}\delta_{\nu}^{\beta}\delta_{\sigma}^{\alpha}\delta_{\mu}^{\rho},\\
\left[\frac{\partial A_{\sigma}^{\rho}}{\partial S_{\nu}^{\mu}}\right]_{\bar{R}_{\rho\sigma}^{\mu\nu}} & =\beta\delta_{\sigma}^{\nu}\delta_{\mu}^{\rho},\\
\left[\frac{\partial A_{\sigma}^{\rho}}{\partial R_{\nu}^{\mu}}\right]_{\bar{R}_{\rho\sigma}^{\mu\nu}} & =\delta_{\sigma}^{\nu}\delta_{\mu}^{\rho}+2\lambda\left(a_{3}\delta_{\mu}^{\rho}\delta_{\sigma}^{\nu}+\frac{b_{2}}{4}\delta_{\mu}^{\nu}\delta_{\sigma}^{\rho}\right).
\end{align}
Using these results, one can calculate the linear order terms in ELA
and EQCA. However, prior to any calculation, it is clear that the
Weyl term, 
\begin{equation}
\left[\frac{\partial\mathcal{L}}{\partial C_{\alpha\beta}^{\mu\nu}}\right]_{\bar{R}_{\rho\sigma}^{\mu\nu}}C_{\alpha\beta}^{\mu\nu}=\sqrt{\det\left(\delta_{\nu}^{\rho}+\gamma\bar{A}_{\nu}^{\rho}\right)}\bar{B}_{\rho}^{\sigma}\left[\frac{\partial A_{\sigma}^{\rho}}{\partial C_{\alpha\beta}^{\mu\nu}}\right]_{\bar{R}_{\rho\sigma}^{\mu\nu}}C_{\alpha\beta}^{\mu\nu},
\end{equation}
and the traceless-Ricci term
\begin{equation}
\left[\frac{\partial\mathcal{L}}{\partial S_{\nu}^{\mu}}\right]_{\bar{R}_{\rho\sigma}^{\mu\nu}}S_{\nu}^{\mu}=\sqrt{\det\left(\delta_{\nu}^{\rho}+\gamma\bar{A}_{\nu}^{\rho}\right)}\bar{B}_{\rho}^{\sigma}\left[\frac{\partial A_{\sigma}^{\rho}}{\partial S_{\nu}^{\mu}}\right]_{\bar{R}_{\rho\sigma}^{\mu\nu}}S_{\nu}^{\mu},
\end{equation}
yield zero as they involve traces of $C_{\alpha\beta}^{\mu\nu}$ and
$S_{\nu}^{\mu}$. The unique contribution comes from the Ricci term
\begin{equation}
\left[\frac{\partial\mathcal{L}}{\partial R_{\nu}^{\mu}}\right]_{\bar{R}_{\rho\sigma}^{\mu\nu}}\left(R_{\nu}^{\mu}-\bar{R}_{\nu}^{\mu}\right)=\sqrt{\det\left(\delta_{\nu}^{\rho}+\gamma\bar{A}_{\nu}^{\rho}\right)}\bar{B}_{\rho}^{\sigma}\left[\frac{\partial A_{\sigma}^{\rho}}{\partial R_{\nu}^{\mu}}\right]_{\bar{R}_{\rho\sigma}^{\mu\nu}}\left(R_{\nu}^{\mu}-\bar{R}_{\nu}^{\mu}\right),
\end{equation}
which becomes 
\begin{align}
\left[\frac{\partial\mathcal{L}}{\partial R_{\nu}^{\mu}}\right]_{\bar{R}_{\rho\sigma}^{\mu\nu}}\left(R_{\nu}^{\mu}-\bar{R}_{\nu}^{\mu}\right)= & \left(\gamma R-4\lambda\right)\frac{1}{2}\left(1+\bar{a}\right)\left[1+2\lambda\left(a_{3}+b_{2}\right)\right].
\end{align}
Adding the background value of $\mathcal{L}$,
\begin{equation}
\overline{\mathcal{L}}=\frac{2}{\gamma}\left[\left(1+\bar{a}\right)^{2}-\left(\lambda_{0}+1\right)\right],
\end{equation}
to this result yields the ELA given in (\ref{eq:ELA-BI}).

Moving to the second order derivatives of $\mathcal{L}$ which can
be calculated by using 
\begin{equation}
\partial^{2}\left(\sqrt{\det\left(\delta_{\nu}^{\rho}+\gamma A_{\nu}^{\rho}\right)}\right)=\frac{\gamma}{2}\sqrt{\det\left(\delta_{\nu}^{\rho}+\gamma A_{\nu}^{\rho}\right)}\left[B_{\gamma}^{\lambda}\partial^{2}A_{\lambda}^{\gamma}-\gamma B_{\theta}^{\lambda}B_{\gamma}^{\tau}\left(\partial A_{\tau}^{\theta}\right)\partial A_{\lambda}^{\gamma}+\frac{\gamma}{2}\left(B_{\gamma}^{\lambda}\partial A_{\lambda}^{\gamma}\right)^{2}\right],\label{del_square}
\end{equation}
where the second order derivatives of $A_{\sigma}^{\rho}$ are needed.
First, the second derivative of $A_{\sigma}^{\rho}$ with respect
to the Weyl tensor is
\begin{equation}
\left[\frac{\partial^{2}A_{\sigma}^{\rho}}{\partial C_{\alpha\beta}^{\mu\nu}\partial C_{\lambda\tau}^{\eta\theta}}\right]_{\bar{R}_{\rho\sigma}^{\mu\nu}}=\gamma a_{1}\delta_{\sigma}^{\alpha}\delta_{\theta}^{\beta}\delta_{\mu}^{\lambda}\delta_{\nu}^{\tau}\delta_{\eta}^{\rho}+\gamma\delta_{\eta}^{\alpha}\delta_{\theta}^{\beta}\delta_{\nu}^{\tau}\left(a_{1}\delta_{\sigma}^{\lambda}\delta_{\mu}^{\rho}+\frac{b_{1}}{2}\delta_{\mu}^{\lambda}\delta_{\sigma}^{\rho}\right),
\end{equation}
where the result does not have the symmetries of the Weyl tensor on
the left-hand side. However, note that the result becomes symmetric
accordingly when it is multiplied with $C_{\alpha\beta}^{\mu\nu}C_{\lambda\tau}^{\eta\theta}$
in finding the final contribution to the expression
\begin{equation}
\left[\frac{\partial^{2}A_{\sigma}^{\rho}}{\partial C_{\alpha\beta}^{\mu\nu}\partial C_{\lambda\tau}^{\eta\theta}}\right]_{\bar{R}_{\rho\sigma}^{\mu\nu}}C_{\alpha\beta}^{\mu\nu}\partial C_{\lambda\tau}^{\eta\theta}.
\end{equation}
Then, the other derivatives can be calculated as
\begin{equation}
\left[\frac{\partial^{2}A_{\sigma}^{\rho}}{\partial C_{\alpha\beta}^{\mu\nu}\partial R_{\theta}^{\eta}}\right]_{\bar{R}_{\rho\sigma}^{\mu\nu}}=\gamma a_{2}\delta_{\eta}^{\beta}\delta_{\nu}^{\theta}\delta_{\sigma}^{\alpha}\delta_{\mu}^{\rho},\qquad\left[\frac{\partial^{2}A_{\sigma}^{\rho}}{\partial S_{\nu}^{\mu}\partial S_{\beta}^{\alpha}}\right]_{\bar{R}_{\rho\sigma}^{\mu\nu}}=\gamma a_{4}\left(\delta_{\sigma}^{\nu}\delta_{\mu}^{\beta}\delta_{\alpha}^{\rho}+\delta_{\alpha}^{\nu}\delta_{\sigma}^{\beta}\delta_{\mu}^{\rho}\right)+\frac{\gamma b_{3}}{2}\delta_{\alpha}^{\nu}\delta_{\mu}^{\beta}\delta_{\sigma}^{\rho},
\end{equation}
\begin{equation}
\left[\frac{\partial^{2}A_{\sigma}^{\rho}}{\partial R_{\nu}^{\mu}\partial R_{\beta}^{\alpha}}\right]_{\bar{R}_{\rho\sigma}^{\mu\nu}}=\gamma a_{3}\left(\delta_{\sigma}^{\nu}\delta_{\mu}^{\beta}\delta_{\alpha}^{\rho}+\delta_{\alpha}^{\nu}\delta_{\sigma}^{\beta}\delta_{\mu}^{\rho}\right)+\frac{\gamma b_{2}}{2}\delta_{\alpha}^{\nu}\delta_{\mu}^{\beta}\delta_{\sigma}^{\rho}.
\end{equation}
It is clear that the remaining ones are just zero 
\begin{equation}
\frac{\partial^{2}A_{\sigma}^{\rho}}{\partial C_{\alpha\beta}^{\mu\nu}\partial S_{\theta}^{\eta}}=0,\qquad\frac{\partial^{2}A_{\sigma}^{\rho}}{\partial S_{\nu}^{\mu}\partial R_{\beta}^{\alpha}}=0.
\end{equation}
Using these results in (\ref{del_square}) let us compute the second
order contributions to the EQCA term by term. First, the Weyl square
term takes the form
\begin{align}
\frac{1}{2}\left[\frac{\partial^{2}\mathcal{L}}{\partial C_{\alpha\beta}^{\mu\nu}C_{\lambda\tau}^{\eta\theta}}\right]_{\bar{R}_{\rho\sigma}^{\mu\nu}}C_{\alpha\beta}^{\mu\nu}C_{\lambda\tau}^{\eta\theta}= & \frac{1}{2}\sqrt{\det\left(\delta_{\nu}^{\mu}+\gamma\bar{A}_{\nu}^{\mu}\right)}\nonumber \\
 & \times\Biggl\{\bar{B}_{\rho}^{\sigma}\left[\frac{\partial^{2}A_{\sigma}^{\rho}}{\partial C_{\alpha\beta}^{\mu\nu}\partial C_{\lambda\tau}^{\eta\theta}}\right]_{\bar{R}_{\rho\sigma}^{\mu\nu}}-\gamma\bar{B}_{\zeta}^{\sigma}\left[\frac{\partial A_{\epsilon}^{\zeta}}{\partial C_{\lambda\tau}^{\eta\theta}}\right]_{\bar{R}_{\rho\sigma}^{\mu\nu}}\bar{B}_{\rho}^{\epsilon}\left[\frac{\partial A_{\sigma}^{\rho}}{\partial C_{\alpha\beta}^{\mu\nu}}\right]_{\bar{R}_{\rho\sigma}^{\mu\nu}}\nonumber \\
 & \phantom{x\Biggl\{}+\frac{\gamma}{2}\bar{B}_{\rho}^{\sigma}\left[\frac{\partial A_{\sigma}^{\rho}}{\partial C_{\alpha\beta}^{\mu\nu}}\right]_{\bar{R}_{\rho\sigma}^{\mu\nu}}\bar{B}_{\zeta}^{\epsilon}\left[\frac{\partial A_{\epsilon}^{\zeta}}{\partial C_{\lambda\tau}^{\eta\theta}}\right]_{\bar{R}_{\rho\sigma}^{\mu\nu}}\Biggr\} C_{\alpha\beta}^{\mu\nu}C_{\lambda\tau}^{\eta\theta},
\end{align}
which then yields 
\begin{equation}
\frac{1}{2}\left[\frac{\partial^{2}\mathcal{L}}{\partial C_{\alpha\beta}^{\mu\nu}C_{\lambda\tau}^{\eta\theta}}\right]_{\bar{R}_{\rho\sigma}^{\mu\nu}}C_{\alpha\beta}^{\mu\nu}C_{\lambda\tau}^{\eta\theta}=\frac{1}{2}\gamma^{2}\left(1+\bar{a}\right)\left(a_{1}+b_{1}\right)C_{\rho\sigma}^{\mu\nu}C_{\mu\nu}^{\rho\sigma}.
\end{equation}
Then, the term involving the square of the traceless-Ricci tensor
has the form
\begin{align}
\frac{1}{2}\frac{\partial^{2}\mathcal{L}}{\partial S_{\nu}^{\mu}\partial S_{\beta}^{\alpha}}S_{\nu}^{\mu}S_{\beta}^{\alpha}= & \frac{1}{2}\sqrt{\det\left(\delta_{\nu}^{\mu}+\gamma\bar{A}_{\nu}^{\mu}\right)}\nonumber \\
 & \times\Biggl\{\bar{B}_{\rho}^{\sigma}\left[\frac{\partial^{2}A_{\sigma}^{\rho}}{\partial S_{\nu}^{\mu}\partial S_{\beta}^{\alpha}}\right]_{\bar{R}_{\rho\sigma}^{\mu\nu}}-\gamma\bar{B}_{\zeta}^{\sigma}\left[\frac{\partial A_{\epsilon}^{\zeta}}{\partial S_{\beta}^{\alpha}}\right]_{\bar{R}_{\rho\sigma}^{\mu\nu}}\bar{B}_{\rho}^{\epsilon}\left[\frac{\partial A_{\sigma}^{\rho}}{\partial S_{\nu}^{\mu}}\right]_{\bar{R}_{\rho\sigma}^{\mu\nu}}\nonumber \\
 & \phantom{\times\Biggl[}+\frac{\gamma}{2}\bar{B}_{\rho}^{\sigma}\left[\frac{\partial A_{\sigma}^{\rho}}{\partial S_{\nu}^{\mu}}\right]_{\bar{R}_{\rho\sigma}^{\mu\nu}}\bar{B}_{\zeta}^{\epsilon}\left[\frac{\partial A_{\epsilon}^{\zeta}}{\partial S_{\beta}^{\alpha}}\right]_{\bar{R}_{\rho\sigma}^{\mu\nu}}\Biggr\} S_{\nu}^{\mu}S_{\beta}^{\alpha},
\end{align}
yielding
\begin{equation}
\frac{1}{2}\frac{\partial^{2}\mathcal{L}}{\partial S_{\nu}^{\mu}\partial S_{\beta}^{\alpha}}S_{\nu}^{\mu}S_{\beta}^{\alpha}=\gamma^{2}\left[-\frac{1}{4}\beta^{2}+\frac{1}{2}\left(1+\bar{a}\right)\left(a_{4}+b_{3}\right)\right]S_{\mu}^{\nu}S_{\nu}^{\mu}.
\end{equation}
Moving to the Ricci square term which has the form 
\begin{align}
\frac{1}{2}\frac{\partial^{2}\mathcal{L}}{\partial R_{\nu}^{\mu}\partial R_{\beta}^{\alpha}}\left(R_{\nu}^{\mu}-\bar{R}_{\nu}^{\mu}\right)\left(R_{\beta}^{\alpha}-\bar{R}_{\beta}^{\alpha}\right)= & \frac{1}{2}\sqrt{\det\left(\delta_{\nu}^{\mu}+\gamma\bar{A}_{\nu}^{\mu}\right)}\nonumber \\
 & \times\Biggl\{\bar{B}_{\rho}^{\sigma}\left[\frac{\partial^{2}A_{\sigma}^{\rho}}{\partial R_{\nu}^{\mu}\partial R_{\beta}^{\alpha}}\right]_{\bar{R}_{\rho\sigma}^{\mu\nu}}-\gamma\bar{B}_{\zeta}^{\sigma}\left[\frac{\partial A_{\epsilon}^{\zeta}}{\partial R_{\beta}^{\alpha}}\right]_{\bar{R}_{\rho\sigma}^{\mu\nu}}\bar{B}_{\rho}^{\epsilon}\left[\frac{\partial A_{\sigma}^{\rho}}{\partial R_{\nu}^{\mu}}\right]_{\bar{R}_{\rho\sigma}^{\mu\nu}}\nonumber \\
 & \phantom{\times\Biggl[}+\frac{\gamma}{2}\bar{B}_{\rho}^{\sigma}\left[\frac{\partial A_{\sigma}^{\rho}}{\partial R_{\nu}^{\mu}}\right]_{\bar{R}_{\rho\sigma}^{\mu\nu}}\bar{B}_{\zeta}^{\epsilon}\left[\frac{\partial A_{\epsilon}^{\zeta}}{\partial R_{\beta}^{\alpha}}\right]_{\bar{R}_{\rho\sigma}^{\mu\nu}}\Biggr\}\nonumber \\
 & \times\left(R_{\nu}^{\mu}-\bar{R}_{\nu}^{\mu}\right)\left(R_{\beta}^{\alpha}-\bar{R}_{\beta}^{\alpha}\right),
\end{align}
becomes
\begin{align}
\frac{1}{2}\frac{\partial^{2}\mathcal{L}}{\partial R_{\nu}^{\mu}\partial R_{\beta}^{\alpha}}\left(R_{\nu}^{\mu}-\bar{R}_{\nu}^{\mu}\right)\left(R_{\beta}^{\alpha}-\bar{R}_{\beta}^{\alpha}\right)= & -\left(\gamma R-2\lambda-\frac{\gamma^{2}}{2\lambda}R_{\mu}^{\nu}R_{\nu}^{\mu}+\frac{\gamma^{2}}{2\lambda}S_{\mu}^{\nu}S_{\nu}^{\mu}\right)\nonumber \\
 & \phantom{-}\times\frac{\lambda}{2}\left[2\left(1+\bar{a}\right)\left(a_{3}+b_{2}\right)+\left(2\lambda\left(a_{3}+b_{2}\right)+1\right)^{2}\right]\nonumber \\
 & -\frac{\gamma^{2}\left(\left(2a_{3}\lambda+1\right)^{2}-2\left(1+\bar{a}\right)\left(a_{3}+b_{2}\right)\right)}{4}S_{\mu}^{\nu}S_{\nu}^{\mu},
\end{align}
after using 
\begin{equation}
R^{2}=4\left(R_{\mu}^{\nu}R_{\nu}^{\mu}-S_{\mu}^{\nu}S_{\nu}^{\mu}\right).
\end{equation}
Then, the first two cross terms yield zero as
\begin{align}
\frac{\partial^{2}\mathcal{L}}{\partial C_{\alpha\beta}^{\mu\nu}\partial S_{\theta}^{\eta}}C_{\alpha\beta}^{\mu\nu}S_{\theta}^{\eta}= & \sqrt{\det\left(\delta_{\nu}^{\mu}+\gamma\bar{A}_{\nu}^{\mu}\right)}\nonumber \\
 & \times\Biggl\{\bar{B}_{\rho}^{\sigma}\left[\frac{\partial^{2}A_{\sigma}^{\rho}}{\partial C_{\alpha\beta}^{\mu\nu}\partial S_{\theta}^{\eta}}\right]_{\bar{R}_{\rho\sigma}^{\mu\nu}}-\gamma\bar{B}_{\zeta}^{\sigma}\left[\frac{\partial A_{\epsilon}^{\zeta}}{\partial S_{\theta}^{\eta}}\right]_{\bar{R}_{\rho\sigma}^{\mu\nu}}\bar{B}_{\rho}^{\epsilon}\left[\frac{\partial A_{\sigma}^{\rho}}{\partial C_{\alpha\beta}^{\mu\nu}}\right]_{\bar{R}_{\rho\sigma}^{\mu\nu}}\nonumber \\
 & \phantom{\times\Biggl[}+\frac{\gamma}{2}\bar{B}_{\rho}^{\sigma}\left[\frac{\partial A_{\sigma}^{\rho}}{\partial C_{\alpha\beta}^{\mu\nu}}\right]_{\bar{R}_{\rho\sigma}^{\mu\nu}}\bar{B}_{\zeta}^{\epsilon}\left[\frac{\partial A_{\epsilon}^{\zeta}}{\partial S_{\theta}^{\eta}}\right]_{\bar{R}_{\rho\sigma}^{\mu\nu}}\Biggr\} C_{\alpha\beta}^{\mu\nu}S_{\theta}^{\eta},
\end{align}
\begin{equation}
\frac{\partial^{2}\mathcal{L}}{\partial C_{\alpha\beta}^{\mu\nu}\partial S_{\theta}^{\eta}}C_{\alpha\beta}^{\mu\nu}S_{\theta}^{\eta}=0,
\end{equation}
and
\begin{align}
\frac{\partial^{2}\mathcal{L}}{\partial C_{\alpha\beta}^{\mu\nu}\partial R_{\theta}^{\eta}}C_{\alpha\beta}^{\mu\nu}\left(R_{\theta}^{\eta}-\bar{R}_{\theta}^{\eta}\right)= & \sqrt{\det\left(\delta_{\nu}^{\mu}+\gamma\bar{A}_{\nu}^{\mu}\right)}\nonumber \\
 & \times\Biggl\{\bar{B}_{\rho}^{\sigma}\left[\frac{\partial^{2}A_{\sigma}^{\rho}}{\partial C_{\alpha\beta}^{\mu\nu}\partial R_{\theta}^{\eta}}\right]_{\bar{R}_{\rho\sigma}^{\mu\nu}}-\gamma\bar{B}_{\zeta}^{\sigma}\left[\frac{\partial A_{\epsilon}^{\zeta}}{\partial R_{\theta}^{\eta}}\right]_{\bar{R}_{\rho\sigma}^{\mu\nu}}\bar{B}_{\rho}^{\epsilon}\left[\frac{\partial A_{\sigma}^{\rho}}{\partial C_{\alpha\beta}^{\mu\nu}}\right]_{\bar{R}_{\rho\sigma}^{\mu\nu}}\nonumber \\
 & \phantom{\times\Biggl[}+\frac{\gamma}{2}\bar{B}_{\rho}^{\sigma}\left[\frac{\partial A_{\sigma}^{\rho}}{\partial C_{\alpha\beta}^{\mu\nu}}\right]_{\bar{R}_{\rho\sigma}^{\mu\nu}}\bar{B}_{\zeta}^{\epsilon}\left[\frac{\partial A_{\epsilon}^{\zeta}}{\partial R_{\theta}^{\eta}}\right]_{\bar{R}_{\rho\sigma}^{\mu\nu}}\Biggr\} C_{\alpha\beta}^{\mu\nu}\left(R_{\theta}^{\eta}-\bar{R}_{\theta}^{\eta}\right),
\end{align}
\begin{equation}
\frac{\partial^{2}\mathcal{L}}{\partial C_{\alpha\beta}^{\mu\nu}\partial R_{\theta}^{\eta}}C_{\alpha\beta}^{\mu\nu}\left(R_{\theta}^{\eta}-\bar{R}_{\theta}^{\eta}\right)=0.
\end{equation}
Lastly, the nonzero cross term is
\begin{align}
\frac{\partial^{2}\mathcal{L}}{\partial S_{\nu}^{\mu}\partial R_{\beta}^{\alpha}}S_{\nu}^{\mu}\left(R_{\beta}^{\alpha}-\bar{R}_{\beta}^{\alpha}\right)= & \sqrt{\det\left(\delta_{\nu}^{\mu}+\gamma\bar{A}_{\nu}^{\mu}\right)}\nonumber \\
 & \times\Biggl\{\bar{B}_{\rho}^{\sigma}\left[\frac{\partial^{2}A_{\sigma}^{\rho}}{\partial S_{\nu}^{\mu}\partial R_{\beta}^{\alpha}}\right]_{\bar{R}_{\rho\sigma}^{\mu\nu}}-\gamma\bar{B}_{\zeta}^{\sigma}\left[\frac{\partial A_{\epsilon}^{\zeta}}{\partial R_{\beta}^{\alpha}}\right]_{\bar{R}_{\rho\sigma}^{\mu\nu}}\bar{B}_{\rho}^{\epsilon}\left[\frac{\partial A_{\sigma}^{\rho}}{\partial S_{\nu}^{\mu}}\right]_{\bar{R}_{\rho\sigma}^{\mu\nu}}\nonumber \\
 & \phantom{\times\Biggl[}+\frac{\gamma}{2}\bar{B}_{\rho}^{\sigma}\left[\frac{\partial A_{\sigma}^{\rho}}{\partial S_{\nu}^{\mu}}\right]_{\bar{R}_{\rho\sigma}^{\mu\nu}}\bar{B}_{\zeta}^{\epsilon}\left[\frac{\partial A_{\epsilon}^{\zeta}}{\partial R_{\beta}^{\alpha}}\right]_{\bar{R}_{\rho\sigma}^{\mu\nu}}\Biggr\} S_{\nu}^{\mu}\left(R_{\beta}^{\alpha}-\bar{R}_{\beta}^{\alpha}\right),
\end{align}
\begin{equation}
\frac{\partial^{2}\mathcal{L}}{\partial S_{\nu}^{\mu}\partial R_{\beta}^{\alpha}}S_{\nu}^{\mu}\left(R_{\beta}^{\alpha}-\bar{R}_{\beta}^{\alpha}\right)=-\frac{1}{2}\gamma^{2}\beta\left(1+2\lambda a_{3}\right)S_{\mu}^{\nu}S_{\nu}^{\mu},
\end{equation}
after using
\begin{equation}
R_{\mu}^{\nu}S_{\nu}^{\mu}=S_{\mu}^{\nu}S_{\nu}^{\mu}.
\end{equation}
Adding all these second order contributions to the ELA yield the EQCA
given in (\ref{eq:EQCA-BI}).

\section{Field Equations\label{sec:Field-Equations}}

In this part we will derive the field equations by minimizing the
action 
\begin{align}
I= & \frac{2}{\kappa\gamma}\int d^{4}x\left[\sqrt{-\text{det}\left(g_{\mu\nu}+\gamma A_{\mu\nu}\right)}-\left(\lambda_{0}+1\right)\sqrt{-\text{det}g}\right],\label{generic_BI_action}
\end{align}
where we shall work in the Riemann--Ricci--curvature-scalar curvature
basis.

\begin{align}
A_{\mu\nu}= & \left(\beta+1\right)R_{\mu\nu}-\frac{\beta}{4}g_{\mu\nu}R+c_{1}g_{\mu\nu}R^{2}+c_{2}RR_{\mu\nu}+c_{3}g_{\mu\nu}R_{\sigma\rho}^{2}\nonumber \\
 & +c_{4}R_{\phantom{\sigma}\mu}^{\sigma}R_{\nu\sigma}+c_{5}R_{\mu\sigma\nu\rho}R^{\sigma\rho}+c_{6}g_{\mu\nu}R_{\sigma\rho\alpha\beta}^{2}+c_{7}R_{\mu}^{\phantom{\mu}\sigma\rho\tau}R_{\nu\sigma\rho\tau}.
\end{align}
One can eliminate $c_{6}$ or $c_{7}$ one in favor of the other but
we will keep it this way.

The variation of the action is
\begin{align}
\delta I= & \frac{2}{\kappa\gamma}\int d^{4}x\sqrt{-\text{det}g}\left\{ \left(-\frac{1}{2}g_{\alpha\beta}\delta g^{\alpha\beta}\right)\left[\sqrt{\text{det}\left(\delta_{\mu}^{\nu}+\gamma A_{\mu}^{\nu}\right)}-\left(\lambda_{0}+1\right)\right]\right.\nonumber \\
 & \left.+\left[\delta\sqrt{\text{det}\left(\delta_{\mu}^{\nu}+\gamma A_{\mu}^{\nu}\right)}\right]\right\} .
\end{align}
The first term is already in the desired form. On the other hand,
the second term can be analyzed by using (\ref{eq:First_diff_of_I-plus-A}),
that is
\begin{equation}
\delta\sqrt{\det\left(\delta_{\mu}^{\nu}+\gamma A_{\mu}^{\nu}\right)}=\frac{\gamma}{2}\sqrt{\det\left(\delta_{\mu}^{\nu}+\gamma A_{\mu}^{\nu}\right)}B_{\beta}^{\alpha}\delta A_{\alpha}^{\beta},
\end{equation}
where $B\equiv\left(\delta+\gamma A\right)^{-1}$. For notational
convenience, let us define $\mathcal{E}_{\beta}^{\alpha}\equiv\frac{\gamma}{2}\sqrt{\det\left(\delta_{\mu}^{\nu}+\gamma A_{\mu}^{\nu}\right)}B_{\beta}^{\alpha}$.
Then, after considering the variations of the curvature terms in $\delta A_{\alpha}^{\beta}$,
a lengthy computation yields the field equations as 

\begin{align}
-\frac{1}{2}g_{\alpha\beta}\left[\sqrt{\text{det}\left(\delta_{\mu}^{\nu}+\gamma A_{\mu}^{\nu}\right)}-\left(\lambda_{0}+1\right)\right]\nonumber \\
+\left(\beta+1\right)\left(\mathcal{E}_{\beta}^{\mu}R_{\mu\alpha}-\nabla_{\mu}\nabla_{\beta}\mathcal{E}_{\alpha}^{\mu}+\frac{1}{2}\square\mathcal{E}_{\beta\alpha}+\frac{1}{2}g_{\alpha\beta}\nabla_{\mu}\nabla_{\nu}\mathcal{E}^{\mu\nu}\right)-\frac{\beta}{4}\left(\mathcal{E}R_{\alpha\beta}-\nabla_{\beta}\nabla_{\alpha}\mathcal{E}+g_{\alpha\beta}\square\mathcal{E}\right)\nonumber \\
+c_{1}\left(2\mathcal{E}RR_{\alpha\beta}-2\nabla_{\alpha}\nabla_{\beta}\left(\mathcal{E}R\right)+2g_{\alpha\beta}\square\left(\mathcal{E}R\right)\right)+c_{2}\left(\mathcal{E}_{\nu}^{\mu}R_{\mu}^{\nu}R_{\alpha\beta}+\mathcal{E}_{\beta}^{\mu}RR_{\alpha\mu}\right)\nonumber \\
+c_{2}\left[-\nabla_{\alpha}\nabla_{\beta}\left(\mathcal{E}_{\nu}^{\mu}R_{\mu}^{\nu}\right)+g_{\alpha\beta}\square\left(\mathcal{E}_{\nu}^{\mu}R_{\mu}^{\nu}\right)-\nabla_{\mu}\nabla_{\beta}\left(\mathcal{E}_{\alpha}^{\mu}R\right)+\frac{1}{2}\square\left(\mathcal{E}_{\alpha\beta}R\right)+\frac{1}{2}\nabla_{\mu}\nabla_{\lambda}\left(\mathcal{E}^{\mu\lambda}R\right)g_{\alpha\beta}\right]\nonumber \\
+c_{3}\left(-2\nabla_{\sigma}\nabla_{\beta}\left(\mathcal{E}R_{\alpha}^{\sigma}\right)+\square\left(\mathcal{E}R_{\alpha\beta}\right)+\nabla_{\rho}\nabla_{\sigma}\left(\mathcal{E}R^{\sigma\rho}\right)g_{\alpha\beta}+2\mathcal{E}R_{\alpha\rho}R_{\beta}^{\rho}\right)\nonumber \\
+c_{4}\left[\left(\mathcal{E}_{\nu}^{\mu}R_{\alpha\mu}R_{\phantom{\nu}\beta}^{\nu}+\mathcal{E}_{\alpha}^{\mu}R_{\phantom{\sigma}\mu}^{\sigma}R_{\beta\sigma}\right)-\nabla_{\mu}\nabla_{\beta}\left(\mathcal{E}_{\alpha\nu}R^{\nu\mu}\right)-\nabla_{\mu}\nabla_{\beta}\left(\mathcal{E}_{\nu}^{\mu}R_{\phantom{\nu}\alpha}^{\nu}\right)\right.\nonumber \\
\left.+\square\left(\mathcal{E}_{\beta\nu}R_{\phantom{\nu}\alpha}^{\nu}\right)+\frac{1}{2}\nabla_{\mu}\nabla_{\sigma}\left(\mathcal{E}_{\nu}^{\mu}R^{\nu\sigma}\right)g_{\alpha\beta}+\frac{1}{2}\nabla_{\sigma}\nabla_{\nu}\left(\mathcal{E}^{\mu\nu}R_{\phantom{\sigma}\mu}^{\sigma}\right)g_{\alpha\beta}\right]\nonumber \\
+c_{5}\left[-\mathcal{E}_{\alpha\nu}R_{\beta\sigma}^{\phantom{\beta\sigma}\nu\rho}R_{\rho}^{\sigma}+\mathcal{E}_{\beta}^{\mu}R_{\mu\sigma\alpha}^{\phantom{\mu\sigma\alpha}\rho}R_{\rho}^{\sigma}+\mathcal{E}_{\nu}^{\mu}R_{\mu\sigma\phantom{\nu}\beta}^{\phantom{\mu\sigma}\nu}R_{\alpha}^{\sigma}+\mathcal{E}_{\nu}^{\mu}R_{\mu\alpha}^{\phantom{\mu\sigma}\nu\rho}R_{\beta\rho}\right.\nonumber \\
-\nabla_{\sigma}\nabla_{\lambda}\left(\mathcal{E}_{\alpha}^{\lambda}R_{\beta}^{\sigma}\right)+\frac{1}{2}\nabla_{\lambda}\nabla_{\sigma}\left(\mathcal{E}^{\lambda\sigma}R_{\alpha\beta}\right)+\frac{1}{2}\nabla_{\sigma}\nabla_{\lambda}\left(\mathcal{E}_{\alpha\beta}R^{\lambda\sigma}\right)\nonumber \\
\left.-\frac{1}{2}\nabla_{\gamma}\nabla_{\beta}\left(\mathcal{E}_{\nu}^{\mu}R_{\mu\phantom{\gamma\nu}\alpha}^{\phantom{\mu}\gamma\nu}\right)-\frac{1}{2}\nabla_{\rho}\nabla_{\beta}\left(\mathcal{E}_{\nu}^{\mu}R_{\mu\alpha}^{\phantom{\mu\alpha}\nu\rho}\right)+\frac{1}{2}\square\left(\mathcal{E}_{\nu}^{\mu}R_{\mu\alpha\phantom{\nu}\beta}^{\phantom{\mu\alpha}\nu}\right)+\frac{1}{2}\nabla_{\rho}\nabla_{\gamma}\left(\mathcal{E}_{\nu}^{\mu}R_{\mu}^{\phantom{\mu}\gamma\nu\rho}\right)g_{\alpha\beta}\right]\nonumber \\
+c_{6}\left[-\mathcal{E}R_{\alpha\rho\tau\gamma}R_{\beta}^{\phantom{\beta}\rho\tau\gamma}+\mathcal{E}R_{\sigma\beta}^{\phantom{\sigma\beta}\rho\pi}R_{\phantom{\sigma}\alpha\rho\pi}^{\sigma}+\mathcal{E}R_{\sigma\phantom{\tau}\alpha}^{\phantom{\sigma}\tau\phantom{\alpha}\pi}R_{\phantom{\sigma}\tau\beta\pi}^{\sigma}+\mathcal{E}R_{\sigma\phantom{\gamma\pi}\alpha}^{\phantom{\sigma}\gamma\pi}R_{\phantom{\sigma}\gamma\pi\beta}^{\sigma}\right.\nonumber \\
\left.+\nabla_{\lambda}\nabla_{\tau}\left(\mathcal{E}R_{\phantom{\lambda}\alpha\phantom{\tau}\beta}^{\lambda\phantom{\alpha}\tau}\right)+\nabla_{\rho}\nabla_{\gamma}\left(\mathcal{E}R_{\beta\phantom{\rho}\alpha}^{\phantom{\beta}\rho\phantom{\alpha}\gamma}\right)-\nabla_{\pi}\nabla_{\lambda}\left(\mathcal{E}R_{\beta\phantom{\pi\lambda}\alpha}^{\phantom{\beta}\pi\lambda}\right)+\nabla_{\lambda}\nabla_{\pi}\left(\mathcal{E}R_{\phantom{\lambda}\alpha\phantom{\pi}\beta}^{\lambda\phantom{\alpha}\pi}\right)\right]\nonumber \\
+c_{7}\left[-\mathcal{E}_{\alpha\nu}R_{\beta\sigma\rho\tau}R^{\nu\sigma\rho\tau}+\mathcal{E}_{\nu}^{\mu}R_{\mu\beta\rho\tau}R_{\phantom{\nu}\alpha}^{\nu\phantom{\alpha}\rho\tau}+\mathcal{E}_{\nu}^{\mu}R_{\mu\sigma\alpha\tau}R_{\phantom{\nu\sigma}\beta}^{\nu\sigma\phantom{\beta}\tau}+\mathcal{E}_{\nu}^{\mu}R_{\mu\sigma\rho\alpha}R_{\phantom{\nu\sigma\rho}\beta}^{\nu\sigma\rho}\right.\nonumber \\
\left.+2\nabla_{\lambda}\nabla_{\pi}\left(\mathcal{E}_{\alpha}^{\mu}R_{\mu\phantom{\lambda}\beta}^{\phantom{\mu}\lambda\phantom{\beta}\pi}\right)+2\nabla_{\lambda}\nabla_{\pi}\left(\mathcal{E}_{\beta}^{\mu}R_{\mu\alpha}^{\phantom{\mu\alpha}\lambda\pi}\right)-2\nabla_{\lambda}\nabla_{\pi}\left(\mathcal{E}^{\lambda\mu}R_{\mu\alpha\beta}^{\phantom{\mu\alpha\beta}\pi}\right)\right] & =0,
\end{align}
where $\mathcal{E}=g_{\mu\nu}\mathcal{E}^{\mu\nu}$.

For the sake of comparison with the equivalent linear action technique
let us find the maximally symmetric vacuum using the field equations.
Note that $\bar{\mathcal{E}}_{\beta}^{\alpha}=\bar{e}\delta_{\beta}^{\alpha}$
with $\bar{e}=\frac{\gamma}{2}\left(1+\bar{a}\right)$. In the calculations
below, all the tensor quantities are evaluated at the their background
values.

$c_{7}$ contribution:
\begin{align}
-\mathcal{E}_{\alpha\nu}R_{\beta\sigma\rho\tau}R^{\nu\sigma\rho\tau}+\mathcal{E}_{\nu}^{\mu}R_{\mu\beta\rho\tau}R_{\phantom{\nu}\alpha}^{\nu\phantom{\alpha}\rho\tau}+\mathcal{E}_{\nu}^{\mu}R_{\mu\sigma\alpha\tau}R_{\phantom{\nu\sigma}\beta}^{\nu\sigma\phantom{\beta}\tau}+\mathcal{E}_{\nu}^{\mu}R_{\mu\sigma\rho\alpha}R_{\phantom{\nu\sigma\rho}\beta}^{\nu\sigma\rho}= & \frac{4\Lambda^{2}}{3}\bar{e}\bar{g}_{\alpha\beta}.
\end{align}

$c_{6}$ contribution:
\begin{align}
-\mathcal{E}R_{\phantom{\alpha}\rho\tau\gamma}^{\alpha}R^{\beta\rho\tau\gamma}+\mathcal{E}R_{\sigma\beta}^{\phantom{\sigma\beta}\rho\pi}R_{\phantom{\sigma}\alpha\rho\pi}^{\sigma}+\mathcal{E}R_{\sigma\phantom{\tau}\alpha}^{\phantom{\sigma}\tau\phantom{\alpha}\pi}R_{\phantom{\sigma}\tau\beta\pi}^{\sigma}+\mathcal{E}R_{\sigma\phantom{\gamma\pi}\alpha}^{\phantom{\sigma}\gamma\pi}R_{\phantom{\sigma}\gamma\pi\beta}^{\sigma}= & \frac{16\Lambda^{2}}{3}\bar{e}\bar{g}_{\alpha\beta}.
\end{align}

$c_{5}$ contribution:
\begin{align}
-\mathcal{E}_{\alpha\nu}R_{\beta\sigma}^{\phantom{\beta\sigma}\nu\rho}R_{\rho}^{\sigma}+\mathcal{E}_{\beta}^{\mu}R_{\mu\sigma\alpha}^{\phantom{\mu\sigma\alpha}\rho}R_{\rho}^{\sigma}+\mathcal{E}_{\nu}^{\mu}R_{\mu\sigma\phantom{\nu}\beta}^{\phantom{\mu\sigma}\nu}R_{\alpha}^{\sigma}+\mathcal{E}_{\nu}^{\mu}R_{\mu\alpha}^{\phantom{\mu\sigma}\nu\rho}R_{\beta\rho}= & 2\Lambda^{2}\bar{e}\bar{g}_{\alpha\beta}.
\end{align}

$c_{4}$ contribution:
\begin{equation}
\left(\mathcal{E}_{\nu}^{\mu}R_{\alpha\mu}R_{\phantom{\nu}\beta}^{\nu}+\mathcal{E}_{\alpha}^{\mu}R_{\phantom{\sigma}\mu}^{\sigma}R_{\beta\sigma}\right)=2\bar{e}R_{\sigma\beta}R_{\alpha}^{\sigma}=2\Lambda^{2}\bar{e}\bar{g}_{\alpha\beta}.
\end{equation}

$c_{3}$ contribution:
\begin{equation}
8\bar{e}R_{\alpha\rho}R_{\beta}^{\rho}=8\Lambda^{2}\bar{e}\bar{g}_{\alpha\beta}.
\end{equation}

$c_{2}$ contribution:
\begin{equation}
\mathcal{E}_{\nu}^{\mu}R_{\mu}^{\nu}R_{\alpha\beta}+\mathcal{E}_{\beta}^{\mu}RR_{\alpha\mu}=2\bar{e}RR_{\alpha\beta}=8\Lambda^{2}\bar{e}\bar{g}_{\alpha\beta}.
\end{equation}

$c_{1}$ contribution:
\begin{equation}
2\mathcal{E}RR_{\alpha\beta}=32\Lambda^{2}\bar{e}\bar{g}_{\alpha\beta}.
\end{equation}
 Using the conversion relations between the bases (\ref{conversion_relations})
and 
\begin{equation}
\gamma\bar{A}_{\sigma}^{\rho}\equiv\delta_{\sigma}^{\rho}\bar{a}=\delta_{\sigma}^{\rho}\lambda\left[1+\lambda\left(a_{3}+b_{2}\right)\right],
\end{equation}
and after defining $c\equiv a_{3}+b_{2}$, 
\begin{equation}
\bar{e}=\frac{\gamma}{2}\left(1+\lambda+c\lambda^{2}\right),
\end{equation}
one arrives at 

\begin{equation}
c^{2}\lambda^{4}+c\lambda^{3}-\lambda+\lambda_{0}=0,
\end{equation}
which is the same as the one found with ELA. Here $\bar{a}$ is defining
in (\ref{a_bar}).


\begin{thebibliography}{References}
\bibitem{Deser_Gibbons} S.~Deser and G.~W.~Gibbons, \emph{``Born-Infeld-Einstein
actions?,''} Class.\ Quant.\ Grav.\ \textbf{15}, L35 (1998).

\bibitem{Eddington} A.~Eddington, \emph{The Mathematical Theory
of General Relativity} (Cambridge University Press, Cambridge, England,
1924).

\bibitem{BI} M.~Born and L.~Infeld, \emph{``Foundations of the
new field theory,''} Proc.\ Roy.\ Soc.\ Lond.\ A \textbf{144},
425 (1934).

\bibitem{Schrodinger} E.~Schrödinger, \emph{Space-Time Structure}
(Cambridge University Press, Cambridge, England, 1960).

\bibitem{Banados_Eddington} M.~Banados and P.~G.~Ferreira, \emph{``Eddington's
theory of gravity and its progeny,''} Phys.~Rev.~Lett. \textbf{105},
011101 (2010).

\bibitem{Fiorini} F.~Fiorini, \emph{``Nonsingular Promises from
Born-Infeld Gravity,}'' Phys.~Rev.~Lett. \textbf{111}, 041104 (2013).

\bibitem{Delsate_Steinhoff} T.~Delsate and J.~Steinhoff, \emph{``New
insights on the matter-gravity coupling paradigm,''} Phys.~Rev.~Lett.
\textbf{109}, 021101 (2012).

\bibitem{Heisenberg} J.~B.~Jiménez, L.~Heisenberg and G.~J.~Olmo,
\emph{``Infrared lessons for ultraviolet gravity: the case of massive
gravity and Born-Infeld,''} arXiv:1409.0233 {[}hep-th{]}.

\bibitem{Ferreira} P.~P.~Avelino and R.~Z.~Ferreira, \emph{``Bouncing
Eddington-inspired Born-Infeld cosmologies: An alternative to inflation?,''
}Phys.\ Rev.\ D, \textbf{86}, 041501 (2012).

\bibitem{Gullu-BINMG} I.~Gullu, T.~C.~Sisman, B.~Tekin, \emph{``Born-Infeld
extension of new massive gravity,''} Class.\ Quant.\ Grav.\ \textbf{27},
162001 (2010).

\bibitem{Gullu-cfunc} I.~Gullu, T.~C.~Sisman and B.~Tekin, \emph{``c-functions
in the Born-Infeld extended New Massive Gravity,''} Phys.\ Rev.\ D
\textbf{82}, 024032 (2010).

\bibitem{Gullu-UniBI} I.~Gullu, T.~C.~Sisman and B.~Tekin, \emph{``Unitarity
analysis of general Born-Infeld gravity theories,''} Phys.\ Rev.\ D
\textbf{82}, 124023 (2010).

\bibitem{Gullu-Horava} I.~Gullu, T.~C.~Sisman and B.~Tekin, \emph{``Born-Infeld-Horava
gravity,''} Phys.\ Rev.\ D \textbf{81}, 104018 (2010).

\bibitem{Nam} S.~Nam, J.~D.~Park and S.~H.~Yi, \emph{``AdS
Black Hole Solutions in the Extended New Massive Gravity,}'' JHEP
\textbf{1007}, 058 (2010).

\bibitem{BHT-PRL} E.~Bergshoeff, O.~Hohm and P.~K.~Townsend,
\emph{``Massive Gravity in Three Dimensions,''} Phys.\ Rev.\ Lett.,
\textbf{102}, 201301 (2009).

\bibitem{BHT-PRD} E. Bergshoeff, O. Hohm and P. K. Townsend, \emph{``More
on Massive 3D Gravity,''} Phys.\ Rev.\ D \textbf{79}, 124042 (2009).

\bibitem{Gullu-Canonical} I.~Gullu, T.~C.~Sisman and B.~Tekin,
\emph{``Canonical Structure of Higher Derivative Gravity in 3D,''}
Phys.\ Rev.\ D \textbf{81}, 104017 (2010).

\bibitem{Sinha} A.~Sinha, \emph{``On the new massive gravity and
AdS/CFT,''} JHEP \textbf{1006}, 061 (2010).

\bibitem{Paulos} M.~F.~Paulos, \emph{``New massive gravity extended
with an arbitrary number of curvature corrections,''} Phys.\ Rev.\ D
\textbf{82}, 084042 (2010).

\bibitem{Jatkar} D.~P.~Jatkar and A.~Sinha, \emph{``New Massive
Gravity and ${\rm AdS}_{4}$ counterterms,''} Phys.\ Rev.\ Lett.\ \textbf{106},
171601 (2011).

\bibitem{Ozkan} E.~Bergshoeff and M.~Ozkan, ``\emph{3D Born-Infeld
Gravity and Supersymmetry,}'' JHEP \textbf{08},\textbf{ }149 (2014).

\bibitem{Gullu_n} I.~Gullu, T.~C.~Sisman and B.~Tekin,\emph{
}to appear soon.

\bibitem{Lovelock}  D.~Lovelock, \emph{``The Einstein tensor and
its generalizations,''} J.\ Math.\ Phys.\ \textbf{12}, 498 (1971).

\bibitem{Comelli} D.~Comelli, \emph{``Born-Infeld type gravity,''}
Phys.\ Rev.\ D\textbf{ 72}, 064018 (2005).

\bibitem{Yi} S.~-H.~Yi, \emph{``Born-Infeld Type Extension of
(Non-)Critical Gravity,''} Phys.~Rev.~D \textbf{86}, 024023 (2012).

\bibitem{Paulos_Tolley} M.~F.~Paulos and A.~J.~Tolley, \emph{``Massive
Gravity Theories and limits of Ghost-free Bigravity models,''} JHEP
\textbf{1209}, 002 (2012).

\bibitem{Sisman-Thesis} T.~C.~Sisman, ``\emph{Born-Infeld gravity
theories in $D$-dimensions,}'' PhD thesis, METU, (2012).

\bibitem{Deser-Nepo} S.~Deser and R.~I.~Nepomechie, \emph{``Gauge
Invariance Versus Masslessness in De Sitter Space,''} Annals Phys.\ \textbf{154},
396 (1984).

\bibitem{Hindawi} A.~Hindawi, B.~A.~Ovrut and D.~Waldram, \emph{``Nontrivial
vacua in higher derivative gravitation,''} Phys.\ Rev.\ D \textbf{53},
5597 (1996).

\bibitem{Gullu_Tekin} I.~Gullu and B.~Tekin, \emph{``Massive Higher
Derivative Gravity in D-dimensional Anti-de Sitter Spacetimes,''}
Phys.~Rev.~D \textbf{80}, 064033 (2009).

\bibitem{Sisman-AllUniD} T.~C.~Sisman, I.~Gullu and B.~Tekin,
\emph{``All unitary cubic curvature gravities in D dimensions,''}
Class.\ Quant.\ Grav.\ \textbf{28}, 195004 (2011).

\bibitem{Metsaev_Tseytlin} R.~R.~Metsaev and A.~A.~Tseytlin,
``\emph{Curvature Cubed Terms in String Theory Effective Actions,}''
Phys. Lett. B \textbf{185}, 52 (1987).

\bibitem{Gullu-AllUni3D} I.~Gullu, T.~C.~Sisman, B.~Tekin, \emph{``All
Bulk and Boundary Unitary Cubic Curvature Theories in Three Dimensions,''}
Phys.\ Rev.\ D\textbf{ 83}, 024033 (2011).

\bibitem{Senturk} C.~Senturk, T.~C.~Sisman and B.~Tekin, \emph{``Energy
and Angular Momentum in Generic F(Riemann) Theories,''} Phys.\ Rev.\ D
\textbf{86}, 124030 (2012).

\bibitem{Abbott-Deser} L.~F.~Abbott and S.~Deser, ``\emph{Stability
of Gravity with a Cosmological Constant,}'' Nucl. Phys. B \textbf{195},
76 (1982).

\bibitem{GursesPRL} M.~Gurses, S.~Hervik, T.~C.~Sisman, and B.~Tekin,
``\emph{Anti--de Sitter-Wave Solutions of Higher Derivative Theories},''
Phys.\ Rev.\ Lett.\ \textbf{111}, 101101 (2013).

\bibitem{Gurses-Spherical} M.~Gurses, T.~C.~Sisman, and B.~Tekin,
``\emph{New Exact Solutions of Quadratic Curvature Gravity},'' Phys.\ Rev.\ D
\textbf{86}, 024009 (2012).

\bibitem{Gurses-AdSandpp} M.~Gurses, T.~C.~Sisman, and B.~Tekin,
``\emph{AdS-plane wave and pp-wave solutions of generic gravity theories},''
arXiv:1407.5301 {[}hep-th{]}.

\bibitem{Deser_Tekin-PRL} S.~Deser and B.~Tekin, \emph{``Gravitational
energy in quadratic curvature gravities,''} Phys.\ Rev.\ Lett.\ \textbf{89},
101101 (2002).

\bibitem{Deser_Tekin} S.~Deser and B.~Tekin, \emph{``Energy in
generic higher curvature gravity theories,''} Phys.\ Rev.\ D\textbf{
67}, 084009 (2003).

\bibitem{Dengiz} S.~Dengiz and B.~Tekin, \emph{``Higgs Mechanism
for New Massive Gravity and Weyl Invariant Extensions of Higher Derivative
Theories,''} Phys.\ Rev.\ D \textbf{84}, 024033 (2011).

\bibitem{Dengiz_Tanhayi} M.~R.~Tanhayi, S.~Dengiz and B.~Tekin,
``\emph{Weyl-Invariant Higher Curvature Gravity Theories in n Dimensions,}''
Phys. Rev. D \textbf{85}, 064016 (2012).

\bibitem{Stelle} K.~S.~Stelle, \emph{``Renormalization of Higher
Derivative Quantum Gravity,''} Phys.\ Rev.\ D\textbf{ 16}, 953
(1977).\end{thebibliography}
\end{document}